\documentclass[a4paper,fleqn,usenatbib]{mnras}

\usepackage{mathptmx}

\usepackage[T1]{fontenc}
\usepackage{ae,aecompl}

\usepackage{graphicx}	
\usepackage{amsmath}	
\usepackage{amssymb}	

\def\hpc{$h^{-1}$Mpc }

\def\hpcv{$ h^{-1}$Mpc, }

\def\mpcoh{\,h^{-1}{\rm Mpc}}
\def\kms{\,{\rm km\, s}^{-1}}
\def\dpmos{DP\textsc{mock }}
\def\dsmos{DS\textsc{mock }}

\def\drmos{DREF\textsc{mock }}
\def\dpmo{DP\textsc{mock}}
\def\dsmo{DS\textsc{mock}}

\def\drmo{DREF\textsc{mock}}
\def\dpmosp{DP\textsc{mocks }}
\def\dsmosp{DS\textsc{mocks }}
\def\rawdsmosp{\textsc{raw}DS\textsc{mocks }}
\def\drmosp{DREF\textsc{mocks }}
\def\dpmop{DP\textsc{mocks}}
\def\dsmop{DS\textsc{mocks}}
\def\rawdsmop{\textsc{raw}DS\textsc{mocks}}
\def\drmop{DREF\textsc{mocks}}

\title[Environment with photometry ad spectroscopy]{Measuring galaxy environment with the synergy of future photometric and spectroscopic surveys}

\author[O. Cucciati et al.]{O. Cucciati,$^{1,2}$\thanks{E-mail: olga.cucciati@oabo.inaf.it} 
F. Marulli,$^{1,2,3}$ 
A. Cimatti,$^{2}$ 
A.I. Merson,$^{4}$ 
P. Norberg,$^{5}$ \newauthor  
L. Pozzetti,$^{1}$ 
C.M. Baugh$^{5}$
and E.Branchini$^{6,7,8}$ \\
$^{1}$INAF-Osservatorio Astronomico di Bologna, via Ranzani 1, 40127 Bologna, Italy\\
$^{2}$Dipartimento di Fisica e Astronomia - Universit\'a di Bologna, viale Berti Pichat 6/2, I-40127 Bologna, Italy\\
$^{3}$INFN, Sezione di Bologna, viale Berti Pichat 6/2, I-40127 Bologna, Italy\\
$^{4}$Department of Physics \& Astronomy, University College London, London, United Kingdom, WC1E 6BT\\
$^{5}$ICC, Department of Physics, Durham University, South Road, Durham DH1 3LE, UK\\
$^{6}$Dipartimento di Matematica e Fisica, Universit\'a degli Studi Roma Tre, via della Vasca Navale 84, 00146, Roma, Italy\\
$^{7}$INFN, Sezione di Roma Tre, via della Vasca Navale 84, I-00146 Roma, Italy\\ 
$^{8}$INAF - Osservatorio Astronomico di Roma, via Frascati 33, I-00040 Monte Porzio Catone (RM), Italy
}

\date{Accepted XXX. Received YYY; in original form ZZZ}

\pubyear{2016}

\begin{document}
\label{firstpage}
\pagerange{\pageref{firstpage}--\pageref{lastpage}}
\maketitle

\begin{abstract}
  We exploit the synergy between low-resolution spectroscopy and
  photometric redshifts to study environmental effects on galaxy
  evolution in slitless spectroscopic surveys from space.  As a test
  case, we consider the future Euclid Deep survey ($\sim40$deg$^2$), which
  combines a slitless spectroscopic survey limited at H$\alpha$ flux
  $\geq5\times 10^{-17}$ erg cm$^{-2}$ s$^{-1}$ and a photometric
  survey limited in H-band ($H\leq26$). We use Euclid-like galaxy mock
  catalogues, in which we anchor the photometric redshifts to the 3D
  galaxy distribution of the available spectroscopic redshifts. We
  then estimate the local density contrast by counting objects in
  cylindrical cells with radius from 1 to 10 \hpcv over the
  redshift range $0.9<z<1.8$.  We compare this density field with the
  one computed in a mock catalogue with the same depth as the Euclid
  Deep survey (H$=26$) but without redshift measurement errors. We find
  that our method successfully separates high from low density
  environments (the last from the first quintile of the density
  distribution), with higher efficiency at low redshift and
  large cells: the fraction of low density regions mistaken by high
  density peaks is $<1\%$ for all scales and redshifts explored, but
  for scales of 1 \hpc for which is a few percent. These
  results show that we can efficiently study environment in
  photometric samples if spectroscopic information is available for a
  smaller sample of objects that sparsely samples the same volume.  We
  demonstrate that these studies are possible in the Euclid Deep
  survey, i.e. in a redshift range in which environmental effects are
  different from those observed in the local universe, hence providing
  new constraints for galaxy evolution models.
\end{abstract}




\section{Introduction}\label{intro}

Environment is known to play a role in galaxy evolution, 
  especially at relatively low redshift ($z \lesssim 1$) and for galaxies
  with small to intermediate stellar mass ($\log(\mathcal{M}/\mathcal{M}_\odot)\lesssim 10.5$, see
  e.g. \citealp{bolzonella10} and \citealp{davidzon15}). There
are still many open debates on this subject, ranging from the relation
between the 3D galaxy distribution and the underlying dark matter (DM)
structures, to the actual physical processes that shape galaxy
properties, which take place in different environments and on
different time scales. Only through observations at different epochs
we can robustly derive a coherent picture of galaxy evolution.

Ideally, we
need to probe from large to small scales, over a large span in
redshift, and with large statistics, to allow us to robustly measure
galaxy properties in different environments.  Several physical
processes that shape galaxy properties take place in galaxy groups and
clusters (e.g. galaxy-galaxy merging, \citealp{toomre1972}, ram
pressure stripping of gas, \citealp{gunn_gott1972}, strangulation,
\citealp{larson1980}, harassment, \citealp{moore1996} and so on),
so the study of local, high-density environment on small scales is
crucial. Moreover, the cosmic web is continuously evolving with time,
and each galaxy can live in very different environments during its
life-time. As a consequence, we need to be able to study how
environment evolves to understand how it shapes
galaxy evolution, and at the same time we need to study how the
properties of the overall galaxy population change with time. For
instance, it would be important to understand if and how we can link
the peak of the star formation rate density ($z=1.5-2$, see
e.g. \citealp{cucciati12_SFRD} and \citealp{madau_dickinson14_CSFH})
with the onset of the relation between star formation rate (SFR) and
local density (\citealp{cucciati06} and \citealp{lin16} at $z\gtrsim1.2$, \citealp{elbaz2007} at
$z\sim1.1$, but see also \citealp{kodama07} and \citealp{spitler12}
for the identification of red massive galaxies in high redshift
clusters). Finally, we need large galaxy samples to remove the
degeneracies among the many parameters that regulate the galaxy
evolution processes (e.g. the complex interplay between local
environment, stellar mass and star formation, like for instance in
\citealp{peng2010_picture}, \citealp{McNaught_Roberts14_GAMA_LF} and
references therein).

In this framework the ability of identifying and isolate a
representative sample of high density structures is of
paramount importance. This would simultaneously require high accuracy
in the redshift measurement, large galaxy density and large
volumes. These requirements cannot be met by standard spectroscopic
surveys performed with current instruments, since the required
observation time would be prohibitively large. In addition, since we
aim to trace galaxy evolution, we should target galaxies at relatively
high redshift. To optimise the targeting of such galaxies using
multi-object spectrographs, some kind of target pre-selection in
near-IR bands is required, but ground-based observations in the
near-IR are limited by the atmospheric transmission windows.

A photometric galaxy survey would provide us with large statistics and
a large span in redshift within a much shorter time-scale of
observations. Many studies in the literature make use of photometric
redshifts to derive local environment. These analyses include galaxy
cluster identification (e.g.
\citealp{mazure07,adami10,bellagamba11,jian14_groups}), detailed studies
of extended structures (e.g., \citealp{guzzo2007_COSMOS,cassata07}),
and a broad analysis of the entire range of density enhancements, from
empty regions to the highest density peaks (see
e.g. \citealp{scoville07_env,scoville13_env,darvish15_env,lin16}).  The
price to pay is the much lower ($>1$ dex) redshift precision with respect to
spectroscopic redshifts, which hampers the precision of the 3D
reconstruction of local environment on small scales (e.g.,
\citealp{cooper2005,lai15_photoz} and \citealp{malavasi16}).

Alternatively, one could carry out narrow-band photometric surveys,
with filters aimed to target emission line galaxies in a very narrow
redshift bin (e.g. the HiZELS and NEWFIRM H$\alpha$ surveys, see
\citealp{geach08_HaLF} and \citealp{ly11}).  The main drawbacks of
this approach are the very limited redshift bin and the possible
contamination from other emission lines. 

Slitless spectroscopy of emission line galaxies (ELG) is a compromise
between redshift precision and being able to probe a large volume
(both wide and deep) in a relative short exposure time.  This kind
  of spectroscopy uses a prism or grism as dispersing element, so the
  spectral resolution can be very high. However, all the sources on
  the sky plane are spread out into their spectrum at once, and as a
  consequence the information on the angular position of the sources
  is limited and spectra can partially overlap. Spectra overlap might
  be significant especially in crowded fields, like deep samples, and
  in high density regions, that are important targets of environmental
  studies. The impact of this contamination can be reduced to some
  extent by taking observations of the same field at different
  position angles.

If a survey based on slitless spectroscopy is carried out from
space, the collected data do not suffer from atmospheric
absorption. This is the case, for instance, of the slitless surveys
performed with the HST, based on NICMOS \citep{mccarthy99_EL,
  yan99_HaLF, shim09_HaLF} and on the WFC3. The WFC3 Infrared
Spectroscopic Parallels program (WISP;
\citealp{atek10_WISP,atek11_WISP}) represents a major improvement with
respect to the previous programme based on NICMOS, given that the WFC3
field of view is much larger than that of NICMOS. These HST
observations will be followed, in the future, by slitless
spectroscopic surveys from two other space missions, the European
Space Agency's Euclid \citep{redbook} and the National Aeronautics and
Space Administration's WFIRST (Wide-Field Infrared Survey Telescope,
\citealp{green12_WFIRST,dressler12_WFIRST}).

Space-based slitless spectroscopy of ELGs thus represents a
potential new advancement in environmental studies, although it also
suffers from some drawbacks. For instance, the deeper the observations,
the greater number of spectra which will overlap on the image. At the same
time, keeping the depth fixed, in over-dense regions the sky is more
crowded than in low density regions, increasing the
contamination from adjacent spectra. A way to tackle this problem is
to repeat observations at the same position but with different
position angles, but this comes at the expense of a longer total
exposure time.

A promising way to study local environment on small scales over large
volumes is through the synergy of photometric and spectroscopic galaxy
samples.  For instance, one can anchor the photometric redshifts to a
robustly defined 3D skeleton, built with a (sub)sample of galaxies
with spectroscopic redshifts. A density estimator based on such a
method has been successfully developed for the zCOSMOS
survey \citep{lilly07_zcosmos} by
\cite{kovac2010_density}. \cite{cucciati14} applied a simplified
version of the same method to the VIPERS survey
\citep{guzzo14_vipers,garilli2014_VIPERS}.

The feasibility of such studies has to be evaluated case by case,
according to the selection function of the planned surveys.  For
instance, the zCOSMOS and VIPERS samples are both flux-limited at
$i=22.5$, but the zCOSMOS spectroscopic sample has a higher target
sampling rate, and the zCOSMOS ancillary photometric catalogue has a
smaller photometric redshift error.  For these two reasons, the
analysis performed with the zCOSMOS sample is robust down to smaller
scales than those reached with VIPERS.  The higher sampling rate of the
zCOSMOS survey comes at the price of a much smaller covered area, and
only with a survey as large as VIPERS we can investigate the
properties of rare objects with large statistical significance.

The future Euclid mission will provide the community with both
slitless spectroscopic observations and photometric observations of a
flux limited sample of galaxies.  The slitless spectroscopy will
sample ELGs, while the photometric observation will be limited in
$H-$band.  The improvements with respect to current ground-based
spectroscopic surveys at $z\lesssim1$ consist of larger and deeper
photometric and spectroscopic catalogues, that might allow
environmental studies up to $z\sim2$ on relatively small scales.

Here, we use galaxy mock catalogues which mimic the Euclid Deep
spectroscopic and photometric surveys to assess the possibility of
studying how environment affects galaxy evolution at high redshift by
exploiting the synergy between spectroscopic and photometric
samples. The approach relies on the ability of sampling high vs. low
density regions rather than reconstructing the whole density field and
its correlation properties as in the galaxy clustering studies. In
this work we shall therefore focus on the ability to discriminate high
density environments from the low density ones.

This paper is organised as follows. In Sect.~\ref{euclid} we present
the mock galaxy catalogues that we use in this work, and in
Sect.~\ref{mocks_comp} some of their properties, such as number counts
and their clustering strength. Sect.~\ref{density} describes how we
estimate the density field using both the photometric and
spectroscopic galaxy catalogues. Our results on the reliability of the
environmental parameterisation are detailed in
Sect.~\ref{reconstruction}, and we show a test case in
Sect.~\ref{clusters}. We summarise our results in
Sect.~\ref{summary}.

We use a cosmology based on a $\Lambda$CDM model with
$\Omega_m = 0.272$, $\Omega_{\Lambda}= 0.728$, $H_0=70.4\kms {\rm
  Mpc}^{-1}$, i.e. the cosmology in which our mock galaxy catalogues
are based. Magnitudes are in the AB system \citep{oke74}.


\section{A test case: the Euclid Deep survey}\label{euclid}

Several next-generation cosmological projects are conceived to
comprise both a spectroscopic and a photometric survey, so to
exploit two independent cosmological probes at once: the baryonic
acoustic oscillations (BAO) and weak gravitational lensing (WL),
respectively.

In this respect, such projects naturally provide the two datasets
required for our strategy: a spectroscopic and a photometric
sample. We will consider the Euclid mission as a realistic case to
illustrate the potential of next generation surveys in environmental
studies.

\subsection{The Euclid Deep survey}\label{euclid_survey}

Full details on the Euclid Mission surveys are given in
the Euclid Definition Study Report (\citealp{redbook}, ``Red Book''
from now on).  The pieces of information most relevant for this work
are the following.

The Euclid Mission comprises a Wide and a Deep survey, each of them
based on photometric and spectroscopic (slitless) observations. The Wide survey
will cover a sky area of 15000 deg$^2$, with a flux limit of $H=24$
($5\sigma$ point source) for the NIR photometry and a H$\alpha$ line
flux limit of $\sim3\times10^{-16}$ erg~cm$^{-2}$~s$^{-1}$
($3.5\sigma$ unresolved line flux) for the spectroscopy. Thanks also
to ancillary photometry from other surveys, it will be possible to
compute photometric redshifts ($z_p$) for the sources in the
photometric catalogue. The $z_p$ measurement error is required to be
at most $\sigma_{zp}=0.05(1+z)$, with a maximum 10\% of catastrophic
measurement failures.  The spectroscopic redshifts ($z_s$) of the
spectroscopic survey are required to have a maximum measurement error
of $\sigma_{zs}=0.001(1+z)$.

The Deep survey will be two magnitudes deeper than the wide one, but
will cover a much smaller sky area ($\sim40$ deg$^2$). 
Namely, the photometric deep survey will be flux limited at $H=26$
($5\sigma$ point source) and the H$\alpha$ flux limit for the
spectroscopic part will be $\sim5\times 10^{-17}$
erg~cm$^{-2}$~s$^{-1}$ ($3.5\sigma$ unresolved line flux).  The
required maximum values of photometric and spectroscopic redshift
errors are the same as for the Wide survey.

The upper value for $\sigma_{zp}$, i.e. $0.05(1+z)$, is granted
  by the use of the Euclid photometry in three NIR bands ($Y$, $J$,
  $H$) covering the wavelength range $0.92-2.0 \mu m$, and the use of
  at least four optical bands (from ground-based data, such as the `$g
  r i z$' filter set) covering the range $420-930 nm$. The depth of
  these observations needs to be at least 24 for the three NIR bands,
  and $g=24.4$, $r=24.1$, $i=24.1$ and $z=23.7$ in the optical. These
  requirements apply to the Euclid Wide survey. For the Deep survey,
  they need to be $\sim2$ magnitudes deeper. To keep $\sigma_{zp}$ as
  low as possible, it is also important that the photometry is uniform
  within a single field, and across several fields.  The requirement
  is to reach a relative photometric accuracy of 1.5\% for all
  the galaxies in the survey.

According to the Red Book, the total wavelength range covered by the
slitless spectroscopy is $1.1-2.0 ~\mu m$, and it will allow us the
detection of the H$\alpha$ line over the redshift range $0.7\lesssim z
\lesssim 2.05$. 

In this work, we focus our attention to the Deep
survey in the redshift range $0.9\leq z \leq 1.8$. Basically, the Deep
survey is conceived to help with the calibration of the Wide survey,
which will be used for one of the Euclid main cosmological
probes. Clearly, the Deep survey will also provide the community with
unprecedented data sets for ancillary science. We want to exploit the
full potentiality of such data in the field of environmental studies,
where it is important to probe small scales, best reached with deep
samples.

\begin{figure} \centering
\includegraphics[width=8cm]{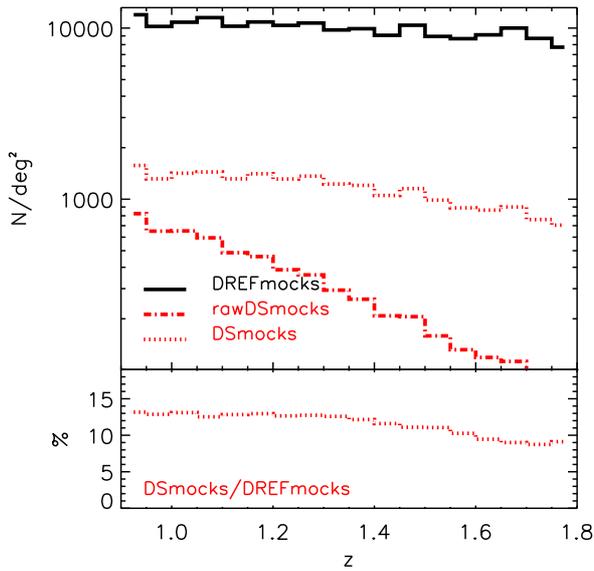}
\caption{{\it Top.} Redshift distribution of the galaxies in the
  \drmop, \rawdsmosp and \dsmop, as detailed in the legend, in bins of
  $\Delta z = 0.05$.  {\it Bottom.} Ratio of the redshift distribution
  of the \dsmosp over the \drmop, expressed in percentage.}
\label{nz_plot} 
\end{figure}

\begin{table*} 
\caption{List of the galaxy mock catalogues used in
this work, with their main properties: the limiting flux in $H$-band or
H$\alpha$ flux, the redshift error added to the peculiar velocities,
and other specific properties. For more details see
Sect.~\ref{mocks}.}  
\label{mocks_tab} 
\centering 
\begin{tabular}{l c c c c c} 
  \hline
  \hline 
  Name & sky area & $H$ band flux limit & H$\alpha$ flux limit & Redshift error & Other properties \\
  &  deg$^2$ &  [mag]                     &  [erg cm$^{-2}$ s$^{-1}$]   &    &  \\    
  \hline								   
  \drmosp  & $2\times2$ &  26  & - & none &   -\\
  \dpmosp  & $2\times2$ &  26  & - & $\sigma = 0.05(1+z_{pec})$ &   10\% of catastrophic failures\\
  \rawdsmosp  & $2\times2$ &  -  & $7\times 10^{-17}$ & $\sigma = 0.001(1+z_{pec})$ &   - \\
  \dsmosp  & $2\times2$ &  -  & $7\times 10^{-17}$ & $\sigma = 0.001(1+z_{pec})$ &   C=P=98\%. Adjusted $n(z)$. \\
  \hline 
\end{tabular} 
\end{table*}

\begin{figure} \centering
\includegraphics[width=\hsize]{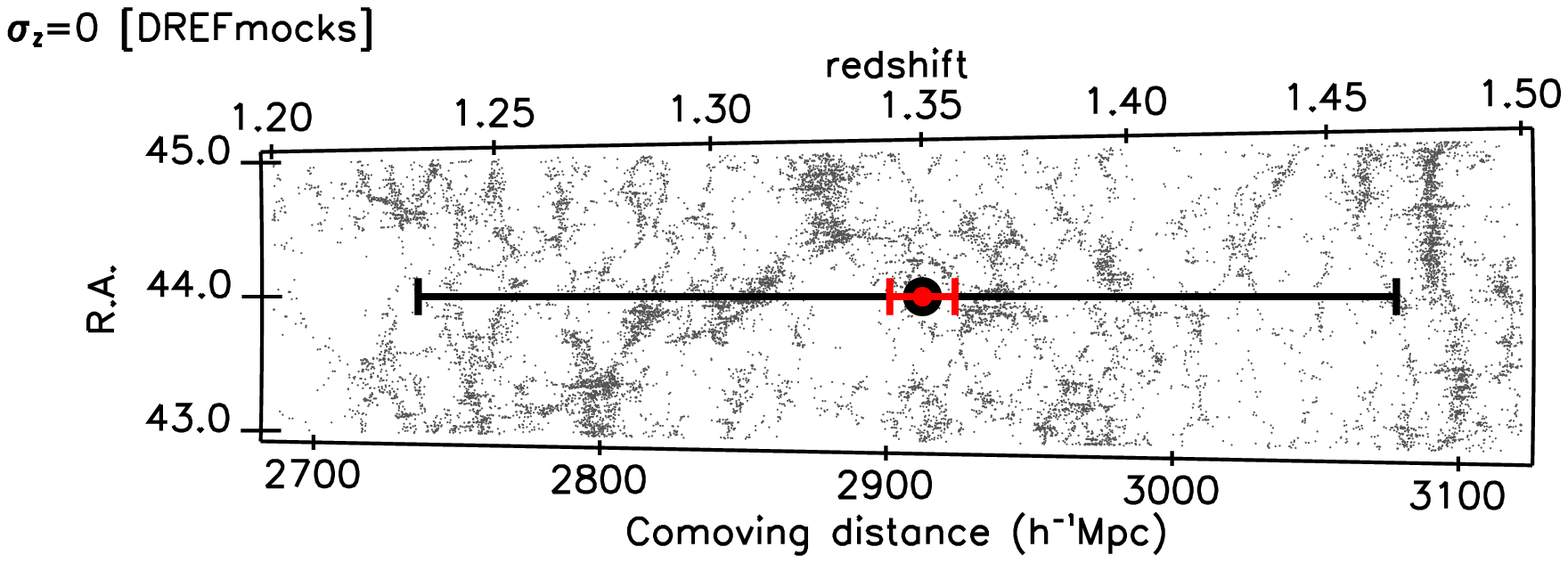}
\includegraphics[width=\hsize]{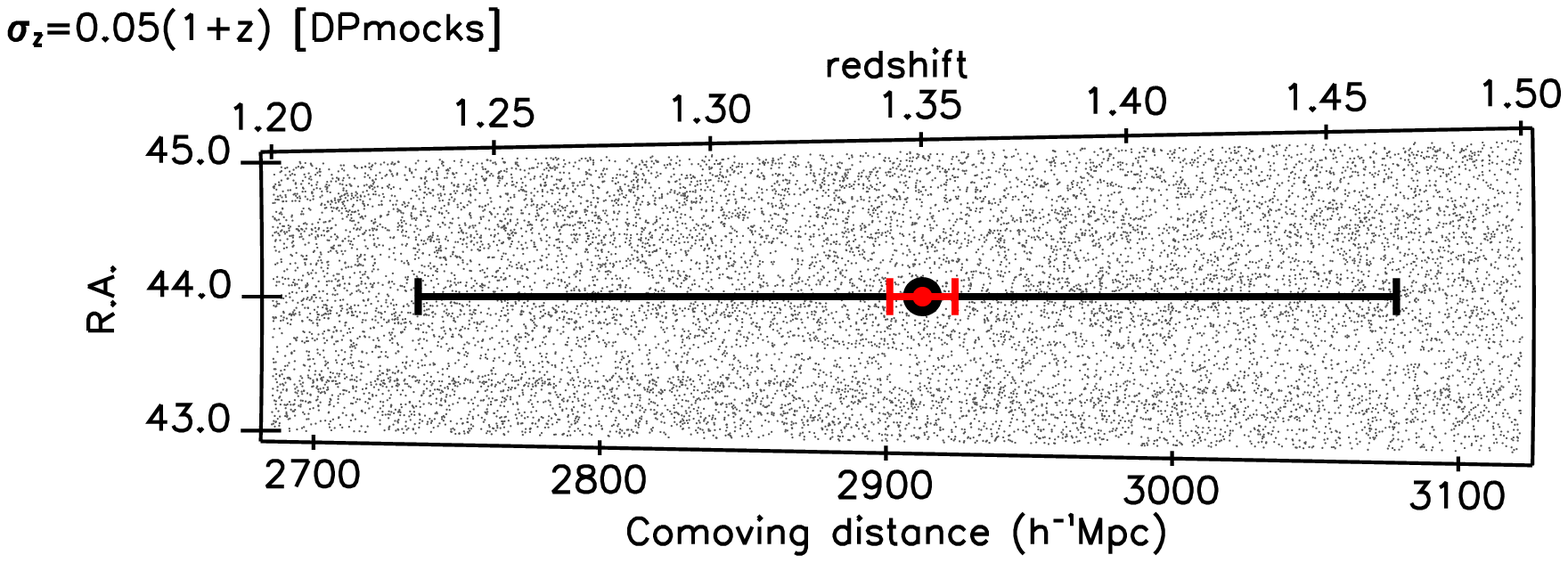}
\includegraphics[width=\hsize]{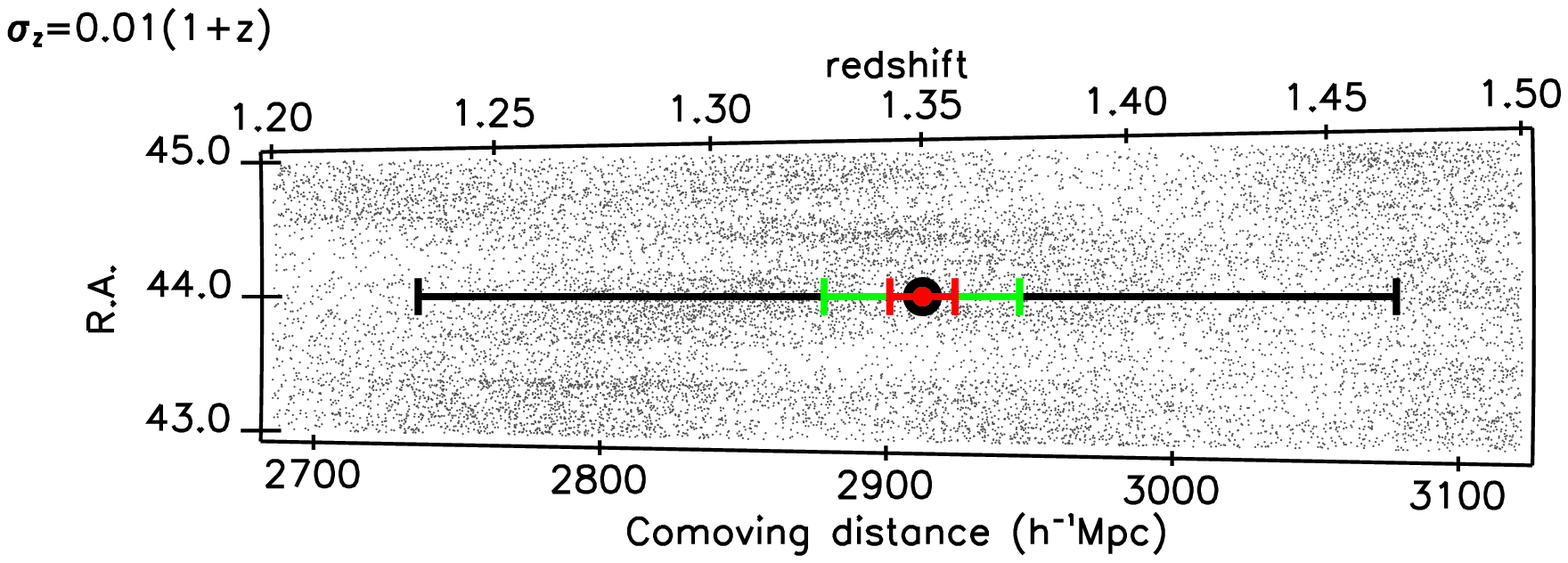}
\includegraphics[width=\hsize]{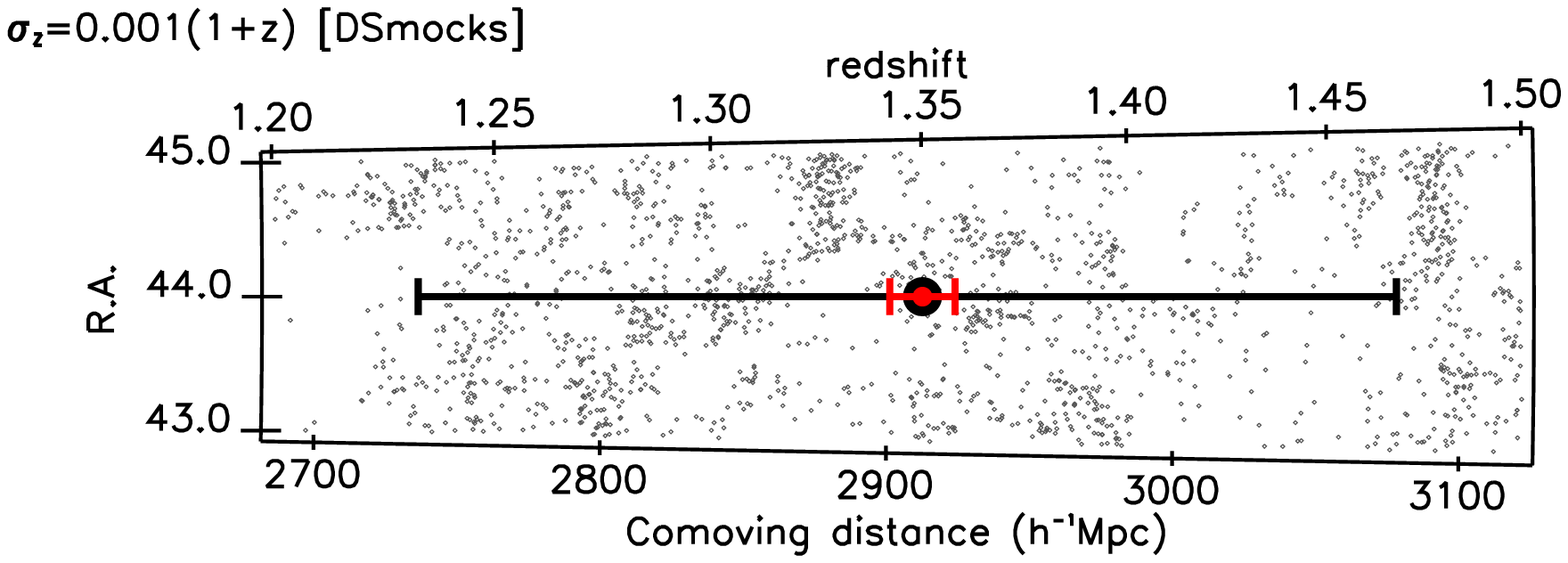}
\caption{Example of $R.A.-z$ distribution of galaxies in one of the lightcones
  used in this work, in the redshift range $1.2<z<1.5$. The projected
  $Dec$ covers a range of $0.5^\circ$. {\it Top:} all galaxies in the
  \drmos (flux limit at $H=26$, 100\% sampling rate, the redshift
  includes cosmological redshift and peculiar velocity). {\it
    Middle-top:} galaxies in the \dpmos obtained from the \drmos in
  the top panel. {\it Middle-bottom:} only for reference, in this
  panel we show the galaxies in the \drmos after adding a photometric
  redshift error of $\sigma = 0.01(1+z)$, i.e. five time less than the
  one we use in our \dpmo. {\it Bottom:} galaxies in the \dsmos
  extracted from the same lightcone as the \drmos in the top panel. In
  all panels, the black horizontal line represents the photometric
  redshift error $\pm\sigma = 0.05(1+z)$ and the red line the
  spectroscopic redshift error $\pm\sigma = 0.001(1+z)$, both
  evaluated at $z=1.35$. In the third panel, the green horizontal line
  represents a redshift error of $\pm\sigma = 0.01(1+z)$.}
\label{coni_deep} 
\end{figure}

\subsection{Mock galaxy samples}\label{mocks}

For our analysis, we use mock galaxy catalogues that mimic the Euclid
Deep spectroscopic and photometric surveys (their flux limit, their
redshift measurement error etc). We will refer to these catalogues as
\dsmosp and \dpmop, respectively, and also as Euclid-like
catalogues. To assess the robustness of our environment
reconstruction, we also need a catalogue as densely populated as the
photometric catalogue, but with no photometric redshift error. The
density estimated in this catalogue will be our reference computation. We call
these reference catalogues \drmop.

We use  mock galaxy samples built with the semi-analytical
model of galaxy formation and evolution described in
\cite{gonzalez_perez14_SAM}.  This model was applied to the dark
matter (DM) halo merging trees derived from a revision of the
Millennium Simulation \citep{springel2005_MILL} run with a cosmology
consistent with the 7 year results of the Wilkinson Microwave
Anisotropy Probe (WMAP7, \citealp{komatsu11_WMAP7}). Namely, the
adopted cosmological model is a $\Lambda$CDM model with $\Omega_m =
0.272$, $\Omega_b = 0.0455$, $h = 0.704$, $\Omega_{\Lambda}= 0.728$, $
n = 0.967$, and $\sigma_8 = 0.810$.  The Millennium Simulation
contains $N = 2160^3$ particles of mass $8.6 \times 10^8 h^{-1} M_
{\odot}$ within a comoving box of size 500 $h^{-1}$Mpc on a side.

The construction of the Euclid-like lightcone are fully described in
\cite{merson13_lightcones}. The lightcones are stored in the Virgo
Consortium
Database\footnote{http://galaxy-catalogue.dur.ac.uk:8080/Millennium/}
\citep{lemson06}. These lightcones contain the galaxy properties we
need for our study, i.e. the observed $H-$band magnitude and the
H$\alpha$ flux. For each galaxy, two redshifts are provided: the
cosmological redshift ($z_{cos}$) and a redshift that includes both
the cosmological redshift and the peculiar velocity ($z_{pec}$). We
are interested in the second one, because we perform our analysis in
redshift space. Two Euclid-like lightcones were available at the
  time of our analysis: one characterised by a wide area, with
  H$\alpha$ and $H-$band limits matching those of the Euclid Wide
  survey, and a deeper one, complete down to H$\alpha$ line flux of
  $\sim3\times10^{-18}$ erg~cm$^{-2}$~s$^{-1}$ and $H=27$ (meaning
  that it comprises all the galaxies which satisfy at least one of the
  two selection criteria), covering a smaller, circular region of
  $\sim20$deg$^2$. Note that the H$\alpha$ line flux and the $H-$band
  limits are both deeper than the selection criteria of the Euclid
  Deep survey. In this work we considered the latter lightcone, that
  we divided into 4 square-shaped independent lightcones of
  $2\times2$ deg$^2$, from which we extracted the mocks used in our
  analyses, as follows.

(i) Galaxy mock catalogues with a flux limit of $H=26$, mimicking the
magnitude limit of the Euclid Deep photometric survey. We consider
them as our reference catalogues, and we call them \drmop. These
catalogues represent the Euclid Deep photometric survey  without
photometric redshift error, and the density field estimated in these
catalogues (in redshift space) sets the reference to assess the
robustness of the density field reconstruction.

(ii) Galaxy mock catalogues with a flux limit of $H=26$, in which
galaxies have photometric redshifts ($z_p$). Namely, we took the
\drmosp and we added to the redshift $z_{pec}$ of each galaxy a random
value extracted from a Gaussian with $\sigma = 0.05(1 +z_{pec})$. We
also added 10\% of catastrophic redshift measurements, by assigning a
random redshift to 10\% of the galaxies (although this is an
simplistic approximation). From now on we call these mock catalogues
\dpmop.

(iii) Galaxy mock catalogues with an H$\alpha$ flux limit of $7\times
10^{-17}$ erg cm$^{-2}$ s$^{-1}$, mimicking the $5\sigma$ flux limit
of the Euclid Deep spectroscopic survey  (more conservative than
  the $3.5\sigma$ described in Sect.~\ref{euclid_survey}, see below
  for the reason of this choice).  Galaxies in this sample will have
spectroscopic redshifts ($z_s$) with a typical measurement error of
$\sigma = 0.001(1 +z)$. To mimic this uncertainty we added to the
redshift $z_{pec}$ of each galaxy a random value extracted from a
Gaussian with $\sigma = 0.001(1 +z_{pec})$. From now on we call these
mock catalogues \rawdsmop.

(iv) Galaxy mock catalogues with H$\alpha$ flux limit as close as
possible to $7\times 10^{-17}$ erg cm$^{-2}$ s$^{-1}$, but with
redshift distribution $n(z)$ adjusted to be consistent with
predictions. Specifically, we take the \rawdsmop, and we modify their $n(z)$
to match the $n(z)$ predicted by \citet[see
Sect.~\ref{mocks_comp_counts}]{pozzetti16_Ha}. This procedure basically
consists in adding in the \rawdsmosp also galaxies with H$\alpha$ flux
below $7\times 10^{-17}$ erg cm$^{-2}$ s$^{-1}$, as described in
details in Sect.~\ref{mocks_comp_counts}. We call these mock
catalogues \dsmop.

From now on, we will use the \drmosp as reference catalogues, and the
\dpmosp and \dsmosp as ``Euclid-like'' mock galaxy catalogues. We note
that we adopted a very simplistic approach to mimic the Euclid
selection function in these catalogues. In particular, it is
  important to explain some of the choices we made in the modelisation
  of our mock catalogues. 

The Euclid photometric survey will be characterised by masked sky
areas, variations of $\sigma_{zp}$ with galaxy luminosity, and so on,
but we do not take them into account.  A full characterisation of
  the photometric redshift probability distribution function (PDF) for
  the Euclid survey is still under study within the Euclid Consortium,
  so we can not mimic, in our mock catalogues, the full $z_P$ PDF nor
  the redshift distribution of the catastrophic $z_P$ errors. For this
  reason, we decided to use a very simplistic approach, assuming the
  PDF to be Gaussian and the catastrophic $z_P$ errors to be $10\%$ at
  all redshifts. This way, we do not take the risk of mimicking a
  sophisticated $z_p$ observing procedure that is not representative
  of the Euclid one.

The spectroscopic sample can be characterised by its
purity and completeness. The completeness, C, is defined as the
fraction of spectroscopic targets for which it is possible to measure
a redshift. The purity, P, is defined as the fraction of real targets
among all the measured redshifts (spurious features due to noise
etc. might be erroneously considered real H$\alpha$ lines).  The
spectroscopic Euclid Deep survey is designed, among other goals, to
help with the computation of C and P for the Wide survey, so its C and
P are foreseen to be very close to 100\% at least for galaxies
brighter than the Euclid Wide spectroscopic survey ($3\times10^{-16}$
erg~cm$^{-2}$~s$^{-1}$).  For galaxies with H$\alpha$ flux fainter
than the Wide survey limit, C and P have not yet been computed, and
they will possibly depend on redshift and H$\alpha$ flux.

For this reason, we decided to adopt, for our \dsmop, a H$\alpha$
flux limit of $7\times10^{-17}$ erg~cm$^{-2}$~s$^{-1}$, which
corresponds to the flux limit of the Deep survey at $\sim5\sigma$,
instead of the nominal one at $3.5\sigma$, and we consider the Deep
survey to have C=P$=98$\% down to the $\sim5\sigma$ flux limit. We
mimicked C and P by removing 2\% of the galaxies (randomly chosen,
irrespectively of their position or flux), and adding a corresponding
number of fake objects (randomly placed in the survey volume).

We verified the robustness of the density field reconstruction in the
case of a brighter H$\alpha$ flux limit for the \dsmop, to understand
what would change in case our assumption of C=P$=98$\% down to
$7\times10^{-17}$ erg~cm$^{-2}$~s$^{-1}$ is too optimistic. In
Appendix \ref{zade_req} we show the results on the density field
reconstruction if we use \dsmosp with H$\alpha$ flux $>10^{-16}$
erg~cm$^{-2}$~s$^{-1}$, and we show that a brighter limit would affect only
marginally our results.

In Table \ref{mocks_tab} we summarise the main properties of the
galaxy mock catalogues listed above.  In Fig.~\ref{nz_plot} we show
the redshift distribution $n(z)$ of our mock catalogues. The density
of galaxies in the \dsmosp ranges between $\lesssim15$\% to $\sim10$\%
of the density of galaxies in the \dpmop, going from $z=0.9$ to
$z=1.8$. We note that, in the redshift range of interest
($0.9<z<1.8$), basically all the galaxies in the \dsmosp also belong
to the \dpmop.

Figure \ref{coni_deep} shows the $R.A.-z$ distribution of galaxies in
one of the lightcones used in this work, in the redshift range
$1.2<z<1.5$. For the given lightcone, we show how galaxies are
distributed in the \drmo, \dpmo, and \dsmo. It is clear that the large
photometric redshift error in the \dpmos smears out the the small- and
large-scale structures visible in the \drmo. For a comparison, in
Fig.~\ref{coni_deep} we also show what would happen to the \drmos when
adding a photometric redshift error of $\sigma_{zp}=0.01(1+z)$, that
roughly corresponds to one of the smallest photometric redshift errors
obtained with real data to date (see e.g. \citealp{ilbert13} and
\citealp{ilbert15} in the COSMOS field, who find
$\sigma_{zp}=0.008(1+z)$ for galaxies with $i<22.5$). At $z\sim1.3$ as
in the Figure, $\sigma_{zp}=0.01(1+z)$ corresponds roughly to the
scale of voids/underdense regions, as  is evident from comparing the
green line in the third panel with the typical dimension of the empty
areas in the top panel. This way, the galaxies in the most over-dense
structures do not mix too much when the photometric redshift error is
added, and it is still possible to identify high density regions (see
also \citealp{malavasi16}).

\begin{figure} \centering
\includegraphics[width=\hsize]{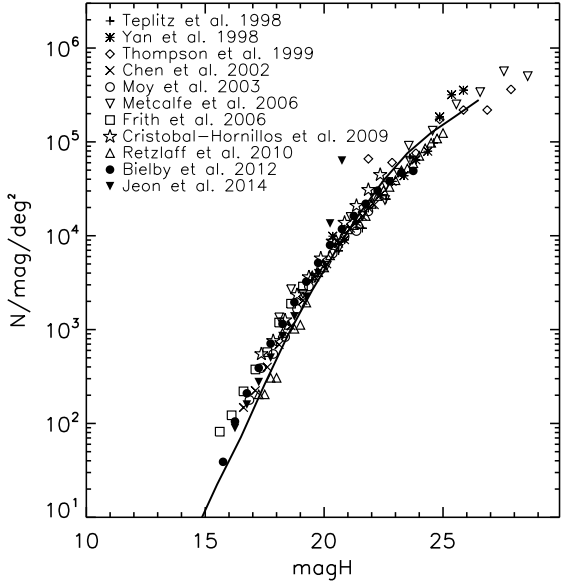} 
\caption{$H-$band number counts in our lightcones (black solid line)
  and in some observational data (symbols as in the legend).}
\label{Hband_counts} 
\end{figure}


\section{Properties of the mock samples}\label{mocks_comp}

In this section we study the number counts, the redshift distribution
$n(z)$ and the clustering strength of the photometric and
spectroscopic mock catalogues and we compare them with available data
in the literature. This comparison is important, since the consistency
between real and mock samples is a key to guarantee the reliability of
our environmental study forecast.

\subsection{Number counts and redshift
distributions}\label{mocks_comp_counts}

Figure \ref{Hband_counts} shows the $H$-band number counts
(N/deg$^{2}$/mag) in our lightcones, compared with some observational
data
\citep{teplitz98_Hband,yan99_HaLF,thompson99_Hband,chen02_Hband,moy03_Hband,metcalfe06_Hband,
  frith06_Hband,cristobal09_Hband,retzlaff10_Hband,bielby12_Hband,jeon14_Hband}. The
counts in the mock catalogues are consistent with the observed galaxy
counts in the range $19\lesssim H \lesssim 23$. Below $H\sim 19$,
predicted counts tend to underestimated the observed ones, up to a
factor of $\lesssim2$ at the brightest magnitudes, but this lack of
objects is mainly related to galaxies at $z\lesssim0.8$, a redshift
range that is not studied in our work. The counts in the mock
catalogues also seem to overestimate the observed counts at
$H\gtrsim23$ by a factor of $\sim50$\%, but at these faint magnitudes
the scatter in the real data is large. Overall, there is a fair
agreement between the counts in the lightcones and in the real data.

The H$\alpha$ counts are not well constrained yet by observations, and
yet the assessment of reliable forecasts of the H$\alpha$ number
counts and redshift distribution is a crucial task for the preparation
of the observational strategy for future missions like Euclid and
WFIRST.  We expect a large number of H$\alpha$ emitters in the
redshift range explored by such missions, given that the cosmic star
formation rate (SFR) was higher in the past,  with a peak at
$z\sim2$ (see e.g. \citealp{cucciati12_SFRD} and
\citealp{madau_dickinson14_CSFH}).  Moreover, the use of star forming
galaxies to study the small scale environment at high redshift, like
in our case, could be particularly effective if it is true (though the
debate is still open in the literature) that SF galaxies reside
preferentially in high densities at $z\gtrsim1-1.5$ \citep{cucciati06,elbaz2007,lin16}.

Several H$\alpha$ samples have been collected in the past years, using
ground-based spectroscopy, grism spectroscopy from space and
narrow-band NIR photometry. We refer the reader to
\cite{pozzetti16_Ha} for the most recent compilation of such samples.
Namely, \cite{pozzetti16_Ha} used these samples to derive three models
of the evolution of the H$\alpha$ luminosity function (LF), to compute
the forecasts for future surveys (see also \citealp{geach10_HaLF} for
a previous modelisation).  We adopt `Model 1' of
\citeauthor{pozzetti16_Ha} as the reference model in our analysis,
because it includes the evolution of both $L^*$ and $\phi^*$, while
$\phi^*$ does not evolve in their `Model 2', and their `Model 3' is
based on a shorter redshift range.

We have compared the $n(z)$ of the galaxies in the \rawdsmosp
($n_{mock}$) with the predicted H$\alpha$ $n(z)$ of `Model 1'
($n_{mod}$).  The \rawdsmosp underestimate the predicted $n(z)$ at all
redshifts (for the same flux limit, i.e. $7\times 10^{-17}$ erg
cm$^{-2}$ s$^{-1}$), by an increasing factor going from $z=0.9$ to
$z=1.8$. The \rawdsmosp contain 50\% (at $z=0.9$) to 20\% (at $z=1.8$)
of the galaxies predicted by Pozzetti's Model 1 (see their
Fig.6). This is true for all H$\alpha$ fluxes above the Euclid flux
limit (see their Figures 4 and 5). We remark that the Euclid
  H$\alpha$ flux limit, even for the Deep survey, corresponds to
  values very close to $L^*$ at $z\sim1.5$, so the underestimate of
  H$\alpha$ counts in the mock catalogues is not (only) due to a
  possibly not-well constrained faint-end slope of the observed
  H$\alpha$ LF (see Fig.~2 of \citealp{pozzetti16_Ha} for the typical
  uncertainty of $\alpha$ in the literature).

To reproduce the predicted H$\alpha$ $n(z)$, we
added galaxies in the \rawdsmosp until $n_{mock}(z)$ and $n_{mod}(z)$ were
the same. Specifically, we picked up these new galaxies from the lightcones from
which the \rawdsmosp were extracted, choosing randomly in $R.A.-Dec$
among the galaxies with H$\alpha$ slightly fainter than the flux limit
of the \rawdsmop. For each redshift bin $z_i$, we added fainter and
fainter galaxies until $n_{mock}(z_i)$ matched $n_{mod}(z_i)$. At the
end of the procedure, we had to add galaxies as faint as $2.5\times
10^{-17}$ erg~cm$^{-2}$~s$^{-1}$ at $z\sim1.0$, and even fainter
galaxies at higher redshifts, down to $1.5\times 10^{-17}$
erg~cm$^{-2}$~s$^{-1}$ at $z=1.8$. 

Following this procedure, we built the \dsmosp from the
\rawdsmop. These modified mock catalogues are those called \dsmosp in
Sect.~\ref{mocks}, in which we subsequently modelled the foreseen
purity and completeness of the Euclid Deep spectroscopic survey, as
described in Sect.~\ref{mocks}. We show the redshift distribution of
the \rawdsmosp and \dsmosp in Fig.~\ref{nz_plot}.

\begin{figure} \centering
\includegraphics[width=\hsize]{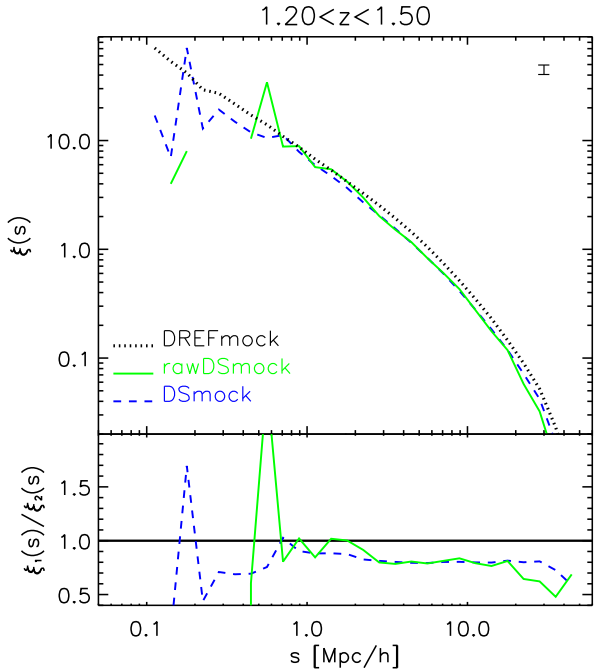}
\caption{{\it Top.} Redshift-space galaxy two-point correlation
  function in different mock catalogues, in the redshift bin
  $1.2<z<1.5$. Different colours and style are for different mock
  catalogues: the black dotted line is for \drmop, the green solid
  line for the \rawdsmop, and the dashed blue line for \dsmop. The
  error bar in the top right corner shows a difference in $\xi(s)$ of
  $\pm10$\%. {\it Bottom.} Ratio $\xi_1 / \xi_2$ between the
  redshift-space galaxy 2PCF in the different mock catalogues
  shown in the top panel. The denominator ($\xi_2$) is always for
  \drmop, and the numerator are the \rawdsmosp and \dsmosp catalogues
  (colour-code as in the top panel).}
\label{clustering_all} 
\end{figure}

\begin{figure} \centering
\includegraphics[width=\hsize]{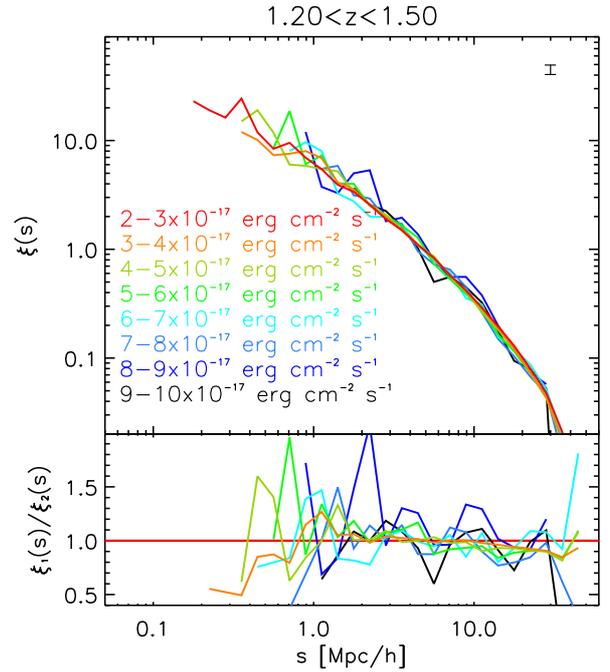} 
\caption{{\it Top.} Redshift-space galaxy two-point correlation
  function in the adopted lightcones (before applying any
  cut in $H-$band or H$\alpha$) for different bins of H$\alpha$ flux
  down to $2\times10^{-17}$ erg~cm$^{-2}$~s$^{-1}$, as in the labels,
  in the redshift bin $1.2<z<1.5$.  Different colours are for
  different H$\alpha$ flux bins. The error bar in the top right corner
  shows a difference in $\xi(s)$ of $\pm10$ \%. {\it Bottom.} Ratio
  $\xi_1 / \xi_2$ between the redshift-space galaxy 2PCF in the
  different mock catalogues shown in the top panel. The denominator
  ($\xi_2$) is always the 2PCF in the faintest flux bin
  ($\sim2-3\times 10^{-17}$ erg~cm$^{-2}$~s$^{-1}$), and the numerator
  is for the other flux bins in the top panel (colour-code as in the
  top panel).}
\label{clustering_bin} 
\end{figure}

\subsection{Two-point correlation function}\label{mocks_comp_clus}

Figure \ref{clustering_all} shows the galaxy two-point correlation
function (2PCF) in the \drmosp and \dsmop. We use \drmosp instead of
\dpmosp for this exercise because our aim is to show the (possibly)
different clustering strength for two samples selected in different
ways ($H-$band or H$\alpha$ flux) rather than to show how photometric
redshift errors would affect the 2PCF (we expect that the small
spectroscopic redshift error does not alter significantly the 2PCF in
the \dsmop). 

The 2PCF has been computed using the CosmoBolognaLib presented in
\cite{marulli16_CBL}. The random samples used to compute the 2PCF have
been built adopting the same geometrical selection of the \dpmosp and
\dsmop, and their specific $n(z)$. The figure shows that the galaxies
in the photometric catalogue are more clustered than those in the
spectroscopic sample, by a factor of $\sim20\%$.  This seems more
evident for scales $r<1 \mpcoh $.  We note that the two samples are
not necessarily expected to have the same clustering (see e.g. the
different clustering of galaxy populations with different stellar mass
or luminosity, like in
\citealp{li06,meneux08,coil08,christodoulou12_clustering,marulli2013_clustering}).

\cite{geach12_haXi} measured the 2PCF of H$\alpha$ emitters in the
Hi-Z Emission Line Survey (HiZELS) at $z\sim2.2$. The HiZELS sample is
complete down to $5\times 10^{-17}$ erg cm$^{-2}$ s$^{-1}$, that
corresponds to $L_{H\alpha} > 10^{42}$ erg cm$^{-2}$ s$^{-1}$ at z =
2.2. They compared their results with the 2PCF of a galaxy sample
extracted from the lightcones described in \cite{merson13_lightcones}.
The lightcones were built with the semi-analytical model of galaxy
formation and evolution described in \cite{lagos12_galform}, embedded
in the dark matter (DM) halo merging trees derived from the Millennium
Simulation \citep{springel2005_MILL}. The sample extracted from the
lightcones had the same selection function as the HiZELS sample.
\cite{geach12_haXi} find that the mock galaxy sample based on the
semi-analytical model, at the same redshift and $L_{H\alpha}$ as
HiZELS, has less clustering than the HiZELS sample on scales $r < 0.5
\mpcoh$, while the two samples have very similar clustering strength
at larger scales.

We measured the 2PCF in the the mock galaxy samples based on the
semi-analytical model of \cite{lagos12_galform}, as in
\citeauthor{geach12_haXi}, but in the same redshift ranges and for the
same H$\alpha$ flux limit as in our \dsmop. We verified that their
2PCF is very similar to the one in our \dsmop, that are based on the
semi-analytical model by \cite{gonzalez_perez14_SAM}. Even if the
redshift explored in \cite{geach12_haXi} is slightly higher than our
maximum redshift ($z=1.8$), their results suggest that the small-scale
clustering in our \dsmosp is too weak with respect to real data.

We remark that the clustering properties of the spectroscopic sample
are relevant for the method we use to estimate the density (see the
discussion in Sect.~\ref{zade} and Appendix \ref{zade_Hband}).  We
verified that the 2PCF of the fainter galaxies that we added in the
\rawdsmosp to match the predicted $n(z)$ is not too different from the
2PCF of the galaxies with H$\alpha$ flux $>7\times 10^{-17}$ erg
cm$^{-2}$ s$^{-1}$. This is shown in Fig.~\ref{clustering_all}, where
we also show the 2PCF for the \rawdsmop, and in
Fig.~\ref{clustering_bin}, where we show the 2PCF for different bins
of H$\alpha$ flux, including fluxes below $7\times 10^{-17}$ erg
cm$^{-2}$ s$^{-1}$. This implies that adding fainter galaxies to reach the
expected number counts does not alter the clustering strength of the
galaxies with brighter H$\alpha$ flux.


\section{Local density and environment}\label{density}

The key quantity that we use to quantify environmental dependencies is
the local density contrast of galaxies, $\delta_N$:
\begin{equation} \displaystyle 
 \delta_N = N/\langle N\rangle -1, 
\label{delta_eq} 
\end{equation}
where $N$ is the number of objects in the volume element, and $\langle
N\rangle$ the mean number of objects at a given redshift. Although
$\delta$ is defined as a `density contrast', we will often use simply
`density', for the sake of simplicity\footnote{For the purposes of
  this paper, the density contrast $\delta$ is simply the way we
  define the environment, we do not need to physically distinguish it
  from the local density.}, and we consider it as a proxy of
environment.

We count objects in cylindrical cells of different sizes, all with the
same half-length of $1000 \kms$ but different radii $R= 1, 2, 3, 5, 8,
10 \mpcoh$ ($R_1, R_2, R_3, R_5, R_8, R_{10}$ from now on). The
lengths of the radii are chosen to span from small scales (like
clusters) to relatively large scales.  Counts are performed by
randomly throwing cells to over-sample the survey volume. A
cylindrical shape of the cells allows us to adopt a symmetric shape on
the sky and an independent size along the line of sight, chosen to
take into account the peculiar velocities of cluster galaxies.

We only consider cells fully contained within the survey area, which
do not require any statistical correction for edge-induced
incompleteness. We will consider the problems related to boundaries
and gaps in the sky coverage in future work. We refer the reader to
e.g.  \cite{cucciati06} for an effective method for boundary
correction in recent spectroscopic surveys, and to \cite{cucciati14}
for some examples of gap-filling methods.

We call `tracer galaxies' (or simply `tracers') the galaxies used to
estimate the local density. The choice of a given tracer 
sample is a compromise between the maximisation of the
number density (the denser the sample, the smaller the scales that
can be reached in the density reconstruction) and the homogeneity of
the sample across the surveyed area and along the explored redshift range.

Both the spectroscopic and the photometric Euclid surveys are flux
limited. As a consequence the mean number density of objects decreases
with redshifts. This is a rather typical situation that, however,
does have an impact in environmental studies since it systematically
modifies the minimum scale-length that we can effectively probe. One
commonly adopted solution is to extract a sub-volume limited catalogue
of objects, for instance a luminosity-limited galaxy sample (see,
e.g., \citealp{cucciati10_zCOSMOS,McNaught_Roberts14_GAMA_LF}). This
has the advantage of providing a homogeneous sample of tracers with
the same mean separation at all redshifts. The drawback is to throw
away information at low redshifts (for instance, many faint galaxies
will not be used as tracers), hampering our ability to probe small
scales. 

In this work, we want to assess the full potential of the data set at
each redshift to investigate environmental effects, so we consider the
whole flux-limited sample. For this reason, we expect that we will be
able to better reconstruct the local density field on small scales at
lower redshift.

Our aim is to reconstruct the density field in the Euclid photometric
survey, minimising the effects of the large photometric redshift
error. For this we exploit the synergy of the photometric and
spectroscopic Euclid surveys. The first one enjoys a large density of
tracers with a large redshift measurement error, the second one is
sparser, but the redshift measurement is much more precise. To assess
how well we can reconstruct environment, we estimate the density
contrast $\delta_N$ in three different mock galaxy catalogues:

\begin{itemize}
\item[-] counts performed in the \drmop, by counting the galaxies
  falling within each cylindrical cell; we remind that in the \drmop,
  galaxy redshifts include peculiar velocities; we call this density
  contrast $\delta_N^R$;
\item[-] counts performed in the \dpmop, by counting the galaxies
  falling within each cylindrical cell; we call this density contrast
  $\delta_N^p$; when compared with $\delta_N^R$, it provides us with an
  estimate of the information lost due to the large photometric
  redshift errors;
\item[-] counts performed in the merged Euclid-like \dpmop+\dsmop, in
  which the spatial distribution of the spectroscopic galaxies is
  exploited to improve the accuracy of the photometric redshift
  estimate (with the ZADE method, see the next section); we call the
  resulting density contrast $\delta_N^E$.
\end{itemize}

\begin{figure*} \centering
\includegraphics[width=5.8cm]{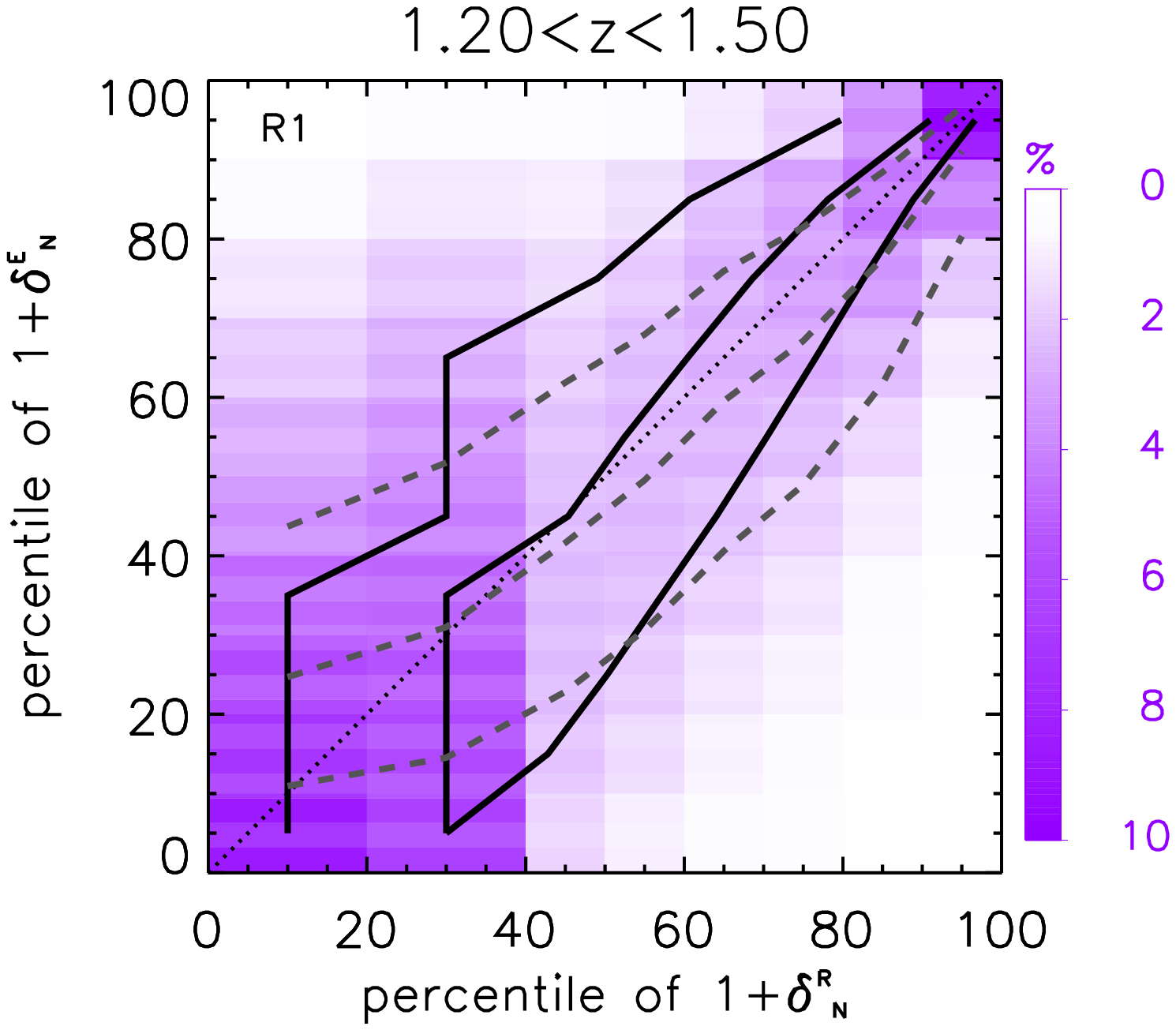}
\includegraphics[width=5.8cm]{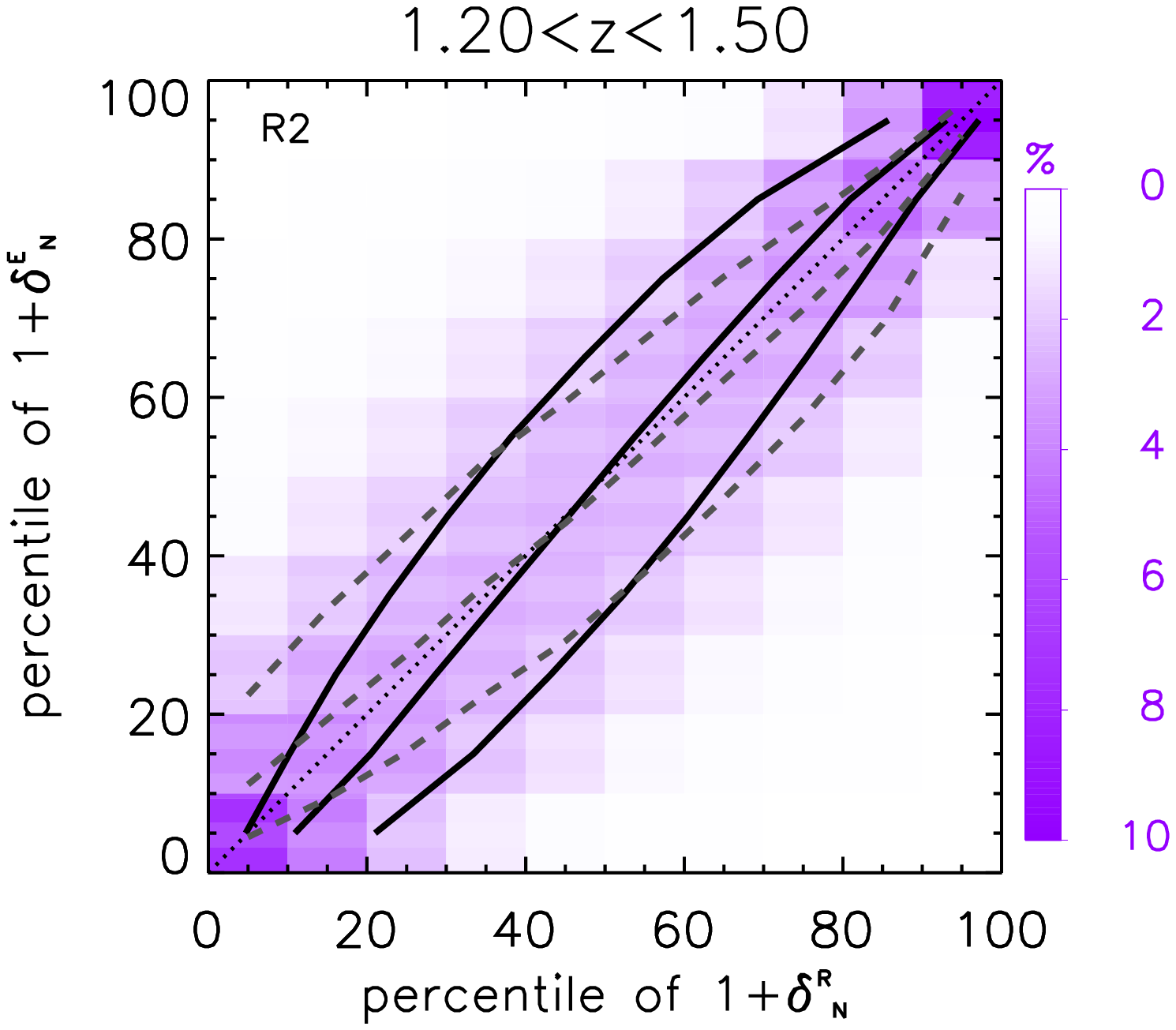}
\includegraphics[width=5.8cm]{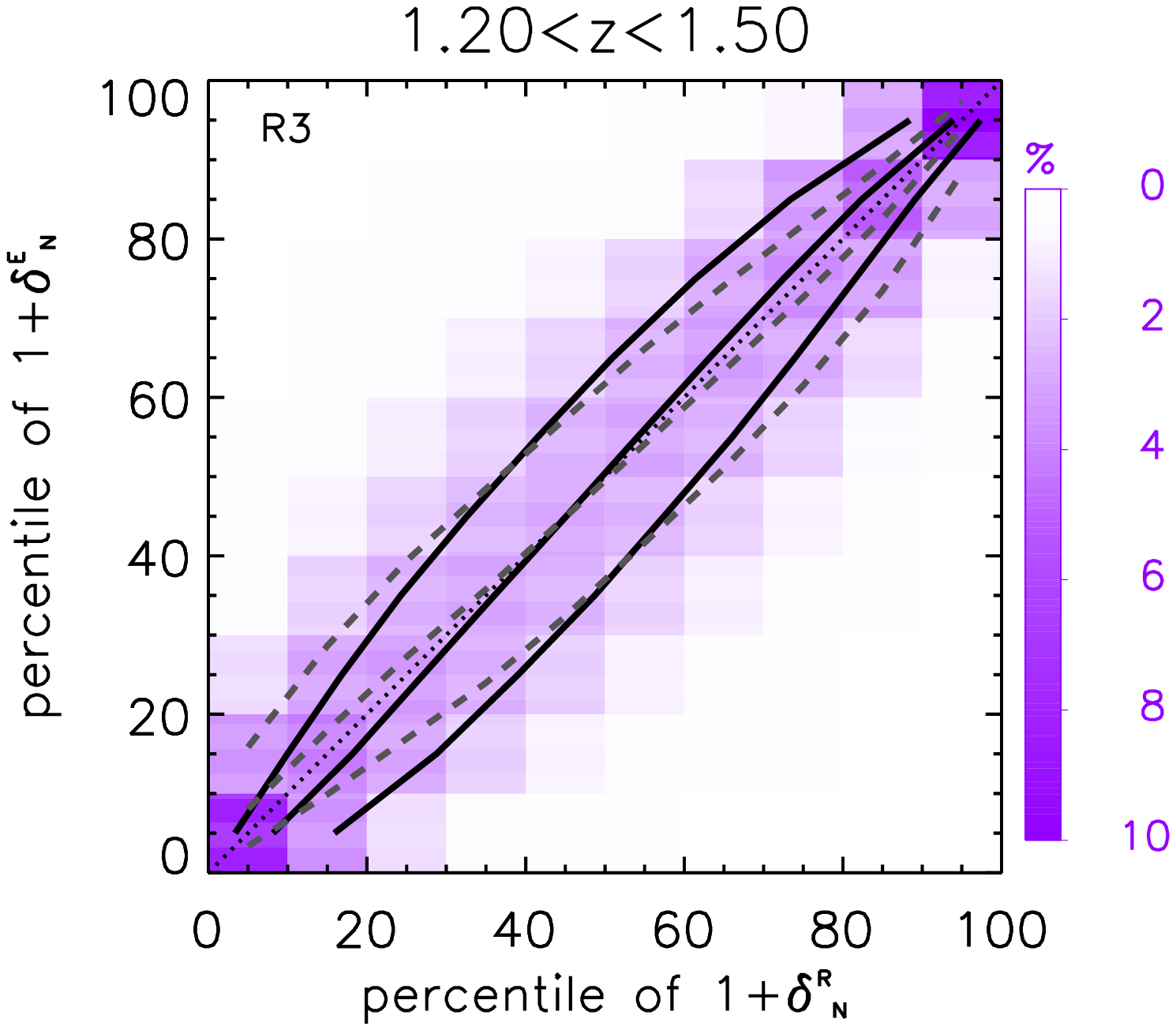}
\includegraphics[width=5.8cm]{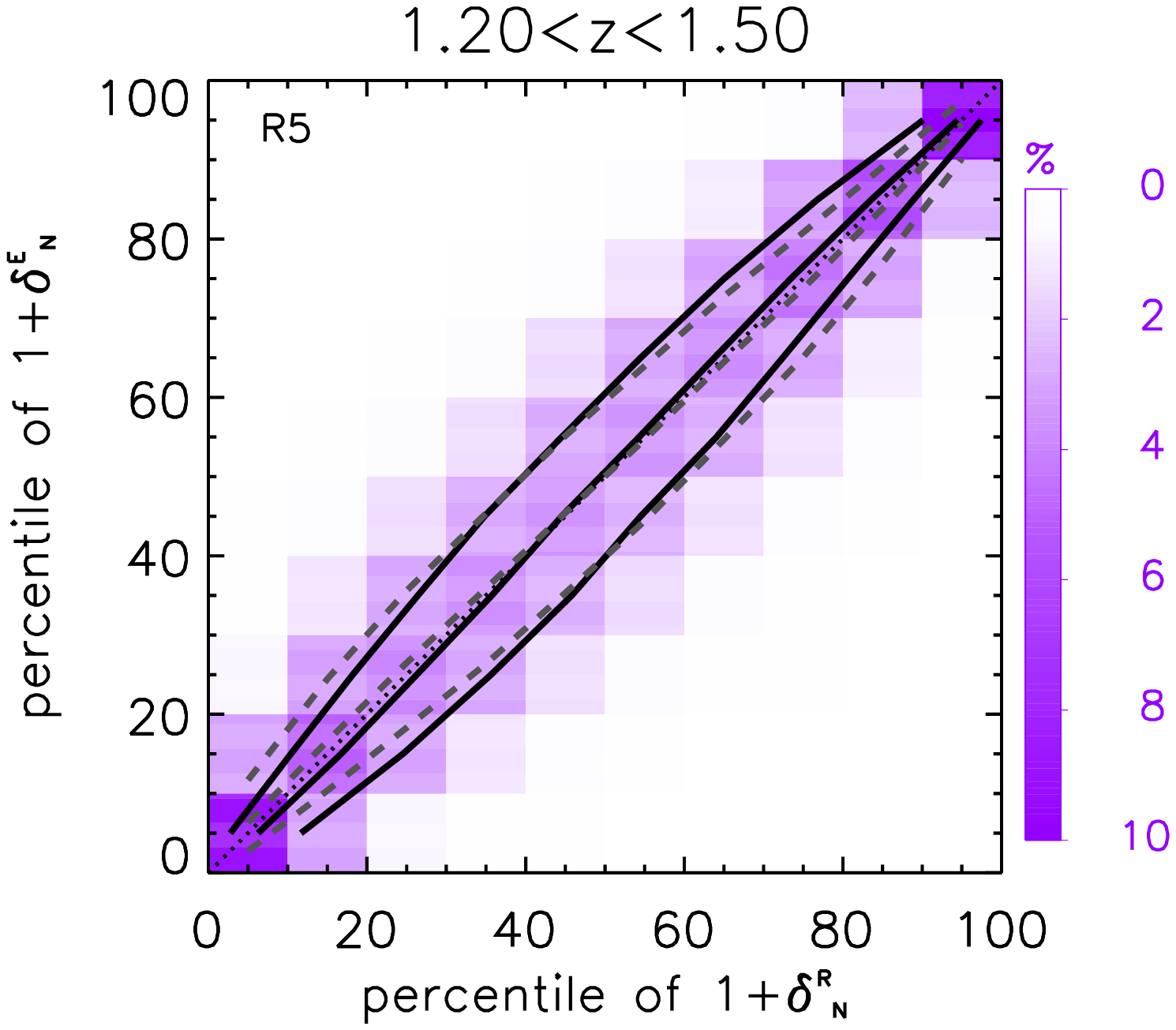}
\includegraphics[width=5.8cm]{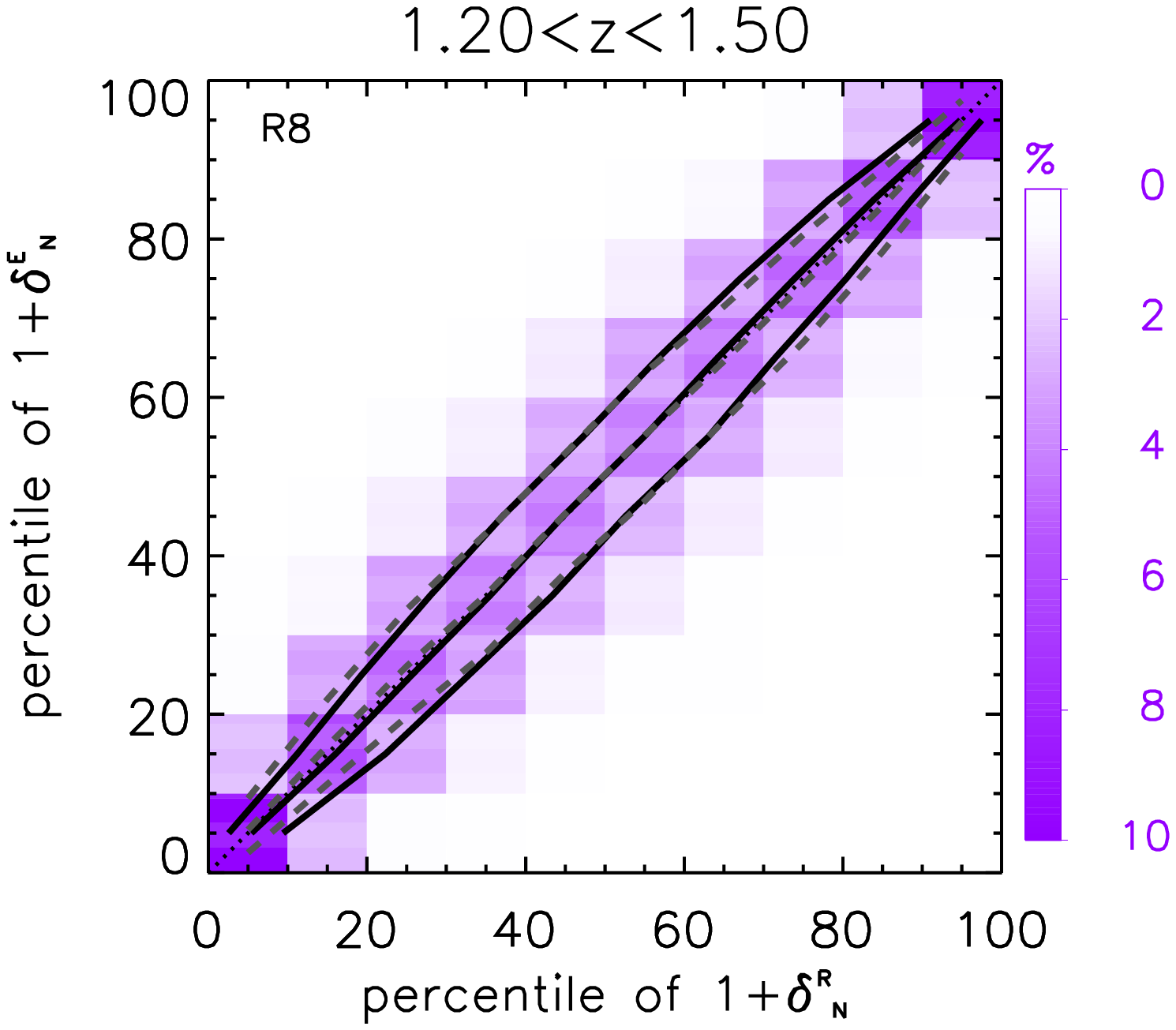}
\includegraphics[width=5.8cm]{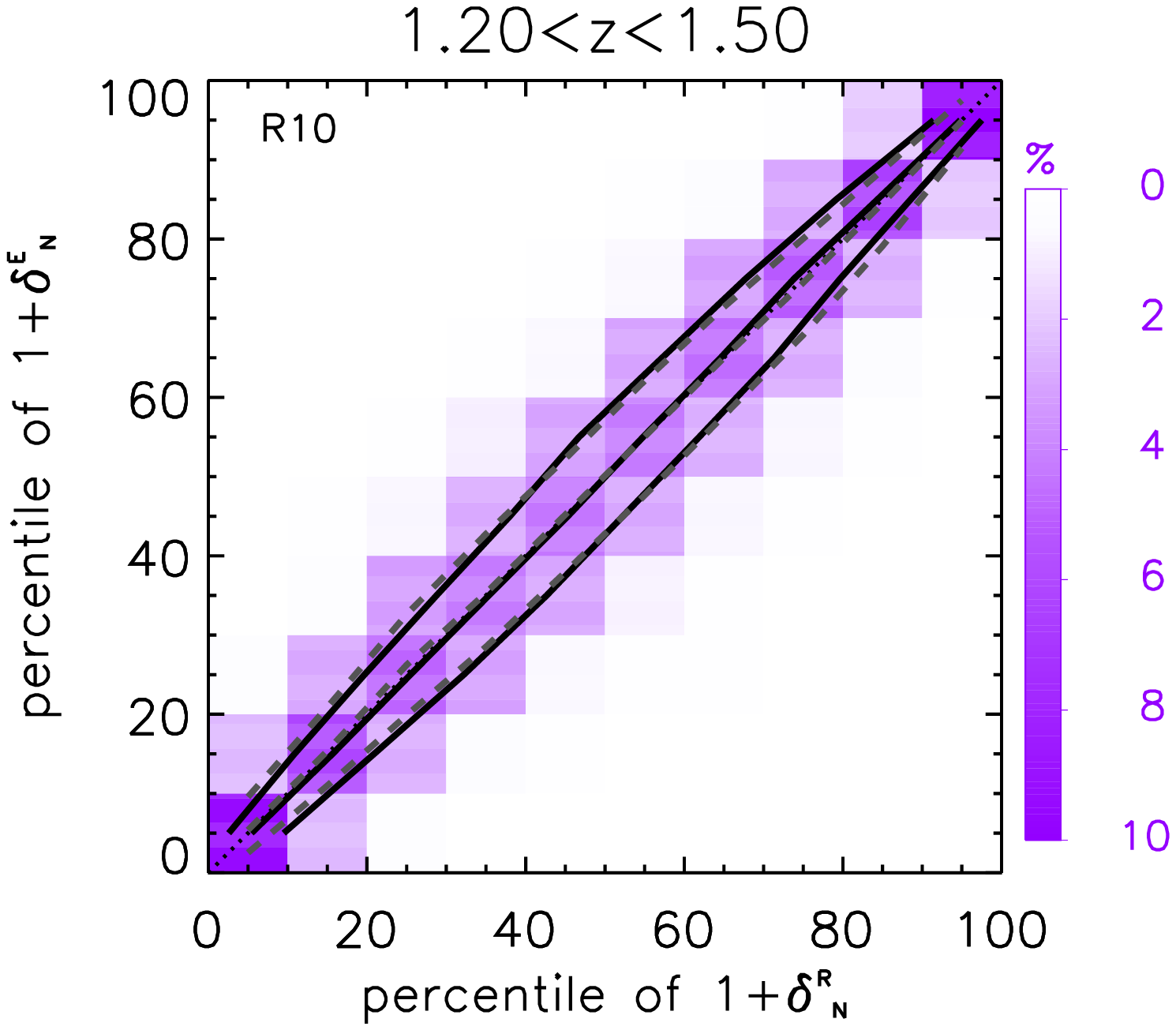}
\caption{Joint probability distribution of $1+\delta_N^R$ (x-axis) and
  $1+\delta_N^E$ (y-axis), in equipopulated bins containing 10\% of
  the density measurements. $1+\delta_N^R$ and $1+\delta_N^E$ on the
  two axes are ranked in increasing order. The colour-coded probability
  value is such that in the generic $\delta_N^{E}$ ($\delta_N^{R}$)
  bin the sum of the probability for $\delta_N^{R}$ ($\delta_N^{E}$)
  is precisely 10\%, so the background colour code ranges from 0\%
  probability (white) to 10\% probability (dark violet) as indicated
  on the vertical bar on the right. The dotted line is the one-to-one
  line, for reference. Black solid lines show, for each bin of the
  y-axis, the 25th, 50th and 75th percentile of the distribution of
  $1+\delta_N^R$ values. Gray dashed lines show, for each bin of the
  x-axis, the 25th, 50th and 75th percentile of the distribution of
  $1+\delta_N^E$ values. For a perfect reconstruction, the joint
  probability function would be everywhere equal to zero but for
  $\delta_N^{R}=\delta_N^{E}$, i.e. along the dotted line in the
  plot. From top-left to bottom-right, the panels refers to different
  radii ($R_1$, $R_2$, $R_3$, $R_5$, $R_8$ and $R_{10}$); all panels
  refer to the redshift range $1.2<z<1.5$ (see
  Fig.~\ref{tails_deep1_otherz} and \ref{tails_deep2_otherz} for other
  redshift bins).}
\label{tails_deep1} 
\end{figure*}

\subsection{The ZADE method}\label{zade}

We use the so called ZADE method to modify the $z_p$ probability
distribution function PDF of the galaxies in the \dpmosp exploiting
the 3D distribution of the galaxies in the \dsmop. The ZADE method was
developed for the zCOSMOS survey \citep{lilly07_zcosmos}, and 
is fully described in \cite{kovac2010_density}. We use a
simplified version of the ZADE approach, that has already been tested
for the VIPERS survey \citep{guzzo14_vipers}  by
\cite{cucciati14}.

In brief, this simplified ZADE method can be broken down into the
following steps.

\begin{itemize}

\item[i)] For each galaxy in the \dpmop, we keep its angular position
  ({\it R.A.} and {\it Dec}), and set the PDF of its photometric
  redshift ($P(z_p)$) to be equal to a normalised Gaussian centred on
  $z_p$ and with standard deviation equal to the 1 $\sigma$ error in
  the photometric redshift, $\sigma_{zp}$=0.05(1+$z$).

\item[ii)] We compute the spectroscopic redshift distribution $n(z_s)$
  of the galaxies in the corresponding \dsmos that fall within a
  cylinder centred on the position of the given photometric galaxy
  ({\it R.A.}, {\it Dec}, $z_p$), with radius $R_{ZADE}$ (see below)
  and half-length equal to $3\sigma_{pz}$.

\item[iii)] We set a new probability distribution function
  PDF$_{ZADE}$ for each galaxy in the \dpmosp equal to
  $AP(z_p)n(z_s)$, where $A$ is a factor to normalise the integral of
  PDF$_{ZADE}$ to unity.  PDF$_{ZADE}$ is characterised by several
  peaks. The value of PDF$_{ZADE}$ at each redshift peak corresponds
  to the weight $w_{ZADE}$ at that given redshift, and for each
  photometric galaxy the sum of all its $w_{ZADE}$ is unity by
  definition.

\end{itemize}

We note that the $n(z_s)$ is quite discontinuous (because of the
  limited number of spectroscopic galaxies) and, for this reason, we
  sample it in discretised bins of $\Delta z=0.003$. For consistency,
  we sample the Gaussian $P(z_p)$ on the same grid.  This way,
  PDF$_{ZADE}$ has the form of a histogram, and we call `peaks' all
  its bins where it is different from zero.  Basically ZADE
transforms a single photometric galaxy into a series of points at the
same {\it R.A.}-{\it Dec} position, and spread along the line of
sight, with each point having a weight given by $w_{ZADE}$. For the
estimation of the density field, the counts in the cylinders are
computed summing up the weights $w_{ZADE}$ of all the peaks that fall
in the given cylinder.

The ZADE approach is particularly suitable to reconstruct the 3D
galaxy distribution if the clustering properties of the spectroscopic
sample used to compute the $n(z_s)$ are the same as the ones in the
photometric sample (see e.g. the zCOSMOS and VIPERS surveys, where the
spectroscopic data set was a random subset of the photometric one). In
the case of Euclid, the photometric and spectroscopic surveys have
different selections (limit in $H$-band and in H$\alpha$ flux,
respectively), so we cannot expect the two samples to have the same
clustering, as discussed in Sect.\ref{mocks_comp_clus}.  

Nevertheless, in the next Sections we demonstrate that the ZADE method
is very effective also in the case of the Euclid Deep survey, with
respect of using photometric redshifts alone. We refer the reader to
Appendix \ref{zade_Hband} for an example of how ZADE would work if we
use, at the place of the \dsmop, a random subsample of the \dpmosp
(but with spectroscopic redshifts at the place of photometric ones).

We conclude with a remark on $R_{ZADE}$. The dimension of $R_{ZADE}$
is chosen as a compromise between the need to minimise the probed
scale length (to use the clustering strength on small scales) and to
maximise the number of spectroscopic galaxies in the cylinders (to
reduce shot noise).  We use $R_{ZADE}=5\mpcoh$, but we have checked
the robustness of our density field reconstruction also by varying
$R_{ZADE}$ between 3 and 10 $\mpcoh$.

\begin{figure} \centering
\includegraphics[width=7cm]{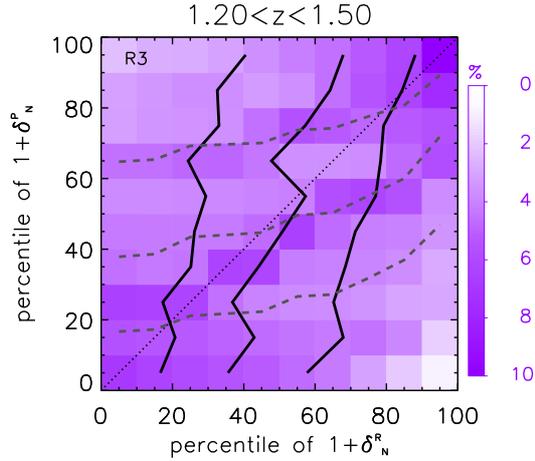}
\caption{As in Fig.~\ref{tails_deep1}, but in this case the
  $y-$axis shows $1+\delta_N^P$. The panel refers to $R_3$
  over the redshift range $1.2<z<1.5$. For a
  perfect reconstruction, the joint probability function would be
  everywhere equal to zero but for $\delta_N^{R}=\delta_N^{P}$,
  i.e. along the dotted line in the plot. }
\label{zphot_tails_deep} 
\end{figure}


\section{Gauging the accuracy of the density field
  reconstruction}\label{reconstruction}

In this Section we assess the ability of our method to trace the
underlying galaxy density. First, we aim at separating low- from
high-density environments to perform comparative studies of galaxy
evolution (Sect.~\ref{low_high} and \ref{PC}). Then we move to the
more challenging task of recovering the local galaxy density at a
given position (Sect.~\ref{zade_results}). We discuss our results in
the framework of environmental studies in Sect.~\ref{discussion}.

More specifically, here we compare the density field $\delta_N^E$ obtained using
the ZADE method in the \dpmop+\dsmosp  to the density
field estimated in the reference mock catalogues $\delta_N^R$.  As an 
additional check,  we
show also the comparison between $\delta_N^{R}$ and the density
$\delta_N^P$ estimated in the \dpmosp. The definition of $\delta_N$ is
the one given in Eq.~\ref{delta_eq}.

\begin{figure} \centering
\includegraphics[width=0.9\linewidth]{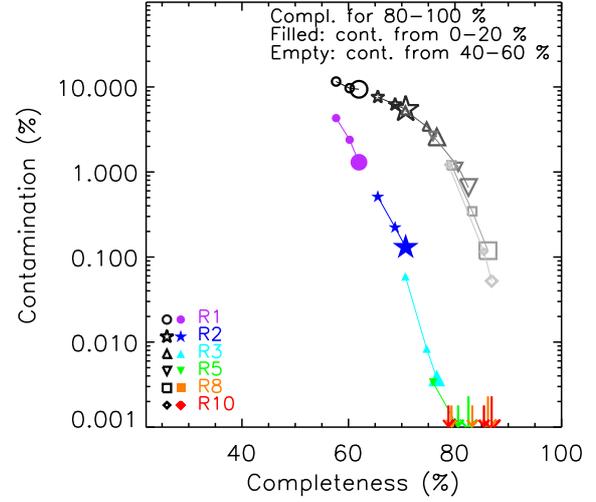}
\includegraphics[width=0.9\linewidth]{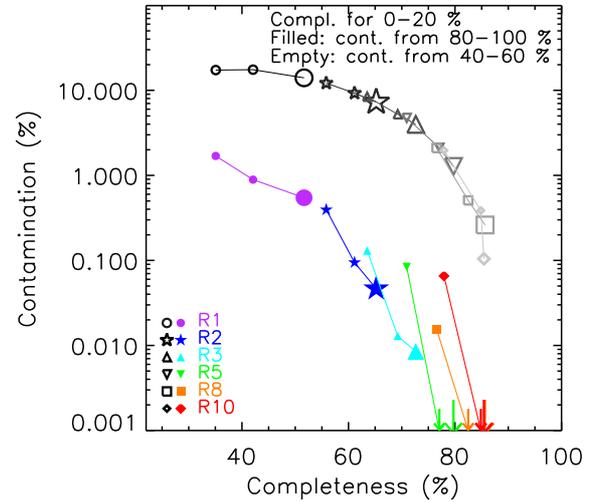}
\caption{Completeness (x-axis) and contamination (y-axis) for our
  density reconstruction (see Sect.~\ref{PC} for the definition of
  completeness and contamination).  Contamination and completeness are
  shown for all cylinder radii (different symbols as in the legend),
  and for three redshift bins ($0.9<z<1.2$, $1.2<z<1.5$,
  $1.5<z<1.8$). The three redshift bins for each radius are connected
  with a line, with the lowest redshift bin indicated with a
  bigger symbol. For each radius there are two series of points
  (filled coloured and empty gray shades) that refer to
  contamination coming from two different percentile ranges, as
  indicated in the panels. {\it Top}.  Completeness for the tail of
  the highest $80^{th}$-$100^{th}$ percentiles and contamination from the
  $0^{th}$-$20^{th}$ percentiles (filled symbols) and from the $40^{th}$-$60^{th}$
  percentile (empty symbols).  {\it Bottom}. As the top panel, but the
  completeness is computed for the lowest densities (0 to 20 \%), and
  its contamination comes from the $40^{th}$-$60^{th}$ percentile range (empty
  symbols) and from the $80^{th}$-$100^{th}$ percentile range (filled symbols).  See
  the text for a caveat on the points for $R_1$ at $1.5<z<1.8$ in the
  top panel. }
\label{CP_deep} 
\end{figure}

\subsection{High vs low density regions}\label{low_high}

To test how well we can separate  high from low
density regions, we do not need to compare $\delta_N^{E}$ and
$\delta_N^{R}$ on a point-by-point basis (as we will do in
Sect.~\ref{zade_results}), but only their ranking. 

For this purpose we computed the probability distribution function
(PDF) of $\delta_N^{E}$ and $\delta_N^{R}$ for the cell counts and
sampled these PDFs in bins containing $10\%$ of the counts. We then
computed the joint probability of $\delta_N^{E}$ and $\delta_N^{R}$
P($\delta_N^{E}$, $\delta_N^{R}$) using the same binning. This
function is shown in Fig.~\ref{tails_deep1}, where, by construction,
$P(\delta_N^{R} \, | \, \delta_N^{E}) = \int P(\delta_N^{E},
\delta_N^{R}) \, \delta_N^{R} \, \mathrm{d} \delta_N^{R} = 10\% =
P(\delta_N^{E} \, | \, \delta_N^{R})$. The colour-coded probability
value is such that in the generic $\delta_N^{E}$ ($\delta_N^{R}$) bin
the sum of the probability for $\delta_N^{R}$ ($\delta_N^{E}$) is
precisely 10\%. For a perfect reconstruction, the joint probability
function would be everywhere equal to zero but for
$\delta_N^{R}=\delta_N^{E}$, i.e. along the dotted line in the plot.

From the plots in Fig.~\ref{tails_deep1} we see that:

\begin{itemize}

\item[-] The correlation between the ranked $\delta_N^R$ values and
  the ranked $\delta_N^E$ values is stronger (both in terms of a
  smaller deviation from the 1:1 correlation and a lower dispersion) for
  larger radii.

\item[-] For $R_1$, at low density the values of $\delta_N^R$ are
  discretised: the radius of the cylinder is very small with respect
  to the mean inter-galaxy separation of the sample, so in many
  cylinders we have 0, 1 or 2 galaxies. This is apparent in
  Fig.~\ref{tails_deep1}, where $\delta_N^R$ is always binned in bins
  of $10\%$. For $R_1$, at low density, we have the
  first bin as large as $\sim20$\% . This is the percentage
  range where $\delta_N^R=0$. This does not happen for $\delta_N^E$,
  because the ZADE method produces objects with fractional weights, so
  their counts result in a continuous $\delta_N^E$ distribution.

\item[-] We also explored the redshift dependence of the correlation
  between $\delta_N^R$ and $\delta_N^E$ (see Appendix
  \ref{zade_otherz}), and we found that the above-mentioned results
  hold at all explored redshifts. Moreover, in general, at fixed
  radius the correlation is stronger for lower redshift, and this is
  especially true for small radii. This happens because at low
  redshift the mean galaxy density is higher (the galaxy sample is
  flux-limited), so the local density reconstruction suffers from a
  lower shot noise. Clearly this effect is more important for small
  radii, where the average counts-in-cells are lower and, as a
  consequence, their relative variation from low to high redshift is
  larger.

\end{itemize}

We notice that the scatter around the $1:1$ correlation in
Fig.~\ref{tails_deep1}, for the lowest\footnote{For all radii but
  $R_1$, because of the discretised density values discussed in the
  text.} and highest densities, is artificially low, simply because
of the finite range of the density values.

To appreciate more the goodness of the density reconstruction based on 
both spectroscopic and photometric redshifts, in
Fig.~\ref{zphot_tails_deep} we show the same as in
Fig.~\ref{tails_deep1} but this time we compare $\delta_N^{R}$ with
the density reconstructed using only photometric redshifts
($\delta_N^{P}$).  The almost uniform colour of
Fig.~\ref{zphot_tails_deep} shows that the large photometric errors
erase any correlation between the rankings, making it impossible to
separate high from low density environments using only a photometric
survey with a photometric redshift  error of
$\sigma_{zp}=0.05(1+z)$.

\subsection{Completeness and contamination}\label{PC}

Figure \ref{tails_deep1} shows  which is the range of $\delta_N^R$
that corresponds to any given selection on $\delta_N^E$. For a more
quantitative analysis, we can choose some specific $\delta_N^E$ ranges
and compute their completeness and contamination, as follows.

For each cylinder radius and redshift, we call $N^E_i$ ($N^R_i$) the
number of cylinders falling in the percentile range $i$ of the
$\delta_N^E$ ($\delta_N^R$) distribution.  Also, we call
$N_{i,j}^{E,R}$ the number of cylinders that fall in the percentile
range $i$ of $\delta_N^E$ and in the percentile $j$ of
$\delta_N^R$. We then select a given percentile range $i$ and define
its completeness and contamination as:
\begin{equation} \displaystyle
\text{completeness} = N_{i,i}^{E,R} / N^R_i
\label{comp_eq} 
\end{equation}
\begin{equation} \displaystyle
\text{contamination} = N_{i,j}^{E,R} / N^E_i  \qquad \text{(with $i \neq j$)}
\label{cont_eq} 
\end{equation}

Namely, the completeness indicates the fraction of cylinders that are
placed in the correct percentile. The contamination shows which
fraction of cylinders belonging to the percentile $i$ of $\delta_N^E$
distribution come from a different percentile of the original
$\delta_N^R$ distribution. Ideally one would like a completeness of
100\% and a contamination equal to zero.

Figure \ref{CP_deep} shows the completeness and the contamination for
our Euclid-like mock catalogues. We analyse three percentile ranges
($i$) of the $\delta_N^E$ distribution: $80^{th}$-$100^{th}$,
$40^{th}$-$60^{th}$, and $0^{th}$-$20^{th}$. Basically, they
correspond to the highest, intermediate and lowest density regimes,
that we call $P_H$, $P_I$, and $P_L$, respectively.  For $P_H$ and
$P_L$, we show the completeness and the contamination coming from the
other density regimes.

\begin{figure} \centering
\includegraphics[width=6cm]{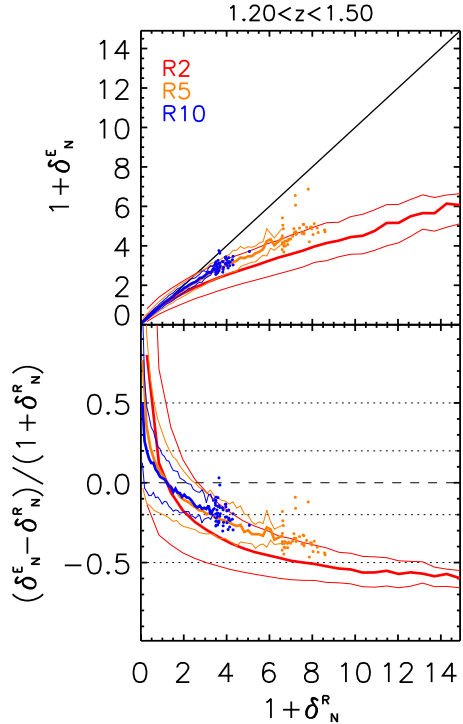}
\caption{Comparison of $\delta_N^R$ and $\delta_N^E$ on a
  cell-by-cell basis. x-axis: density contrast in the reference
  catalogue ($1+\delta_N^R$); y-axis, top panel: density contrast in
  the Euclid-like catalogue ($1+\delta_N^{E}$); y-axis, bottom panel:
  relative difference
  ($(\delta_N^{E}-\delta_N^R)/(1+\delta_N^R)$). The thick lines are
  the median value of the quantity displayed on the y-axis in each
  x-axis bin. Thin lines represent the 16th and 84th percentiles of
  its distribution. Points are single cells, when cells per bin are
  $<20$ (in which case we do not compute a median and
  percentiles).  The solid black line in the top panels is
  the one-to-one line, and the horizontal lines in the bottom panels
  are for reference. Results are shown for different  cylinder
  radii as in the label ($R_2$,  $R_5$ and $R_{10}$) for the redshift bin
  $1.2<z<1.5$. See Fig.\ref{delta_deep_otherz} for other redshift
  bins.}
\label{delta_deep} 
\end{figure}

We find that:
\begin{itemize}
\item[-] for all the density regimes, the completeness
  (/contamination) increases (/decreases) for lower redshift and for
  larger radii; the dependence of the completeness on redshift is
  especially evident at small radii for $P_L$; in all cases, the
  completeness varies  more with  radius (at fixed redshift) than with
  redshift (at fixed radius);
\item[-] at fixed radius and redshift, $P_H$ is the most complete,
  having always a completeness $\gtrsim60\%$ at all radii and redshifts; this is due to
  the smaller random error in the density reconstruction (see
  Fig.~\ref{delta_deep}) for the highest densities; this difference is
  especially true at small radii, where the reconstruction of the low
  densities is more difficult;
\item[-] the contamination of $P_L$ and $P_H$ from the opposite
  density regime ($P_H$ and $P_L$, respectively) is very low, being
  always below 1\% but for $R_1$ for which it gets to a few percent;
  this implies that we can always safely separate $P_L$ from $P_H$;
\item[-] for $R_8$ and $R_{10}$ at all redshifts, and for $R_5$ at the
  lowest redshift, the contamination in $P_H$ and $P_L$ coming from
  $P_I$ is below 2\%; this implies that for these radii and redshift
  ranges it is possible to study environment in three different
  density regimes, instead of selecting only $P_L$ and $P_H$; 
\item[-] for $R_5$, $R_{8}$ and $R_{10}$ the completeness in $P_H$ and
  $P_L$ is very similar (always $\gtrsim75$\%), as can be derived by
  the similar behaviour at the lowest and highest densities in
  Fig.\ref{tails_deep1}; for smaller radii, we still have a good
  completeness for $P_H$ ($\gtrsim55$\%, $\gtrsim60$\%, and
  $\gtrsim70$\% for $R_1$, $R_{2}$ and $R_{3}$, respectively).

\end{itemize}

A caveat about the contamination from $P_L$: for $R_1$ the values of
$\delta_N^R$ at the lowest densities are discretised (see
Sec.~\ref{low_high}), and in the three redshift bins the lowest
$\delta_N^R$ value is taken by 20\%, 20\% and 30\% of the cylinders
(from low to high redshift). This means that the points at $1.5<z<1.8$
for $R_1$ in the top panel of Fig.~\ref{CP_deep} represents a
contamination from the percentile range 0-30\% instead of from the range
0-20\% as all the other points in the Figure.

In summary, we have shown that we can robustly identify high-density
regions and distinguish them from the lowest densities, at all
explored scales and at all explored redshift.  The level of
completeness of the selected sample of high-density regions depends on
the scale and on the redshift, but the contamination from the lowest
densities is always below a few percent for $R_1$ and $<1\%$ for
larger radii.  We identify high densities in a more complete
way than low densities, especially at small scales.  Moreover, for
$R\geq5\mpcoh$ we can separate very well high densities also from
intermediate densities (the contamination is only by a few percent).

\begin{figure} \centering
\includegraphics[width=6cm]{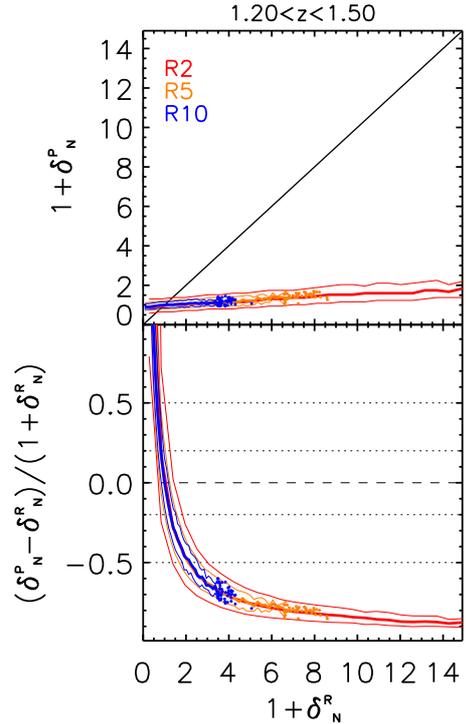}
\caption{ As in Fig.~\ref{delta_deep}, but in this case we compare
  $\delta_N^R$ with $\delta_N^P$.  }
\label{zphot_delta_deep} 
\end{figure}

\subsection{Density-density comparison}\label{zade_results}

In this Section we move to a more demanding test, i.e. we compare the
values of $\delta_N^{E}$ and $\delta_N^{R}$ on a cell-by-cell basis.
The aim of this density-density comparison is to estimate the systematic
deviation from the true value and the random dispersion around the mean
value for the single cells, to quantify the goodness
of the density reconstruction at a local level.

In Fig.~\ref{delta_deep} we compare $\delta_N^{R}$ with
$\delta_N^{E}$, for different cylinder radii for the redshift bin
$1.2<z<1.5$. We show results only for $R_2$, $R_5$ and $R_{10}$ for
the sake of clarity, and the reader can easily extrapolate the results
for the other radii from those shown here, and from the discussion
below. In Appendix \ref{zade_otherz} we show the same results for the
two other redshift bins.

We note that, for $\delta_N^R \rightarrow -1$, the denominator of the
normalised residuals (the variable in $y$-axis in the bottom panels)
approaches zero and residuals rapidly increase.  This is an artifact
related to our definition of residuals, as demonstrated by the upper
panels in which the relation between the two density fields is well
behaved close to zero.

We observe the following.

\begin{itemize}

\item[-] Our density reconstruction $\delta_N^E$ underestimates the
  reference density $\delta_N^R$ at the highest densities, for all
  cylinder radii and at all redshifts. This underestimation ranges
  from $\sim70$\% for the smallest radii ($R_1$ and $R_2$) to
  $\sim20$\% for the largest ones ($R_8$ and $R_{10}$). 

\item[-] The lowest densities are always overestimated, by a
  percentage that rapidly increases for $\delta_N^R \rightarrow -1$
  (as explained above).

\item[-] The systematic error depends on the cylinder radius (it is
  larger for smaller radii) at fixed redshift, and it mildly depends
  on redshift at fixed radius; this indicates that it is more difficult
  to reconstruct the absolute value of the local density at small
  scales.

\item[-] The random error for the highest densities depends very
  mildly (if at all) on the cylinder radius (being possibly slightly
  larger for smaller radii) at fixed redshift, while it depends more
  evidently on redshift at fixed radius (it varies from
  $\lesssim10$\% in the lowest redshift bin to $\sim10-15$\% in the
  highest redshift bin).

\item[-] At the highest densities, the random error is smaller than
  the systematic error at all radii and redshifts, but for $R_8$ and
  for $R_{10}$ at $z>1.5$ where the two errors are similar.

\end{itemize}

\begin{figure*} \centering
\includegraphics[width=5.8cm]{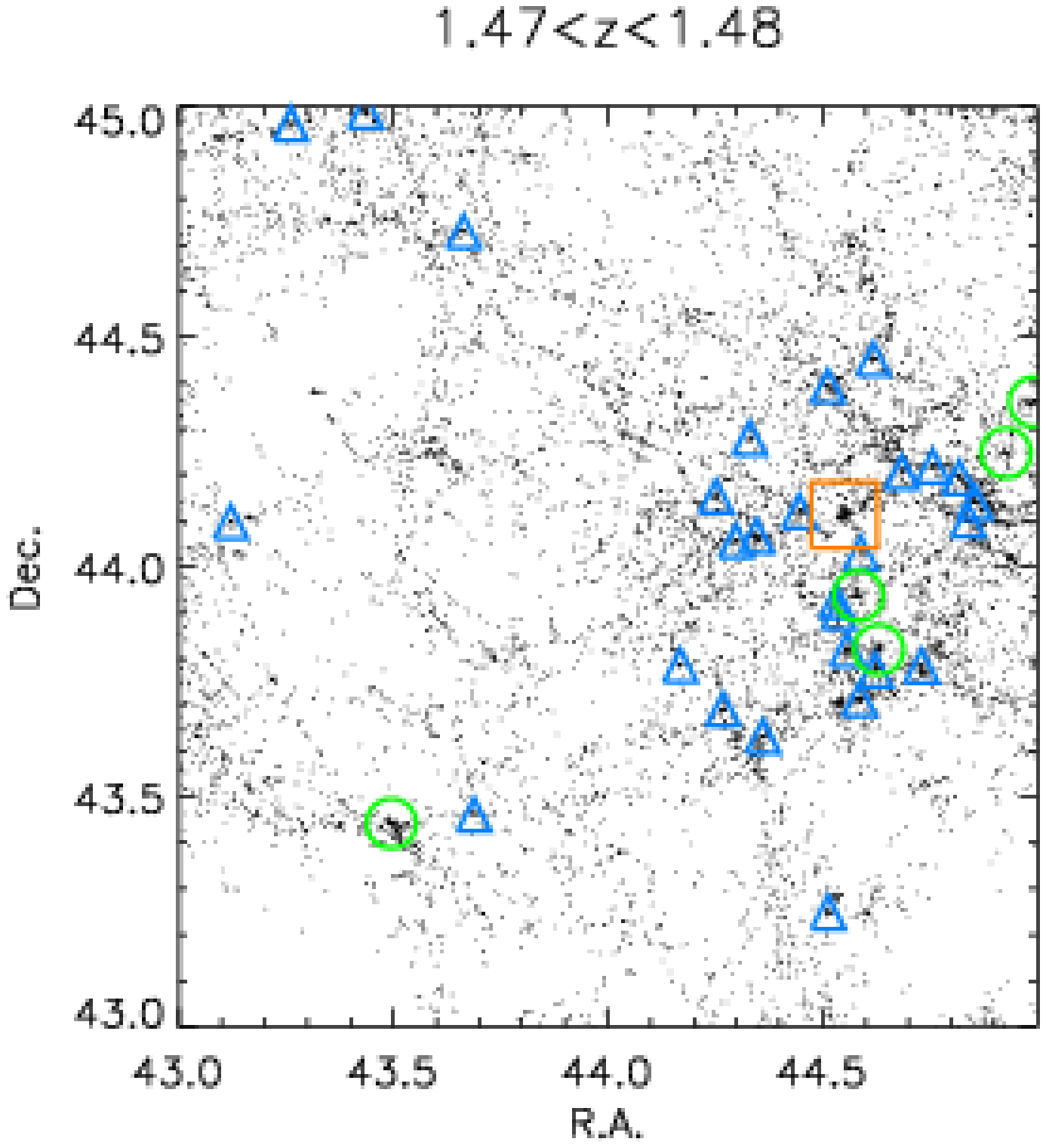}
\includegraphics[width=5.8cm]{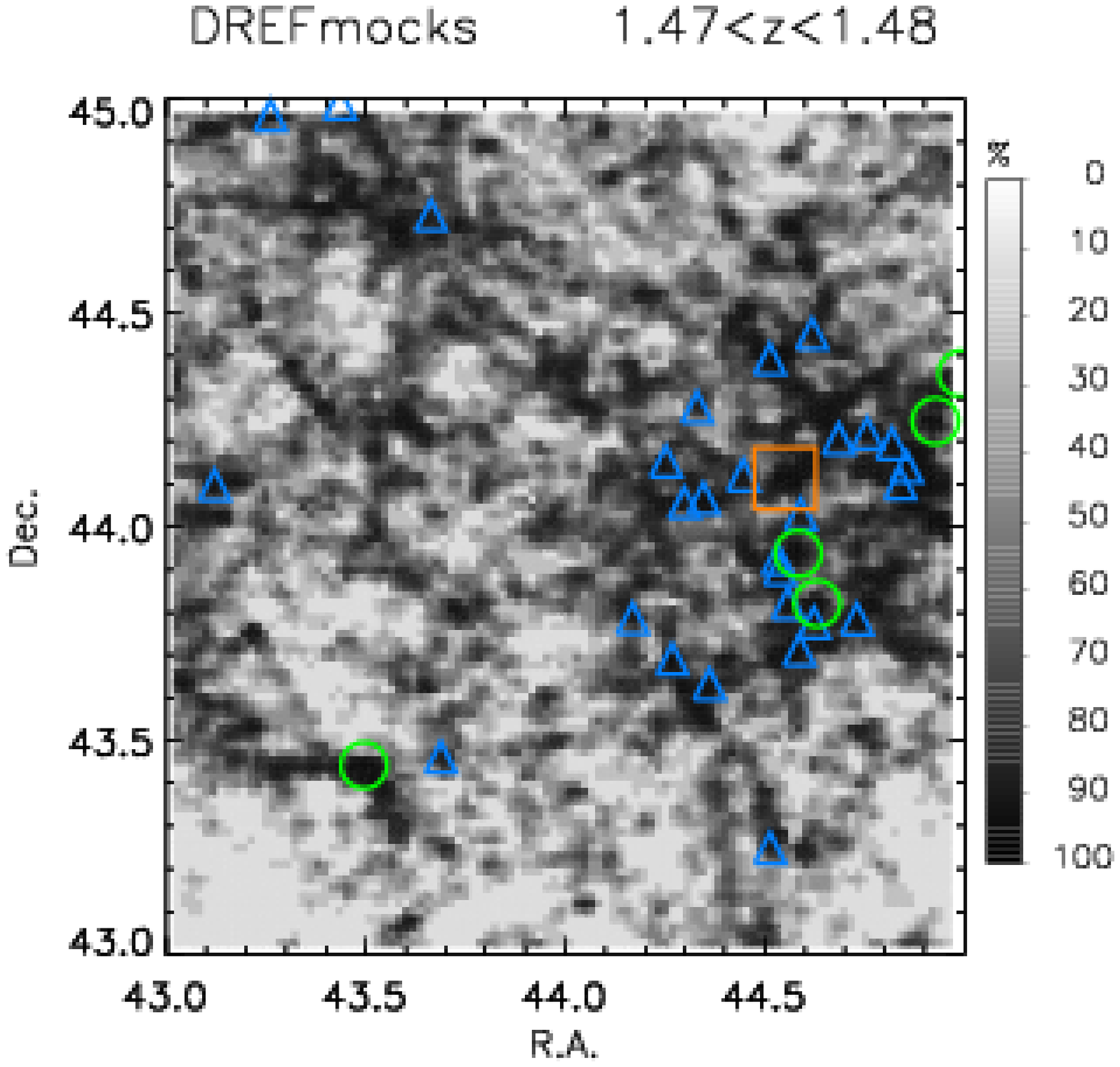}
\includegraphics[width=5.8cm]{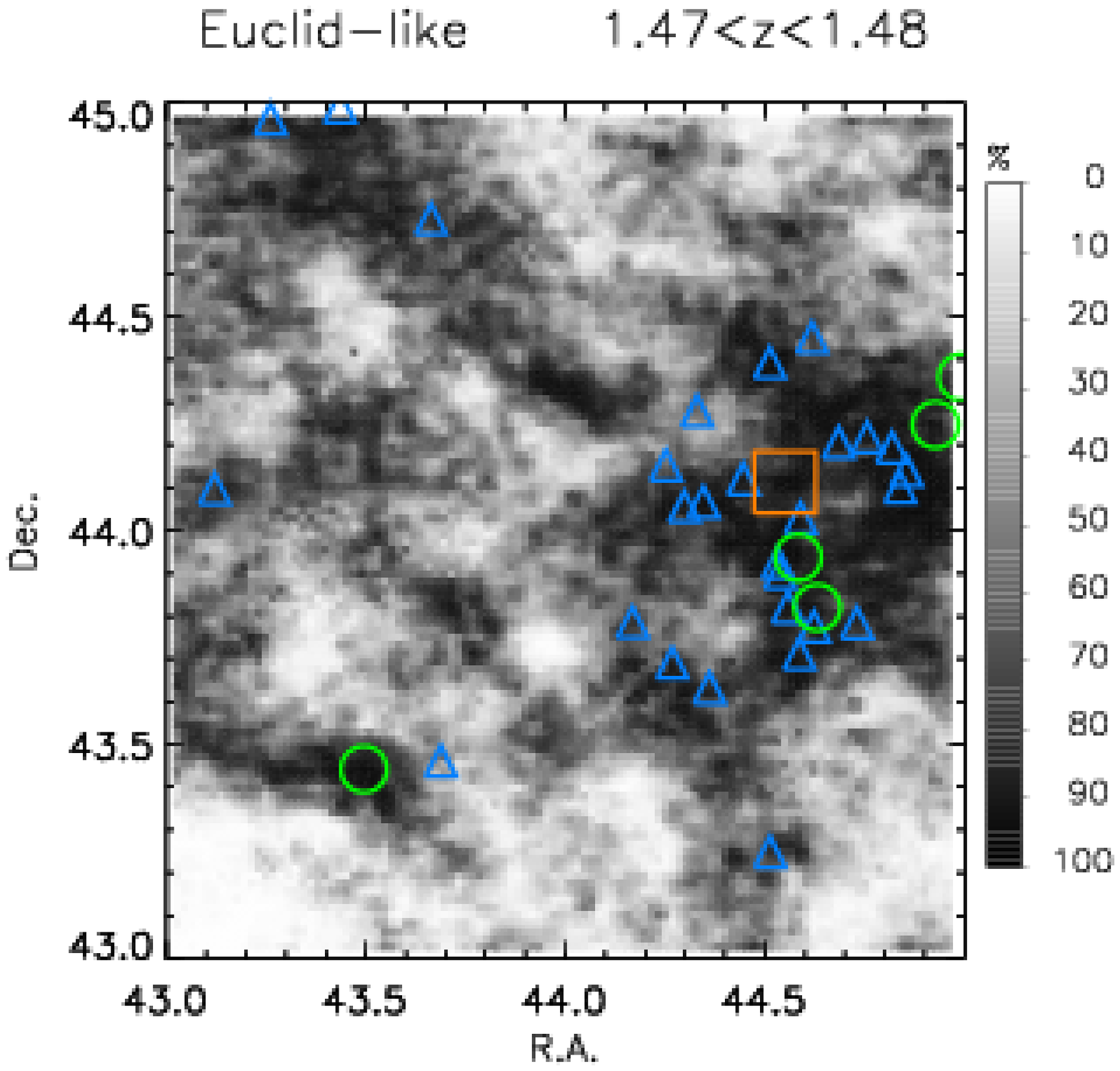}
\caption{ {\it Left.} $R.A.-Dec$ distribution of galaxies in one of
  the \drmop, for the redshift bin $1.47<z<1.48$. Blue triangles,
  green circles and orange squares locate the position of galaxy
  clusters (in this same redshift bin) with total masses of
  $\log(M/M_{\odot})\in$ $[13,13.5)$, $[13.5,14)$, and $>14$,
  respectively. {\it Middle.} In the same $R.A.$, $Dec$ and redshift
  range as in the left panel, the grey-scale colour map represents the
  density field estimated in the \drmos with cylinders of $R=1
  \mpcoh$. The map is computed on a grid of $1 \mpcoh$ per
  element. The density value in each grid element is computed
  averaging the $\delta_N^R$ values of the cylinders falling in that
  given grid element. The grey scale refers to the percentiles of the
  $\delta_N^R$ distribution as in the colour bar.  Triangles, circles
  and squares are as in the left panel. {\it Right.} As in the middle
  panel, but here we use $\delta_N^E$ instead of $\delta_N^R$.}
\label{map_deep}
\end{figure*}

In summary, for $R_8$ and $R_{10}$ we are able to compute the absolute
value of the local density field $\delta_N^{R}$ with a small random
error and an acceptable systematic underestimate of $\sim20\%$. For
smaller scales both types of errors increase and the recovery of the
density is mainly hampered by the random ones in low density regions
and by the systematic ones in the high density peaks.

The underestimate of the highest densities in general, and especially
for the smallest radii, is mainly due to the following reasons: i) the
large photometric error, that spread the ZADE weights along the line
of sight on a large distance; ii) the low density of galaxies in the
\dsmosp with respect to the density of galaxies in the \dpmosp (see
e.g. \citealp{cucciati14}, Appendix A.3, where we verified that the
systematic error in the density reconstruction increases when we
decrease the sampling of the spectroscopic sample); iii) the lower
clustering of the \dsmosp with respect to the \dpmosp (see the effects
described in Appendix \ref{zade_Hband}).

In general, we expect to overestimate low densities and to
underestimate high densities when using tracers with large redshift
error, like photometric redshifts. This can be more easily seen in
Fig.~\ref{zphot_delta_deep}, where we compare $\delta_N^{R}$ and
$\delta_N^{P}$ values on a cell-by-cell basis. 
The large photometric redshift error ($\sigma_{zp}$=0.05(1+$z$))
displaces galaxies far away from their true distance (see the second
panel of Fig.~\ref{coni_deep}), making the galaxy distribution almost
homogeneous in the entire explored volume (voids are filled, and high
density regions are depopulated).  As a consequence, the reconstructed
density within cells is everywhere close to the cosmological
mean. This produces a rather small random error, but the reconstructed
$\delta_N^{P}$ is systematically smaller (higher) than the true one
for over-dense (under-dense) regions, and the mismatch increases with
the distance from $\delta_N^{R}=0$. The same trend is found at all
redshifts and for all cell sizes.

\subsection{Discussion}\label{discussion}

The results shown in Sect.~\ref{low_high} and \ref{PC} imply that
using the Euclid Deep survey it will be possible to perform
comparative studies of galaxy evolution in low and high density
regions from large to small scales up to $z\leq1.8$. In particular,
the identification of small-scale ($R\sim1 h^{-1}$Mpc) high-density
regions, with a very low contamination from other environments, will
allow us to study the regions (like e.g. galaxy clusters and groups)
where several physical processes shaping galaxy evolution normally
occur. Moreover, the high completeness and very low contamination in
identifying the lowest density regions on large scales
($R=10h^{-1}$Mpc) indicates that one can potentially use the Euclid
Deep survey to identify cosmic voids (see
e.g. \citealp{micheletti14_voids} for a systematic search of cosmic
voids in spheres of $R\gtrsim15 Mpc$). 

More generally, we have shown that in a sample like the Euclid Deep
survey it will be possible to identify high densities and low
densities on different scales. The potentiality of this multi-scale
approach is already been demonstrated in several work in the
literature. On one side, correlating galaxy properties with the local
density on different scales is an effective way to understand how
galaxy evolution depends on the assembly of the large scale structure
(see e.g. \citealp{wilman10, fossati15} and references therein). On
the other side, a multi-scale approach can also be used to identify
the different components of the cosmic web (voids, filaments, walls
and clusters, see
e.g. \citealp{aragon_calvo10_multiscale,smith12_multiscale}).  

Clearly, each science case will require a
dedicated feasibility study. For instance, the minimum (maximum) level
of completeness (contamination) required for any given science case
has to be quantified case by case (see e.g. \citealp{cucciati12_MILL}
and \citealp{malavasi16}, where they study whether the contamination
in the density field reconstruction hampers the detection of the
colour-density relation or of the different shape of the galaxy
stellar mass function in low and high densities). 

The precise computation of the absolute value of the local density
field is a task much more difficult than the identification of the
lowest and highest densities, especially when we lack a precise
measurement of the 3D position of the tracers of the density field
itself. Nevertheless, our tests show that we are able to measure the
value of $\delta_N$ on scales of $\geq 8\mpcoh$ up to $z\sim1.8$. 

This achievement, for instance, will allow us to study the evolution
of the galaxy bias on such scales, as compared with surveys at lower
redshift (see e.g. \citealp{diporto14}, \citealp{bel15},
\citealp{cappi15} and references therein). Another application of a
precise local density value is the detection of clusters in their
phase of build-up, although by now this kind of detection has been
systematically tested in the literature at higher redshifts and on
slightly larger scales than those investigate here ($\sim10-17\mpcoh$,
$z\gtrsim2$, see e.g.  \citealp{chiang13,chiang14}).

We defer to a future work a detailed analysis of the feasibility of
specific environmental studies with the Euclid Deep survey.  As a very
simple test case, we show in Sect.~\ref{clusters} the potential
ability of our method of detecting the regions where the most massive
clusters lie.


\section{A test case: the relation between the local environment and galaxy clusters}\label{clusters}

The main goal of this paper is to distinguish in a robust way the
highest density peaks from the low density regions on both small and
large scales, so the precise identification of clusters, voids, and
all the intermediate structures that form the cosmic web is beyond our
aim. Nevertheless it is interesting to see how the density
reconstructed with our mock catalogues compares with some of the
typical components of the large scale structure of the universe. Among
these components, galaxy clusters are certainly more easy to locate
with a precise 3D position, rather than filaments and voids, at least
in simulated DM/galaxy catalogues. Moreover, Fig.~\ref{CP_deep} shows
that our identification of small-scale ($R=1-2 h^{-1}$Mpc) high
density peaks is robust. 

In Fig. \ref{map_deep} we show where (relatively massive) clusters are
located in $R.A.-Dec$, in a small redshift bin ($1.47<z<1.48$) of one
of our light-cones, compared to the position of galaxies in the \drmos
and to the density fields $\delta_N^R$ and $\delta_N^E$.  The redshift
bin has been chosen to be at relatively high redshift and to include
at least one of the most massive clusters ($\log(M/M_{\odot}) > 14$),
together with other over-dense regions and voids.

\begin{figure} \centering
\includegraphics[width=\hsize]{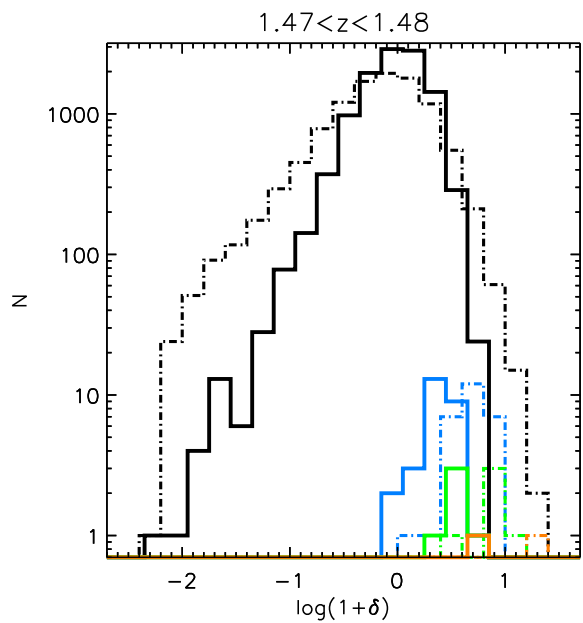}
\caption{Distribution of densities from the density maps presented in
  Fig.~\ref{map_deep}: the black dot-dashed histogram is for the \drmos
  ($\delta_N^R$) and the black solid histogram is for the
  \dpmo$+$\dsmos ($\delta_N^E$).  Blue, green and orange histograms
  represent the density distribution of the map elements of
  Fig.~\ref{map_deep} where the three classes of galaxy clusters fall
  (see the caption of Fig.~\ref{map_deep} for the definition of the
  three classes, here we use the same colour code). Also for the
  clusters, the dot-dashed histograms refer to $\delta_N^R$ and the solid
  histograms to $\delta_N^E$. }
\label{dens_distrib_deep}
\end{figure}

Qualitatively, we see that clusters, and especially the most massive
ones, fall in the highest density regions, and this is true for both
$\delta_N^R$ and $\delta_N^E$. In this example, the density is
estimated on a scale of $R=1\mpcoh$ (the smallest scale we study in
this work, to better compare with the typical dimension of galaxy
clusters). More quantitatively, we see in Fig.~\ref{dens_distrib_deep}
that the values of local density evaluated at the positions where
galaxy clusters are located fall in the highest density tail of the
total density distribution for both $\delta_N^R$ and $\delta_N^E$, and
there is even some correlation between the mass of the clusters and the
local density value where they reside. 

From Fig.~\ref{dens_distrib_deep}, it is also evident that there are
regions (map elements) with the same measured density as those where
clusters reside, but where there are no clusters. This is both due to
the contamination of our density reconstruction and to the fact that
our method is not fine-tuned to identify clusters as the highest
density peaks. Nevertheless this simple sanity check shows the
potentiality of future galaxy surveys in the study of the large scale
structure of the universe.

\section{Summary and conclusions}\label{summary}

We studied the feasibility of environmental studies in slitless
spectroscopic surveys from space. In particular, we want to exploit
the synergy between spectroscopic and photometric samples.  As a test
case, we use galaxy mock catalogues which mimic the future Euclid Deep
surveys, namely a spectroscopic survey limited in H$\alpha$ flux and a
photometric survey limited in $H-$band.  Our main goal is to verify
that it will be possible to disentangle the highest from the lowest
density regions, by means of the parameterisation of the local
environment. We anchored the photometric redshifts of the photometric
sample to the 3D skeleton of the spectroscopic sample, and we measured
the local environment on different scales, from 1 to $10 \mpcoh$, in
different redshift bins in the range $0.9<z<1.8$. We found that:

\begin{itemize}

\item[-] We are able to disentangle the highest densities from the
  lowest densities at all explored redshifts, and for all radii.  

\item[-] For $R_8$ and $R_{10}$ at all redshifts, and for $R_5$ at
  $0.9<z<1.2$, it is also possible to select an intermediate-density
  regime, robustly separated from the highest and lowest density
  tails.

\item[-] When we select the tail of highest densities, the
  contamination from poorly-measured low densities decreases for larger
  radii and at lower redshift, while the completeness increases.
  Completeness is $\gtrsim60\%$ if we select the tails of the
  highest 20\% using $R_1$. This percentage becomes 70\%
  and 80\% for $R_3$ and $R_8$, respectively.  The
  contamination is always $\lesssim1\%$, at all redshifts and for all
  radii, with the exception of $R_1$ for which it 
  reaches a few percent at the highest redshift.

\item[-] When we move to the more demanding task of computing the
  precise density values, our density reconstruction under-predicts
  galaxy counts in the highest densities, for all cylinder radii and
  at all redshifts, by a factor ranging from $\sim70$\% for the
  smallest radii ($R_1$ and $R_2$) to $\sim20$\% for the largest ones
  ($R_8$ and $R_{10}$). Moreover, the error budget at low density is
  dominated by the random error, while at the highest densities it is
  dominated by the systematic error. This is true for all radii, with
  the exception of $R_8$ and $R_{10}$ at high density, where the
  random and systematic error are comparable.

\item[-] As a qualitative test case, we verified that in a given
redshift bin at $z\sim1.5$, chosen to include at least a small sample
of relatively-high-mass galaxy clusters ($\log(M/M_{\odot}) > 13.5$), the galaxy
clusters reside in the highest tail of the density distribution, when
the density is estimated on a scale of $R=1\mpcoh$.

\end{itemize}

Our results show that environmental studies can be efficiently
performed in photometric samples if spectroscopic information is
available for a smaller sample of objects that sparsely samples the
same volume. Namely, we are able to robustly identify the highest
density peaks from the lowest density regions using slitless
spectroscopy coupled with a deep photometric sample, like in the
Euclid Deep survey.

In particular, the redshift range probed by this survey ($0.7\lesssim
z \lesssim 2$) covers the epoch when the relation between local
environment and galaxy properties has been observed to change with
respect to the local universe (\citealp{cucciati06} and
\citealp{lin16} at $z\gtrsim1.2$, \citealp{elbaz2007} at
$z\gtrsim1$). Provided that the spectroscopic redshift information is
crucial to study small-scale environment, only relatively small
ground-based spectroscopic surveys have been able so far to study
galaxy evolution at $z>1$ (e.g. VVDS and DEEP2), while the largest
ones are too shallow (see e.g. zCOSMOS and VIPERS) to probe these
epochs. The possibility to verify these earlier findings at $z>1$ with
more statistics and on smaller scales will be one of the most
important scientific results from the ancillary science foreseen for
the Euclid survey.


\section*{Acknowledgements}

OC and AC acknowledge the support from grants ASI-INAF I/023/12/0
``Attivit\`a relative alla fase B2/C per la missione Euclid'' and MIUR
PRIN 2010-2011 ``The dark Universe and the cosmic evolution of baryons:
from current surveys to Euclid''. OC acknowledges the support from
grant PRIN INAF 2014 ``VIPERS: unveiling the combined evolution of
galaxies and large-scale structure at $0.5<z<1.2$''. OC thanks Sandro
Bardelli, Fabio Bellagamba, Lauro Moscardini and Mauro Roncarelli, for
useful discussions. The Millennium Simulation databases used in this
paper and the web application providing online access to them were
constructed as part of the activities of the German Astrophysical
Virtual Observatory.



\bibliographystyle{mnras}
\bibliography{biblio} 

\begin{thebibliography}{}
\makeatletter
\relax
\def\mn@urlcharsother{\let\do\@makeother \do\$\do\&\do\#\do\^\do\_\do\%\do\~}
\def\mn@doi{\begingroup\mn@urlcharsother \@ifnextchar [ {\mn@doi@}
  {\mn@doi@[]}}
\def\mn@doi@[#1]#2{\def\@tempa{#1}\ifx\@tempa\@empty \href
  {http://dx.doi.org/#2} {doi:#2}\else \href {http://dx.doi.org/#2} {#1}\fi
  \endgroup}
\def\mn@eprint#1#2{\mn@eprint@#1:#2::\@nil}
\def\mn@eprint@arXiv#1{\href {http://arxiv.org/abs/#1} {{\tt arXiv:#1}}}
\def\mn@eprint@dblp#1{\href {http://dblp.uni-trier.de/rec/bibtex/#1.xml}
  {dblp:#1}}
\def\mn@eprint@#1:#2:#3:#4\@nil{\def\@tempa {#1}\def\@tempb {#2}\def\@tempc
  {#3}\ifx \@tempc \@empty \let \@tempc \@tempb \let \@tempb \@tempa \fi \ifx
  \@tempb \@empty \def\@tempb {arXiv}\fi \@ifundefined
  {mn@eprint@\@tempb}{\@tempb:\@tempc}{\expandafter \expandafter \csname
  mn@eprint@\@tempb\endcsname \expandafter{\@tempc}}}

\bibitem[\protect\citeauthoryear{{Adami} et~al.,}{{Adami}
  et~al.}{2010}]{adami10}
{Adami} C.,  et~al., 2010, \mn@doi [\aap] {10.1051/0004-6361/200913067}, \href
  {http://adsabs.harvard.edu/abs/2010A%26A...509A..81A} {509, A81}

\bibitem[\protect\citeauthoryear{{Arag{\'o}n-Calvo}, {van de Weygaert}  \&
  {Jones}}{{Arag{\'o}n-Calvo} et~al.}{2010}]{aragon_calvo10_multiscale}
{Arag{\'o}n-Calvo} M.~A.,  {van de Weygaert} R.,   {Jones} B.~J.~T.,  2010,
  \mn@doi [\mnras] {10.1111/j.1365-2966.2010.17263.x}, \href
  {http://adsabs.harvard.edu/abs/2010MNRAS.408.2163A} {408, 2163}

\bibitem[\protect\citeauthoryear{{Atek} et~al.,}{{Atek}
  et~al.}{2010}]{atek10_WISP}
{Atek} H.,  et~al., 2010, \mn@doi [\apj] {10.1088/0004-637X/723/1/104}, \href
  {http://adsabs.harvard.edu/abs/2010ApJ...723..104A} {723, 104}

\bibitem[\protect\citeauthoryear{{Atek} et~al.,}{{Atek}
  et~al.}{2011}]{atek11_WISP}
{Atek} H.,  et~al., 2011, \mn@doi [\apj] {10.1088/0004-637X/743/2/121}, \href
  {http://adsabs.harvard.edu/abs/2011ApJ...743..121A} {743, 121}

\bibitem[\protect\citeauthoryear{{Bel} et~al.,}{{Bel} et~al.}{2016}]{bel15}
{Bel} J.,  et~al., 2016, \mn@doi [\aap] {10.1051/0004-6361/201526455}, \href
  {http://adsabs.harvard.edu/abs/2016A%26A...588A..51B} {588, A51}

\bibitem[\protect\citeauthoryear{{Bellagamba}, {Maturi}, {Hamana},
  {Meneghetti}, {Miyazaki}  \& {Moscardini}}{{Bellagamba}
  et~al.}{2011}]{bellagamba11}
{Bellagamba} F.,  {Maturi} M.,  {Hamana} T.,  {Meneghetti} M.,  {Miyazaki} S.,
   {Moscardini} L.,  2011, \mn@doi [\mnras] {10.1111/j.1365-2966.2011.18202.x},
  \href {http://adsabs.harvard.edu/abs/2011MNRAS.413.1145B} {413, 1145}

\bibitem[\protect\citeauthoryear{{Bielby} et~al.,}{{Bielby}
  et~al.}{2012}]{bielby12_Hband}
{Bielby} R.,  et~al., 2012, \mn@doi [\aap] {10.1051/0004-6361/201118547}, \href
  {http://adsabs.harvard.edu/abs/2012A%26A...545A..23B} {545, A23}

\bibitem[\protect\citeauthoryear{{Bolzonella} et~al.,}{{Bolzonella}
  et~al.}{2010}]{bolzonella10}
{Bolzonella} M.,  et~al., 2010, \mn@doi [\aap] {10.1051/0004-6361/200912801},
  \href {http://adsabs.harvard.edu/abs/2010A%26A...524A..76B} {524, A76}

\bibitem[\protect\citeauthoryear{{Cappi} et~al.,}{{Cappi}
  et~al.}{2015}]{cappi15}
{Cappi} A.,  et~al., 2015, \mn@doi [\aap] {10.1051/0004-6361/201525727}, \href
  {http://adsabs.harvard.edu/abs/2015A%26A...579A..70C} {579, A70}

\bibitem[\protect\citeauthoryear{{Cassata} et~al.,}{{Cassata}
  et~al.}{2007}]{cassata07}
{Cassata} P.,  et~al., 2007, \mn@doi [\apjs] {10.1086/516591}, \href
  {http://adsabs.harvard.edu/abs/2007ApJS..172..270C} {172, 270}

\bibitem[\protect\citeauthoryear{{Chen} et~al.,}{{Chen}
  et~al.}{2002}]{chen02_Hband}
{Chen} H.-W.,  et~al., 2002, \mn@doi [\apj] {10.1086/339426}, \href
  {http://adsabs.harvard.edu/abs/2002ApJ...570...54C} {570, 54}

\bibitem[\protect\citeauthoryear{{Chiang}, {Overzier}  \& {Gebhardt}}{{Chiang}
  et~al.}{2013}]{chiang13}
{Chiang} Y.-K.,  {Overzier} R.,   {Gebhardt} K.,  2013, \mn@doi [\apj]
  {10.1088/0004-637X/779/2/127}, \href
  {http://adsabs.harvard.edu/abs/2013ApJ...779..127C} {779, 127}

\bibitem[\protect\citeauthoryear{{Chiang}, {Overzier}  \& {Gebhardt}}{{Chiang}
  et~al.}{2014}]{chiang14}
{Chiang} Y.-K.,  {Overzier} R.,   {Gebhardt} K.,  2014, \mn@doi [\apjl]
  {10.1088/2041-8205/782/1/L3}, \href
  {http://adsabs.harvard.edu/abs/2014ApJ...782L...3C} {782, L3}

\bibitem[\protect\citeauthoryear{{Christodoulou} et~al.,}{{Christodoulou}
  et~al.}{2012}]{christodoulou12_clustering}
{Christodoulou} L.,  et~al., 2012, \mn@doi [\mnras]
  {10.1111/j.1365-2966.2012.21434.x}, \href
  {http://adsabs.harvard.edu/abs/2012MNRAS.425.1527C} {425, 1527}

\bibitem[\protect\citeauthoryear{{Coil} et~al.,}{{Coil} et~al.}{2008}]{coil08}
{Coil} A.~L.,  et~al., 2008, \mn@doi [\apj] {10.1086/523639}, \href
  {http://adsabs.harvard.edu/abs/2008ApJ...672..153C} {672, 153}

\bibitem[\protect\citeauthoryear{{Cooper}, {Newman}, {Madgwick}, {Gerke}, {Yan}
   \& {Davis}}{{Cooper} et~al.}{2005}]{cooper2005}
{Cooper} M.~C.,  {Newman} J.~A.,  {Madgwick} D.~S.,  {Gerke} B.~F.,  {Yan} R.,
   {Davis} M.,  2005, \mn@doi [\apj] {10.1086/432868}, \href
  {http://adsabs.harvard.edu/cgi-bin/nph-bib_query?bibcode=2005ApJ...634..833C&db_key=AST}
  {634, 833}

\bibitem[\protect\citeauthoryear{{Crist{\'o}bal-Hornillos}
  et~al.,}{{Crist{\'o}bal-Hornillos} et~al.}{2009}]{cristobal09_Hband}
{Crist{\'o}bal-Hornillos} D.,  et~al., 2009, \mn@doi [\apj]
  {10.1088/0004-637X/696/2/1554}, \href
  {http://adsabs.harvard.edu/abs/2009ApJ...696.1554C} {696, 1554}

\bibitem[\protect\citeauthoryear{{Cucciati} et~al.,}{{Cucciati}
  et~al.}{2006}]{cucciati06}
{Cucciati} O.,  et~al., 2006, \mn@doi [\aap] {10.1051/0004-6361:20065161},
  \href {http://adsabs.harvard.edu/abs/2006A%26A...458...39C} {458, 39}

\bibitem[\protect\citeauthoryear{{Cucciati} et~al.,}{{Cucciati}
  et~al.}{2010}]{cucciati10_zCOSMOS}
{Cucciati} O.,  et~al., 2010, \mn@doi [\aap] {10.1051/0004-6361/200912585},
  \href {http://adsabs.harvard.edu/abs/2010A%26A...524A...2C} {524, A2}

\bibitem[\protect\citeauthoryear{{Cucciati} et~al.,}{{Cucciati}
  et~al.}{2012a}]{cucciati12_SFRD}
{Cucciati} O.,  et~al., 2012a, \mn@doi [\aap] {10.1051/0004-6361/201118010},
  \href {http://adsabs.harvard.edu/abs/2012A%26A...539A..31C} {539, A31}

\bibitem[\protect\citeauthoryear{{Cucciati} et~al.,}{{Cucciati}
  et~al.}{2012b}]{cucciati12_MILL}
{Cucciati} O.,  et~al., 2012b, \mn@doi [\aap] {10.1051/0004-6361/201219554},
  \href {http://adsabs.harvard.edu/abs/2012A%26A...548A.108C} {548, A108}

\bibitem[\protect\citeauthoryear{{Cucciati} et~al.,}{{Cucciati}
  et~al.}{2014}]{cucciati14}
{Cucciati} O.,  et~al., 2014, \mn@doi [\aap] {10.1051/0004-6361/201423409},
  \href {http://adsabs.harvard.edu/abs/2014A%26A...565A..67C} {565, A67}

\bibitem[\protect\citeauthoryear{{Darvish}, {Mobasher}, {Sobral}, {Scoville}
  \& {Aragon-Calvo}}{{Darvish} et~al.}{2015}]{darvish15_env}
{Darvish} B.,  {Mobasher} B.,  {Sobral} D.,  {Scoville} N.,   {Aragon-Calvo}
  M.,  2015, \mn@doi [\apj] {10.1088/0004-637X/805/2/121}, \href
  {http://adsabs.harvard.edu/abs/2015ApJ...805..121D} {805, 121}

\bibitem[\protect\citeauthoryear{{Davidzon} et~al.,}{{Davidzon}
  et~al.}{2016}]{davidzon15}
{Davidzon} I.,  et~al., 2016, \mn@doi [\aap] {10.1051/0004-6361/201527129},
  \href {http://adsabs.harvard.edu/abs/2016A%26A...586A..23D} {586, A23}

\bibitem[\protect\citeauthoryear{{Di Porto} et~al.,}{{Di Porto}
  et~al.}{2014}]{diporto14}
{Di Porto} C.,  et~al., 2014, preprint, \href
  {http://adsabs.harvard.edu/abs/2014arXiv1406.6692D} {} (\mn@eprint {arXiv}
  {1406.6692})

\bibitem[\protect\citeauthoryear{{Dressler} et~al.,}{{Dressler}
  et~al.}{2012}]{dressler12_WFIRST}
{Dressler} A.,  et~al., 2012, preprint, \href
  {http://adsabs.harvard.edu/abs/2012arXiv1210.7809D} {} (\mn@eprint {arXiv}
  {1210.7809})

\bibitem[\protect\citeauthoryear{{Elbaz} et~al.,}{{Elbaz}
  et~al.}{2007}]{elbaz2007}
{Elbaz} D.,  et~al., 2007, \mn@doi [\aap] {10.1051/0004-6361:20077525}, \href
  {http://adsabs.harvard.edu/abs/2007A%26A...468...33E} {468, 33}

\bibitem[\protect\citeauthoryear{{Fossati} et~al.,}{{Fossati}
  et~al.}{2015}]{fossati15}
{Fossati} M.,  et~al., 2015, \mn@doi [\mnras] {10.1093/mnras/stu2255}, \href
  {http://adsabs.harvard.edu/abs/2015MNRAS.446.2582F} {446, 2582}

\bibitem[\protect\citeauthoryear{{Frith}, {Metcalfe}  \& {Shanks}}{{Frith}
  et~al.}{2006}]{frith06_Hband}
{Frith} W.~J.,  {Metcalfe} N.,   {Shanks} T.,  2006, \mn@doi [\mnras]
  {10.1111/j.1365-2966.2006.10736.x}, \href
  {http://adsabs.harvard.edu/abs/2006MNRAS.371.1601F} {371, 1601}

\bibitem[\protect\citeauthoryear{{Garilli} et~al.,}{{Garilli}
  et~al.}{2014}]{garilli2014_VIPERS}
{Garilli} B.,  et~al., 2014, \mn@doi [\aap] {10.1051/0004-6361/201322790},
  \href {http://adsabs.harvard.edu/abs/2014A%26A...562A..23G} {562, A23}

\bibitem[\protect\citeauthoryear{{Geach}, {Smail}, {Best}, {Kurk}, {Casali},
  {Ivison}  \& {Coppin}}{{Geach} et~al.}{2008}]{geach08_HaLF}
{Geach} J.~E.,  {Smail} I.,  {Best} P.~N.,  {Kurk} J.,  {Casali} M.,  {Ivison}
  R.~J.,   {Coppin} K.,  2008, \mn@doi [\mnras]
  {10.1111/j.1365-2966.2008.13481.x}, \href
  {http://adsabs.harvard.edu/abs/2008MNRAS.388.1473G} {388, 1473}

\bibitem[\protect\citeauthoryear{{Geach} et~al.,}{{Geach}
  et~al.}{2010}]{geach10_HaLF}
{Geach} J.~E.,  et~al., 2010, \mn@doi [\mnras]
  {10.1111/j.1365-2966.2009.15977.x}, \href
  {http://adsabs.harvard.edu/abs/2010MNRAS.402.1330G} {402, 1330}

\bibitem[\protect\citeauthoryear{{Geach}, {Sobral}, {Hickox}, {Wake}, {Smail},
  {Best}, {Baugh}  \& {Stott}}{{Geach} et~al.}{2012}]{geach12_haXi}
{Geach} J.~E.,  {Sobral} D.,  {Hickox} R.~C.,  {Wake} D.~A.,  {Smail} I.,
  {Best} P.~N.,  {Baugh} C.~M.,   {Stott} J.~P.,  2012, \mn@doi [\mnras]
  {10.1111/j.1365-2966.2012.21725.x}, \href
  {http://adsabs.harvard.edu/abs/2012MNRAS.426..679G} {426, 679}

\bibitem[\protect\citeauthoryear{{Gonzalez-Perez}, {Lacey}, {Baugh}, {Lagos},
  {Helly}, {Campbell}  \& {Mitchell}}{{Gonzalez-Perez}
  et~al.}{2014}]{gonzalez_perez14_SAM}
{Gonzalez-Perez} V.,  {Lacey} C.~G.,  {Baugh} C.~M.,  {Lagos} C.~D.~P.,
  {Helly} J.,  {Campbell} D.~J.~R.,   {Mitchell} P.~D.,  2014, \mn@doi [\mnras]
  {10.1093/mnras/stt2410}, \href
  {http://adsabs.harvard.edu/abs/2014MNRAS.439..264G} {439, 264}

\bibitem[\protect\citeauthoryear{{Green} et~al.,}{{Green}
  et~al.}{2012}]{green12_WFIRST}
{Green} J.,  et~al., 2012, preprint, \href
  {http://adsabs.harvard.edu/abs/2012arXiv1208.4012G} {} (\mn@eprint {arXiv}
  {1208.4012})

\bibitem[\protect\citeauthoryear{{Gunn} \& {Gott}}{{Gunn} \&
  {Gott}}{1972}]{gunn_gott1972}
{Gunn} J.~E.,  {Gott} J.~R.~I.,  1972, \apj, \href
  {http://adsabs.harvard.edu/cgi-bin/nph-bib_query?bibcode=1972ApJ...176....1G&db_key=AST}
  {176, 1}

\bibitem[\protect\citeauthoryear{{Guzzo} et~al.,}{{Guzzo}
  et~al.}{2007}]{guzzo2007_COSMOS}
{Guzzo} L.,  et~al., 2007, \mn@doi [\apjs] {10.1086/516588}, \href
  {http://adsabs.harvard.edu/abs/2007ApJS..172..254G} {172, 254}

\bibitem[\protect\citeauthoryear{{Guzzo} et~al.,}{{Guzzo}
  et~al.}{2014}]{guzzo14_vipers}
{Guzzo} L.,  et~al., 2014, \mn@doi [\aap] {10.1051/0004-6361/201321489}, \href
  {http://adsabs.harvard.edu/abs/2014A%26A...566A.108G} {566, A108}

\bibitem[\protect\citeauthoryear{{Ilbert} et~al.,}{{Ilbert}
  et~al.}{2013}]{ilbert13}
{Ilbert} O.,  et~al., 2013, \mn@doi [\aap] {10.1051/0004-6361/201321100}, \href
  {http://adsabs.harvard.edu/abs/2013A%26A...556A..55I} {556, A55}

\bibitem[\protect\citeauthoryear{{Ilbert} et~al.,}{{Ilbert}
  et~al.}{2015}]{ilbert15}
{Ilbert} O.,  et~al., 2015, \mn@doi [\aap] {10.1051/0004-6361/201425176}, \href
  {http://adsabs.harvard.edu/abs/2015A%26A...579A...2I} {579, A2}

\bibitem[\protect\citeauthoryear{{Jeon}, {Im}, {Kang}, {Lee}  \&
  {Matsuhara}}{{Jeon} et~al.}{2014}]{jeon14_Hband}
{Jeon} Y.,  {Im} M.,  {Kang} E.,  {Lee} H.~M.,   {Matsuhara} H.,  2014, \mn@doi
  [\apjs] {10.1088/0067-0049/214/2/20}, \href
  {http://adsabs.harvard.edu/abs/2014ApJS..214...20J} {214, 20}

\bibitem[\protect\citeauthoryear{{Jian} et~al.,}{{Jian}
  et~al.}{2014}]{jian14_groups}
{Jian} H.-Y.,  et~al., 2014, \mn@doi [\apj] {10.1088/0004-637X/788/2/109},
  \href {http://adsabs.harvard.edu/abs/2014ApJ...788..109J} {788, 109}

\bibitem[\protect\citeauthoryear{{Kodama}, {Tanaka}, {Kajisawa}, {Kurk},
  {Venemans}, {De Breuck}, {Vernet}  \& {Lidman}}{{Kodama}
  et~al.}{2007}]{kodama07}
{Kodama} T.,  {Tanaka} I.,  {Kajisawa} M.,  {Kurk} J.,  {Venemans} B.,  {De
  Breuck} C.,  {Vernet} J.,   {Lidman} C.,  2007, \mn@doi [\mnras]
  {10.1111/j.1365-2966.2007.11739.x}, \href
  {http://adsabs.harvard.edu/abs/2007MNRAS.377.1717K} {377, 1717}

\bibitem[\protect\citeauthoryear{{Komatsu} et~al.,}{{Komatsu}
  et~al.}{2011}]{komatsu11_WMAP7}
{Komatsu} E.,  et~al., 2011, \mn@doi [\apjs] {10.1088/0067-0049/192/2/18},
  \href {http://adsabs.harvard.edu/abs/2011ApJS..192...18K} {192, 18}

\bibitem[\protect\citeauthoryear{{Kova{\v c}} et~al.,}{{Kova{\v c}}
  et~al.}{2010}]{kovac2010_density}
{Kova{\v c}} K.,  et~al., 2010, \mn@doi [\apj] {10.1088/0004-637X/708/1/505},
  \href {http://adsabs.harvard.edu/abs/2010ApJ...708..505K} {708, 505}

\bibitem[\protect\citeauthoryear{{Lagos}, {Bayet}, {Baugh}, {Lacey}, {Bell},
  {Fanidakis}  \& {Geach}}{{Lagos} et~al.}{2012}]{lagos12_galform}
{Lagos} C.~d.~P.,  {Bayet} E.,  {Baugh} C.~M.,  {Lacey} C.~G.,  {Bell} T.~A.,
  {Fanidakis} N.,   {Geach} J.~E.,  2012, \mn@doi [\mnras]
  {10.1111/j.1365-2966.2012.21905.x}, \href
  {http://adsabs.harvard.edu/abs/2012MNRAS.426.2142L} {426, 2142}

\bibitem[\protect\citeauthoryear{{Lai} et~al.,}{{Lai}
  et~al.}{2015}]{lai15_photoz}
{Lai} C.-C.,  et~al., 2015, preprint, \href
  {http://adsabs.harvard.edu/abs/2015arXiv150101398L} {} (\mn@eprint {arXiv}
  {1501.01398})

\bibitem[\protect\citeauthoryear{{Larson}, {Tinsley}  \& {Caldwell}}{{Larson}
  et~al.}{1980}]{larson1980}
{Larson} R.~B.,  {Tinsley} B.~M.,   {Caldwell} C.~N.,  1980, \mn@doi [\apj]
  {10.1086/157917}, \href
  {http://adsabs.harvard.edu/cgi-bin/nph-bib_query?bibcode=1980ApJ...237..692L&db_key=AST}
  {237, 692}

\bibitem[\protect\citeauthoryear{{Laureijs} et~al.,}{{Laureijs}
  et~al.}{2011}]{redbook}
{Laureijs} R.,  et~al., 2011, ArXiv:1110.3193, \href
  {http://adsabs.harvard.edu/abs/2011arXiv1110.3193L} {}

\bibitem[\protect\citeauthoryear{{Lemson} \& {Virgo Consortium}}{{Lemson} \&
  {Virgo Consortium}}{2006}]{lemson06}
{Lemson} G.,  {Virgo Consortium} t.,  2006, preprint, \href
  {http://adsabs.harvard.edu/abs/2006astro.ph..8019L} {} (\mn@eprint {arXiv}
  {astro-ph/0608019})

\bibitem[\protect\citeauthoryear{{Li}, {Kauffmann}, {Jing}, {White},
  {B{\"o}rner}  \& {Cheng}}{{Li} et~al.}{2006}]{li06}
{Li} C.,  {Kauffmann} G.,  {Jing} Y.~P.,  {White} S.~D.~M.,  {B{\"o}rner} G.,
  {Cheng} F.~Z.,  2006, \mn@doi [\mnras] {10.1111/j.1365-2966.2006.10066.x},
  \href {http://adsabs.harvard.edu/abs/2006MNRAS.368...21L} {368, 21}

\bibitem[\protect\citeauthoryear{{Lilly} et~al.,}{{Lilly}
  et~al.}{2007}]{lilly07_zcosmos}
{Lilly} S.~J.,  et~al., 2007, \mn@doi [\apjs] {10.1086/516589}, \href
  {http://adsabs.harvard.edu/abs/2007ApJS..172...70L} {172, 70}

\bibitem[\protect\citeauthoryear{{Lin} et~al.,}{{Lin} et~al.}{2016}]{lin16}
{Lin} L.,  et~al., 2016, \mn@doi [\apj] {10.3847/0004-637X/817/2/97}, \href
  {http://adsabs.harvard.edu/abs/2016ApJ...817...97L} {817, 97}

\bibitem[\protect\citeauthoryear{{Ly}, {Lee}, {Dale}, {Momcheva}, {Salim},
  {Staudaher}, {Moore}  \& {Finn}}{{Ly} et~al.}{2011}]{ly11}
{Ly} C.,  {Lee} J.~C.,  {Dale} D.~A.,  {Momcheva} I.,  {Salim} S.,  {Staudaher}
  S.,  {Moore} C.~A.,   {Finn} R.,  2011, \mn@doi [\apj]
  {10.1088/0004-637X/726/2/109}, \href
  {http://adsabs.harvard.edu/abs/2011ApJ...726..109L} {726, 109}

\bibitem[\protect\citeauthoryear{{Madau} \& {Dickinson}}{{Madau} \&
  {Dickinson}}{2014}]{madau_dickinson14_CSFH}
{Madau} P.,  {Dickinson} M.,  2014, \mn@doi [\araa]
  {10.1146/annurev-astro-081811-125615}, \href
  {http://adsabs.harvard.edu/abs/2014ARA%26A..52..415M} {52, 415}

\bibitem[\protect\citeauthoryear{{Malavasi}, {Pozzetti}, {Cucciati}, {Bardelli}
   \& {Cimatti}}{{Malavasi} et~al.}{2016}]{malavasi16}
{Malavasi} N.,  {Pozzetti} L.,  {Cucciati} O.,  {Bardelli} S.,   {Cimatti} A.,
  2016, \mn@doi [\aap] {10.1051/0004-6361/201526718}, \href
  {http://adsabs.harvard.edu/abs/2016A%26A...585A.116M} {585, A116}

\bibitem[\protect\citeauthoryear{{Marulli} et~al.,}{{Marulli}
  et~al.}{2013}]{marulli2013_clustering}
{Marulli} F.,  et~al., 2013, \mn@doi [\aap] {10.1051/0004-6361/201321476},
  \href {http://adsabs.harvard.edu/abs/2013A%26A...557A..17M} {557, A17}

\bibitem[\protect\citeauthoryear{{Marulli}, {Veropalumbo}  \&
  {Moresco}}{{Marulli} et~al.}{2016}]{marulli16_CBL}
{Marulli} F.,  {Veropalumbo} A.,   {Moresco} M.,  2016, \mn@doi [Astronomy and
  Computing] {10.1016/j.ascom.2016.01.005}, \href
  {http://adsabs.harvard.edu/abs/2016A%26C....14...35M} {14, 35}

\bibitem[\protect\citeauthoryear{{Mazure} et~al.,}{{Mazure}
  et~al.}{2007}]{mazure07}
{Mazure} A.,  et~al., 2007, \mn@doi [\aap] {10.1051/0004-6361:20066379}, \href
  {http://adsabs.harvard.edu/abs/2007A%26A...467...49M} {467, 49}

\bibitem[\protect\citeauthoryear{{McCarthy} et~al.,}{{McCarthy}
  et~al.}{1999}]{mccarthy99_EL}
{McCarthy} P.~J.,  et~al., 1999, \mn@doi [\apj] {10.1086/307491}, \href
  {http://adsabs.harvard.edu/abs/1999ApJ...520..548M} {520, 548}

\bibitem[\protect\citeauthoryear{{McNaught-Roberts} et~al.,}{{McNaught-Roberts}
  et~al.}{2014}]{McNaught_Roberts14_GAMA_LF}
{McNaught-Roberts} T.,  et~al., 2014, \mn@doi [\mnras] {10.1093/mnras/stu1886},
  \href {http://adsabs.harvard.edu/abs/2014MNRAS.445.2125M} {445, 2125}

\bibitem[\protect\citeauthoryear{{Meneux} et~al.,}{{Meneux}
  et~al.}{2008}]{meneux08}
{Meneux} B.,  et~al., 2008, \mn@doi [\aap] {10.1051/0004-6361:20078182}, \href
  {http://adsabs.harvard.edu/abs/2008A%26A...478..299M} {478, 299}

\bibitem[\protect\citeauthoryear{{Merson} et~al.,}{{Merson}
  et~al.}{2013}]{merson13_lightcones}
{Merson} A.~I.,  et~al., 2013, \mn@doi [\mnras] {10.1093/mnras/sts355}, \href
  {http://adsabs.harvard.edu/abs/2013MNRAS.429..556M} {429, 556}

\bibitem[\protect\citeauthoryear{{Metcalfe}, {Shanks}, {Weilbacher},
  {McCracken}, {Fong}  \& {Thompson}}{{Metcalfe}
  et~al.}{2006}]{metcalfe06_Hband}
{Metcalfe} N.,  {Shanks} T.,  {Weilbacher} P.~M.,  {McCracken} H.~J.,  {Fong}
  R.,   {Thompson} D.,  2006, \mn@doi [\mnras]
  {10.1111/j.1365-2966.2006.10534.x}, \href
  {http://adsabs.harvard.edu/abs/2006MNRAS.370.1257M} {370, 1257}

\bibitem[\protect\citeauthoryear{{Micheletti} et~al.,}{{Micheletti}
  et~al.}{2014}]{micheletti14_voids}
{Micheletti} D.,  et~al., 2014, \mn@doi [\aap] {10.1051/0004-6361/201424107},
  \href {http://adsabs.harvard.edu/abs/2014A%26A...570A.106M} {570, A106}

\bibitem[\protect\citeauthoryear{{Moore}, {Katz}, {Lake}, {Dressler}  \&
  {Oemler}}{{Moore} et~al.}{1996}]{moore1996}
{Moore} B.,  {Katz} N.,  {Lake} G.,  {Dressler} A.,   {Oemler} A.,  1996,
  \mn@doi [\nat] {10.1038/379613a0}, \href
  {http://adsabs.harvard.edu/cgi-bin/nph-bib_query?bibcode=1996Natur.379..613M&db_key=AST}
  {379, 613}

\bibitem[\protect\citeauthoryear{{Moy}, {Barmby}, {Rigopoulou}, {Huang},
  {Willner}  \& {Fazio}}{{Moy} et~al.}{2003}]{moy03_Hband}
{Moy} E.,  {Barmby} P.,  {Rigopoulou} D.,  {Huang} J.-S.,  {Willner} S.~P.,
  {Fazio} G.~G.,  2003, \mn@doi [\aap] {10.1051/0004-6361:20030245}, \href
  {http://adsabs.harvard.edu/abs/2003A%26A...403..493M} {403, 493}

\bibitem[\protect\citeauthoryear{{Oke}}{{Oke}}{1974}]{oke74}
{Oke} J.~B.,  1974, \mn@doi [\apjs] {10.1086/190287}, \href
  {http://adsabs.harvard.edu/abs/1974ApJS...27...21O} {27, 21}

\bibitem[\protect\citeauthoryear{{Peng} et~al.,}{{Peng}
  et~al.}{2010}]{peng2010_picture}
{Peng} Y.-j.,  et~al., 2010, \mn@doi [\apj] {10.1088/0004-637X/721/1/193},
  \href {http://adsabs.harvard.edu/abs/2010ApJ...721..193P} {721, 193}

\bibitem[\protect\citeauthoryear{{Pozzetti} et~al.,}{{Pozzetti}
  et~al.}{2016}]{pozzetti16_Ha}
{Pozzetti} L.,  et~al., 2016, \mn@doi [\aap] {10.1051/0004-6361/201527081},
  \href {http://adsabs.harvard.edu/abs/2016A%26A...590A...3P} {590, A3}

\bibitem[\protect\citeauthoryear{{Retzlaff}, {Rosati}, {Dickinson}, {Vandame},
  {Rit{\'e}}, {Nonino}, {Cesarsky}  \& {GOODS Team}}{{Retzlaff}
  et~al.}{2010}]{retzlaff10_Hband}
{Retzlaff} J.,  {Rosati} P.,  {Dickinson} M.,  {Vandame} B.,  {Rit{\'e}} C.,
  {Nonino} M.,  {Cesarsky} C.,   {GOODS Team} 2010, \mn@doi [\aap]
  {10.1051/0004-6361/200912940}, \href
  {http://adsabs.harvard.edu/abs/2010A%26A...511A..50R} {511, A50}

\bibitem[\protect\citeauthoryear{{Scoville} et~al.,}{{Scoville}
  et~al.}{2007}]{scoville07_env}
{Scoville} N.,  et~al., 2007, \mn@doi [\apjs] {10.1086/516751}, \href
  {http://adsabs.harvard.edu/abs/2007ApJS..172..150S} {172, 150}

\bibitem[\protect\citeauthoryear{{Scoville} et~al.,}{{Scoville}
  et~al.}{2013}]{scoville13_env}
{Scoville} N.,  et~al., 2013, \mn@doi [\apjs] {10.1088/0067-0049/206/1/3},
  \href {http://adsabs.harvard.edu/abs/2013ApJS..206....3S} {206, 3}

\bibitem[\protect\citeauthoryear{{Shim}, {Colbert}, {Teplitz}, {Henry},
  {Malkan}, {McCarthy}  \& {Yan}}{{Shim} et~al.}{2009}]{shim09_HaLF}
{Shim} H.,  {Colbert} J.,  {Teplitz} H.,  {Henry} A.,  {Malkan} M.,  {McCarthy}
  P.,   {Yan} L.,  2009, \mn@doi [\apj] {10.1088/0004-637X/696/1/785}, \href
  {http://adsabs.harvard.edu/abs/2009ApJ...696..785S} {696, 785}

\bibitem[\protect\citeauthoryear{{Smith}, {Hopkins}, {Hunstead}  \&
  {Pimbblet}}{{Smith} et~al.}{2012}]{smith12_multiscale}
{Smith} A.~G.,  {Hopkins} A.~M.,  {Hunstead} R.~W.,   {Pimbblet} K.~A.,  2012,
  \mn@doi [\mnras] {10.1111/j.1365-2966.2012.20400.x}, \href
  {http://adsabs.harvard.edu/abs/2012MNRAS.422...25S} {422, 25}

\bibitem[\protect\citeauthoryear{{Spitler} et~al.,}{{Spitler}
  et~al.}{2012}]{spitler12}
{Spitler} L.~R.,  et~al., 2012, \mn@doi [\apjl] {10.1088/2041-8205/748/2/L21},
  \href {http://adsabs.harvard.edu/abs/2012ApJ...748L..21S} {748, L21}

\bibitem[\protect\citeauthoryear{{Springel} et~al.,}{{Springel}
  et~al.}{2005}]{springel2005_MILL}
{Springel} V.,  et~al., 2005, \mn@doi [\nat] {10.1038/nature03597}, \href
  {http://adsabs.harvard.edu/abs/2005Natur.435..629S} {435, 629}

\bibitem[\protect\citeauthoryear{{Teplitz}, {Gardner}, {Malumuth}  \&
  {Heap}}{{Teplitz} et~al.}{1998}]{teplitz98_Hband}
{Teplitz} H.~I.,  {Gardner} J.~P.,  {Malumuth} E.~M.,   {Heap} S.~R.,  1998,
  \mn@doi [\apjl] {10.1086/311665}, \href
  {http://adsabs.harvard.edu/abs/1998ApJ...507L..17T} {507, L17}

\bibitem[\protect\citeauthoryear{{Thompson}, {Storrie-Lombardi}, {Weymann},
  {Rieke}, {Schneider}, {Stobie}  \& {Lytle}}{{Thompson}
  et~al.}{1999}]{thompson99_Hband}
{Thompson} R.~I.,  {Storrie-Lombardi} L.~J.,  {Weymann} R.~J.,  {Rieke} M.~J.,
  {Schneider} G.,  {Stobie} E.,   {Lytle} D.,  1999, \mn@doi [\aj]
  {10.1086/300705}, \href {http://adsabs.harvard.edu/abs/1999AJ....117...17T}
  {117, 17}

\bibitem[\protect\citeauthoryear{{Toomre} \& {Toomre}}{{Toomre} \&
  {Toomre}}{1972}]{toomre1972}
{Toomre} A.,  {Toomre} J.,  1972, \apj, \href
  {http://adsabs.harvard.edu/cgi-bin/nph-bib_query?bibcode=1972ApJ...178..623T&db_key=AST}
  {178, 623}

\bibitem[\protect\citeauthoryear{{Wilman}, {Zibetti}  \&
  {Budav{\'a}ri}}{{Wilman} et~al.}{2010}]{wilman10}
{Wilman} D.~J.,  {Zibetti} S.,   {Budav{\'a}ri} T.,  2010, \mn@doi [\mnras]
  {10.1111/j.1365-2966.2010.16845.x}, \href
  {http://adsabs.harvard.edu/abs/2010MNRAS.406.1701W} {406, 1701}

\bibitem[\protect\citeauthoryear{{Yan}, {McCarthy}, {Freudling}, {Teplitz},
  {Malumuth}, {Weymann}  \& {Malkan}}{{Yan} et~al.}{1999}]{yan99_HaLF}
{Yan} L.,  {McCarthy} P.~J.,  {Freudling} W.,  {Teplitz} H.~I.,  {Malumuth}
  E.~M.,  {Weymann} R.~J.,   {Malkan} M.~A.,  1999, \mn@doi [\apjl]
  {10.1086/312099}, \href {http://adsabs.harvard.edu/abs/1999ApJ...519L..47Y}
  {519, L47}

\makeatother
\end{thebibliography}


\appendix


\section{The effects of the different selection function of the
  photometric and spectroscopic catalogues}\label{zade_Hband}

The strength of the ZADE method resides in the fact that the
spectroscopic sample, when it is a random subsample of the photometric
one, has the same clustering properties as the photometric sample,
and so it is the perfect data set to anchor the $z_p$. In the case of
Euclid, the spectroscopic and photometric samples are selected with
two different selection functions, and so they are not supposed to
have the same clustering properties.  This is shown in
Fig.~\ref{clustering_all}, were we plot the value of $\xi(s)$ for the
Euclid spectroscopic and photometric samples.

In this Appendix we show which are the differences, in the density
field reconstruction, when we use a spectroscopic sample like the one
expected in Euclid and another one that is a random subsample of the
photometric one (but with same number of galaxies as in the \dsmop).

We extracted from the \drmosp a random galaxy subsample with the same
redshift distribution $n(z)$ as the \dsmop. In these new catalogues,
the redshift of the galaxies mimics the Euclid spectroscopic redshift,
i.e. it corresponds to the cosmological redshift, plus the peculiar
velocity, plus the spectroscopic measurement error as in the \dsmop.
We applied again ZADE on the \dpmos catalogues, but using these new
catalogues as spectroscopic skeleton, and then we estimated $\delta_N$
using the new ZADE output.

\begin{figure} \centering
\includegraphics[width=7cm]{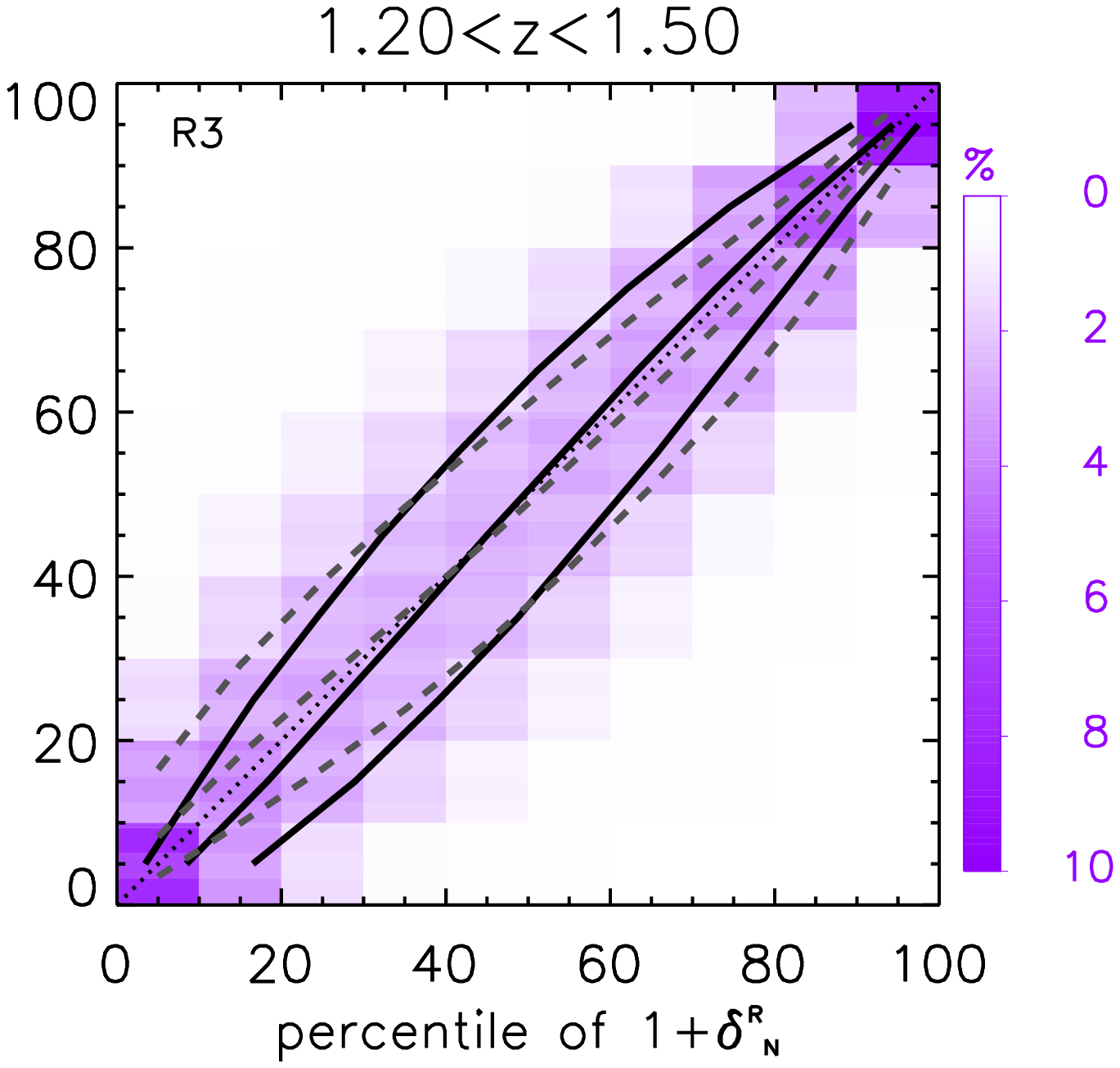}
\caption{As in the top right panel of Fig.~\ref{tails_deep1}, but in this case
the spectroscopic catalogue used for ZADE is a random subsample of the
photometric catalogue. }
\label{hband_deep_mat} 
\end{figure}

\begin{figure} \centering
\includegraphics[width=6cm]{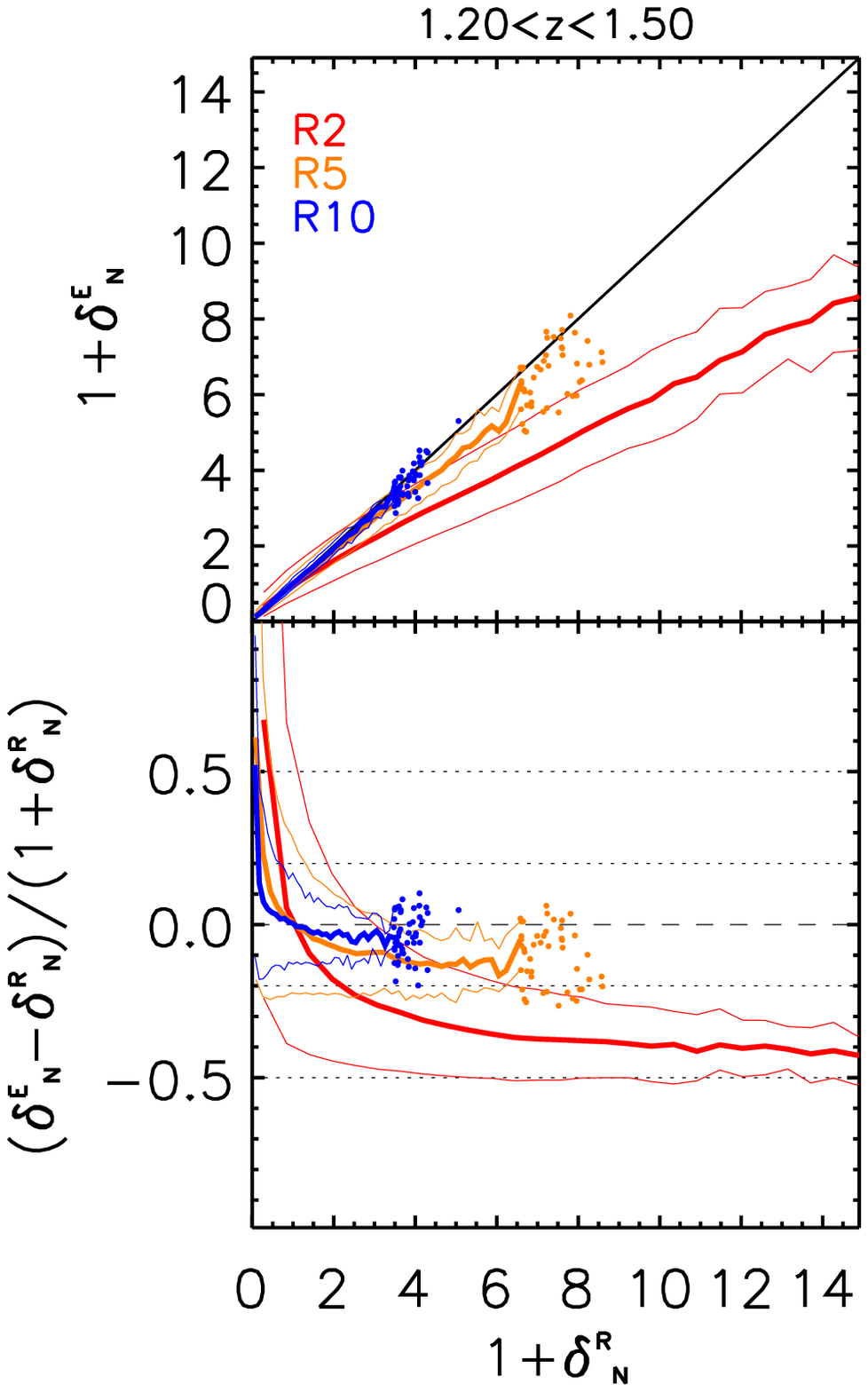}
\caption{As in Fig.~\ref{delta_deep}, but in this case
the spectroscopic catalogue used for ZADE is a random subsample of the
photometric catalogue. }
\label{hband_deep} 
\end{figure}

Fig.~\ref{hband_deep_mat} shows the joint probability (only for $R_3$)
and Fig.~\ref{hband_deep} the cell-by-cell comparison.
Fig.~\ref{hband_deep} shows that if we use these new spectroscopic
catalogues we obtain a better reconstruction of the density field for
the over-dense regions, namely obtaining a much smaller systematic
error (with respect to Fig~\ref{delta_deep}). The results for the
other radii and redshift bins are very similar.  In particular, for
$R>5\mpcoh$ the systematic error in the density reconstruction is very
close to zero for the over-dense regions when using these new
spectroscopic catalogues.

\section{The effects of the variation of the flux limit of the spectroscopic survey}\label{zade_req}

\begin{figure} \centering
\includegraphics[width=7cm]{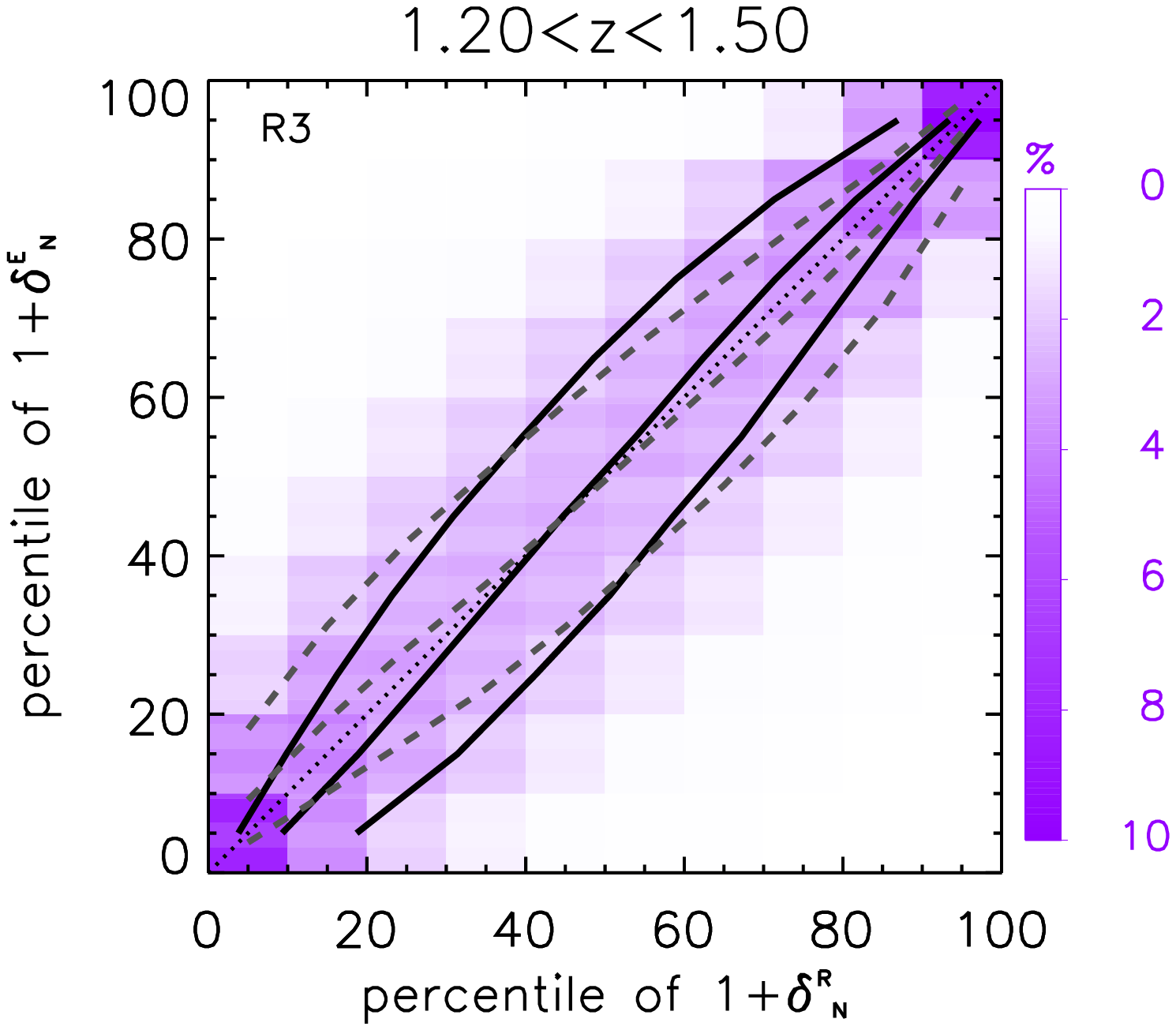}
\caption{As the top right panel of
   Fig.~\ref{tails_deep1}, but in this case the
  \dsmosp have an H$\alpha$ flux limit of $10^{-16}$
  erg~cm$^{-2}$~s$^{-1}$. }
\label{flux10_deep_mat} 
\end{figure}

\begin{figure} \centering
\includegraphics[width=6cm]{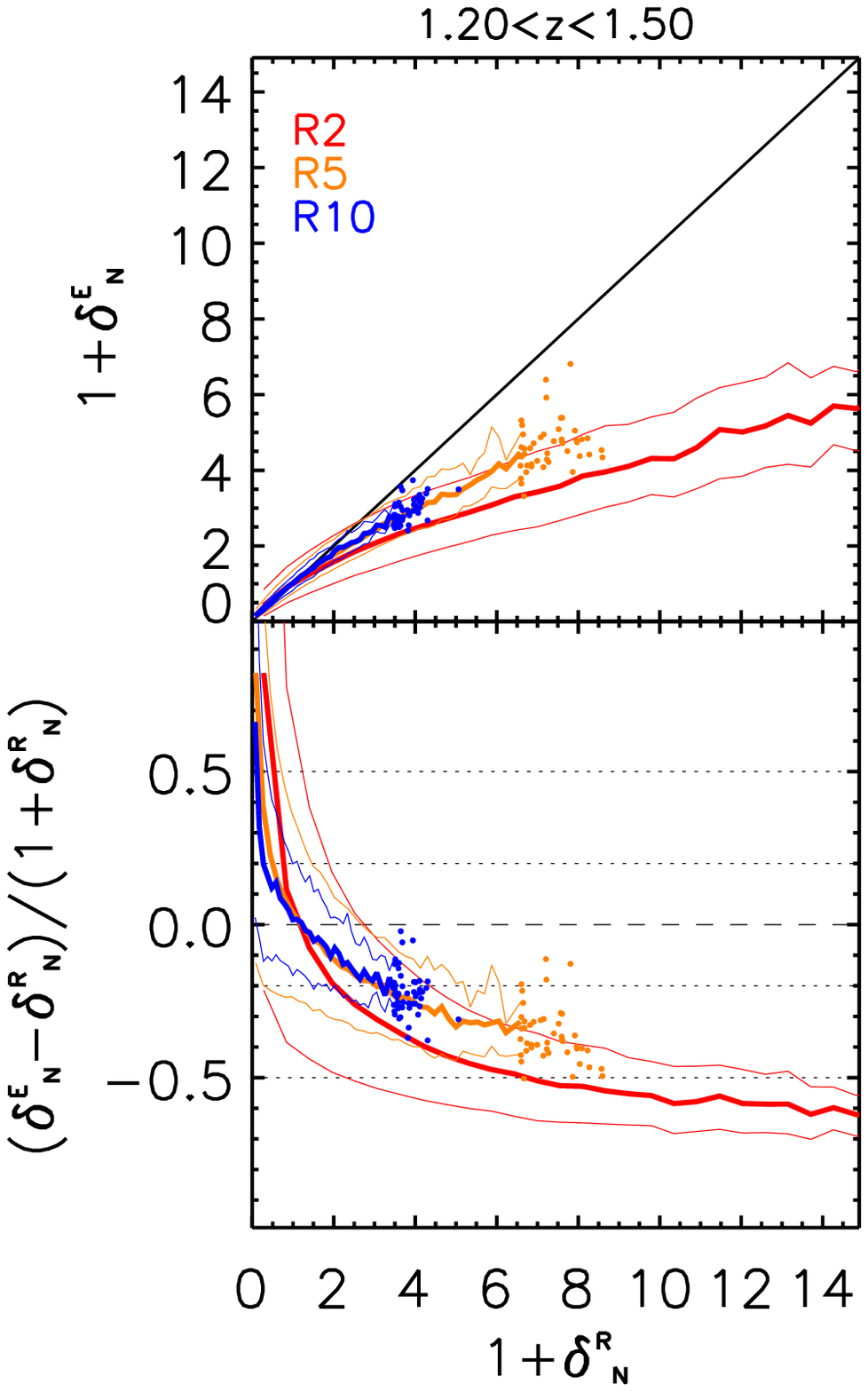}
\caption{As in 
  Fig.~\ref{delta_deep}, but in this case the \dsmosp have an H$\alpha$
  flux limit of $10^{-16}$ erg~cm$^{-2}$~s$^{-1}$. }
\label{flux10_deep} 
\end{figure}

In this Appendix we show how our results would change if we used a
more conservative flux cut in our \dsmop, i.e. $10^{-16}$
erg~cm$^{-2}$~s$^{-1}$ instead of $7\times10^{-17}$
erg~cm$^{-2}$~s$^{-1}$ (Fig.\ref{flux10_deep_mat} and
Fig.\ref{flux10_deep}). This brighter cut corresponds to the flux
limit of the Deep survey at $\sim7\sigma$. To create these brighter
\dsmosp we used the same procedure as for our original \dsmop,
including the mimicking of 2\% of purity and completeness.

Comparing the results in Fig.\ref{flux10_deep} with those in
Fig.~\ref{delta_deep}, we can see that the density field
reconstruction is only slightly hampered by the use of the brighter
cut. Namely, the brighter cut causes a slightly larger random error in
the density reconstruction, while the systematic error remains
similar. This result holds at all explored redshifts, and also for the
other cylinder radii not showed in Fig.\ref{flux10_deep}.

\section{Density field reconstruction in other redshift bins}\label{zade_otherz}

\begin{figure} \centering
\includegraphics[width=4.1cm]{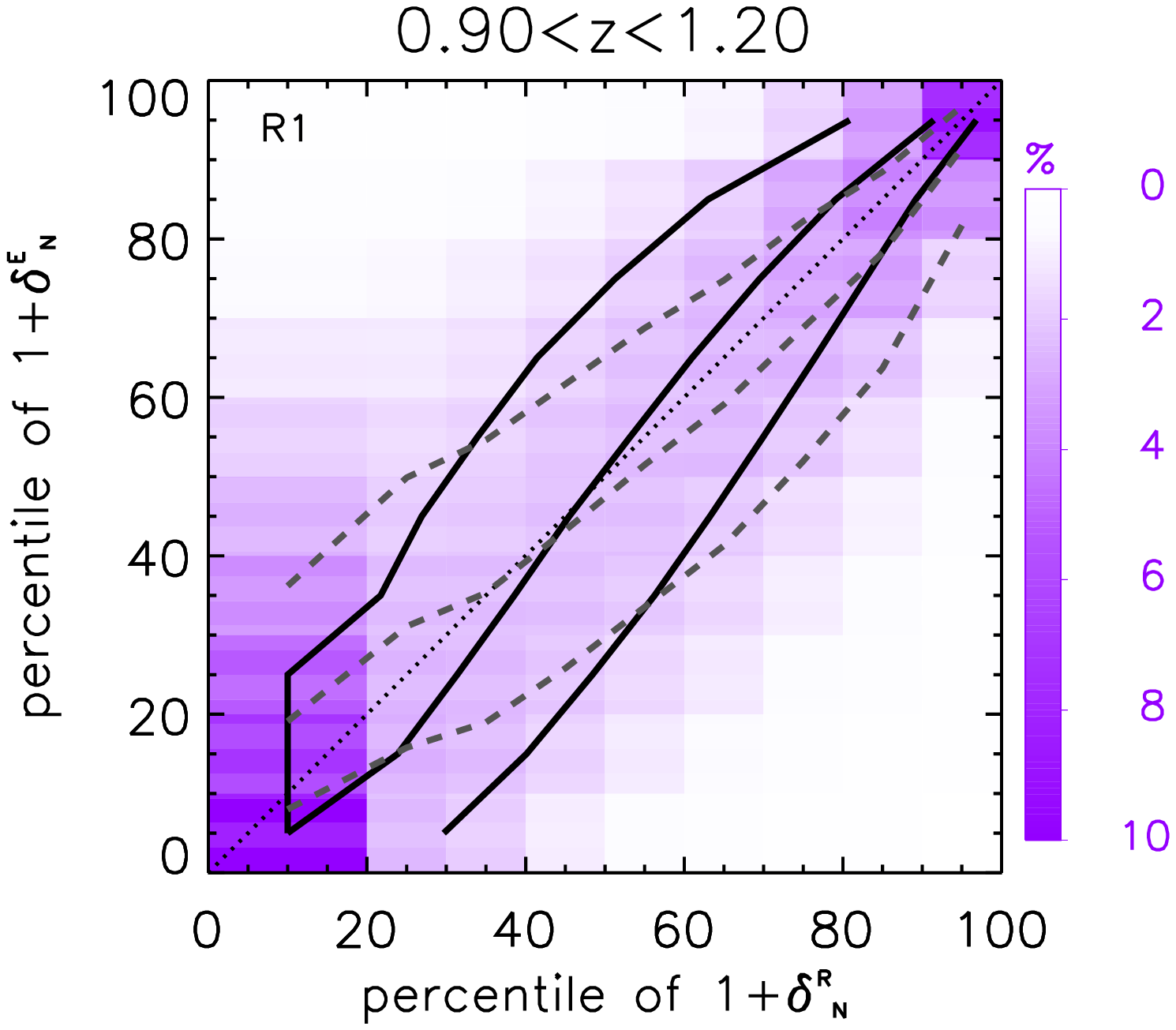}
\includegraphics[width=4.1cm]{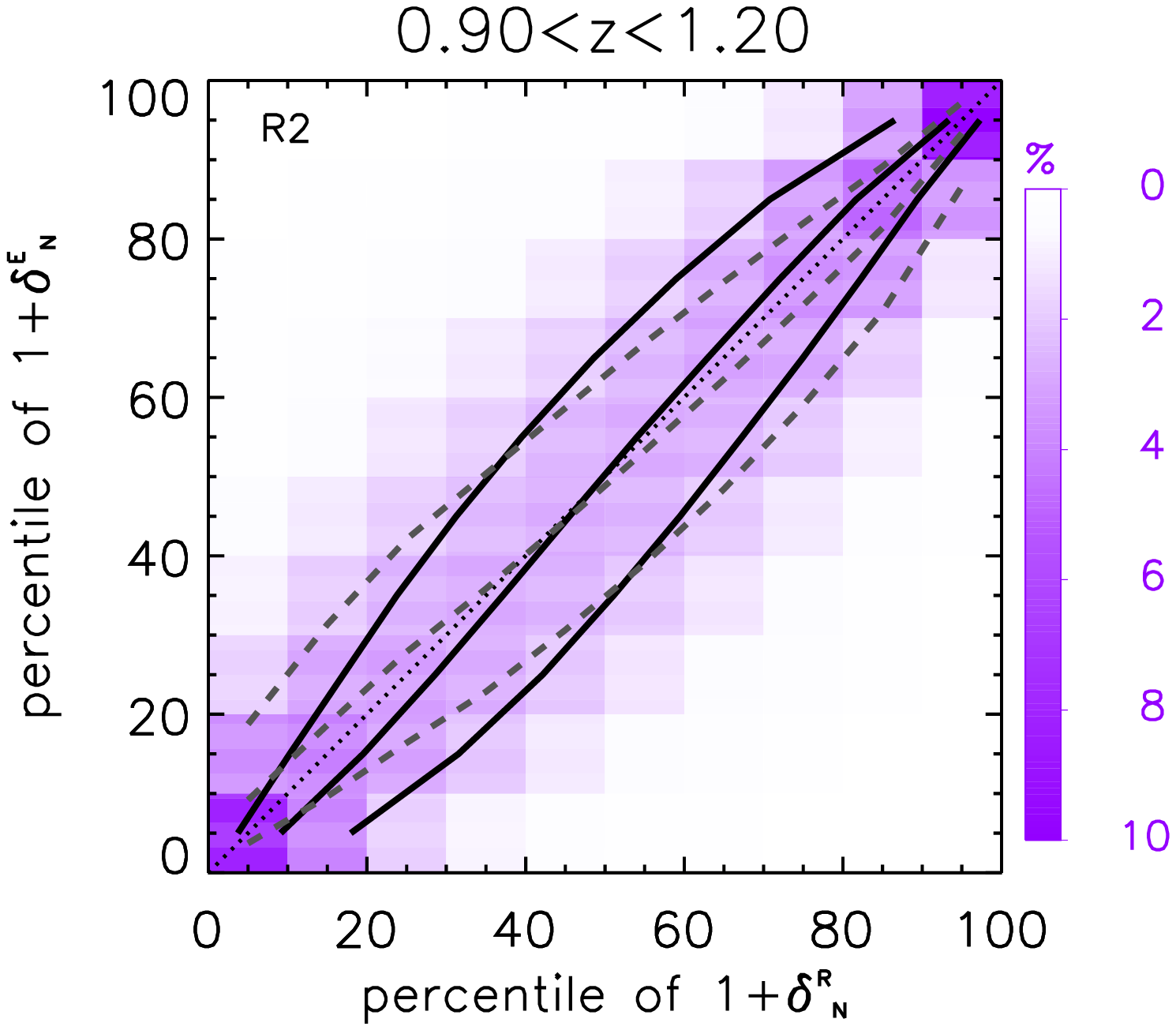}
\includegraphics[width=4.1cm]{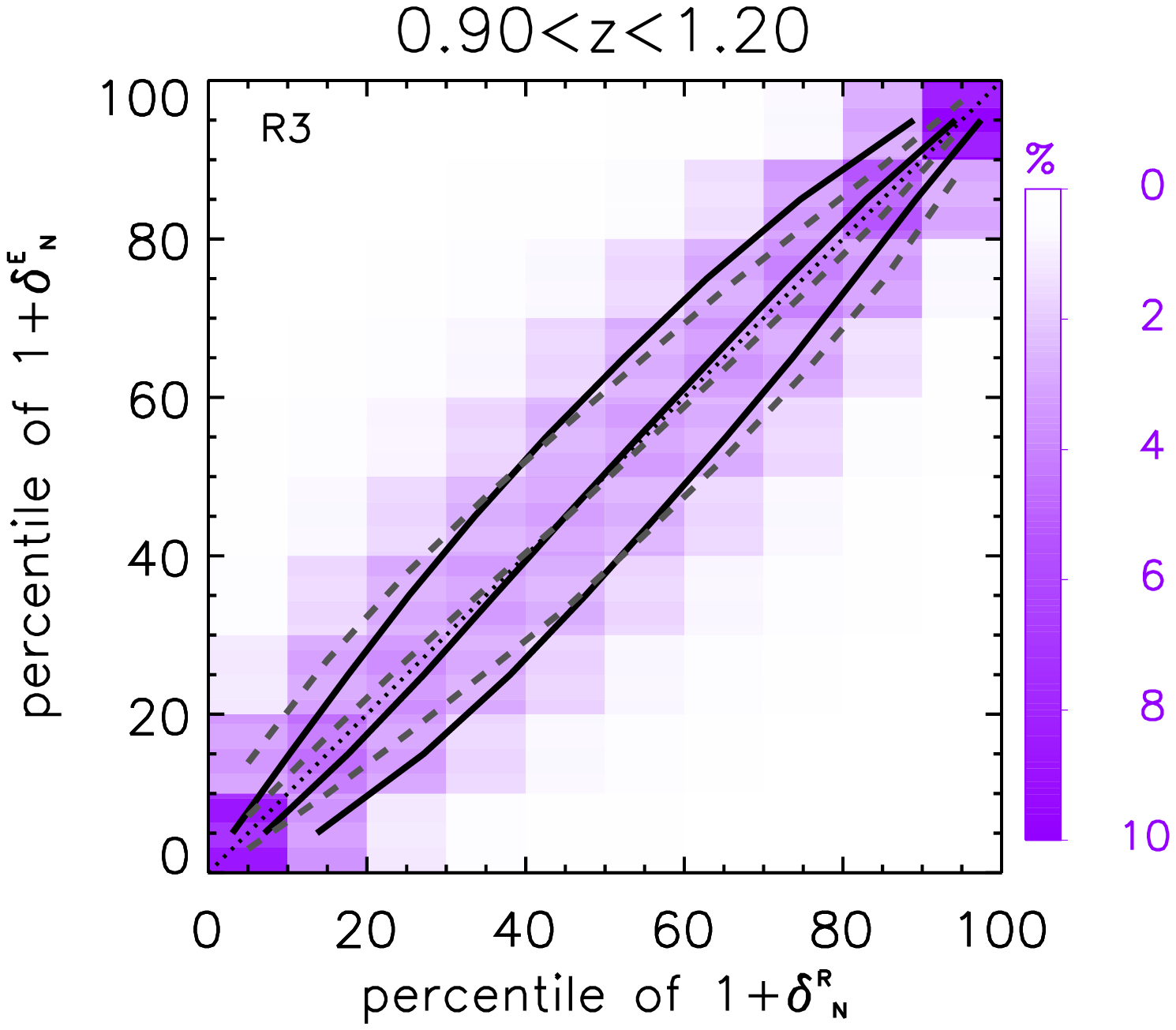}
\includegraphics[width=4.1cm]{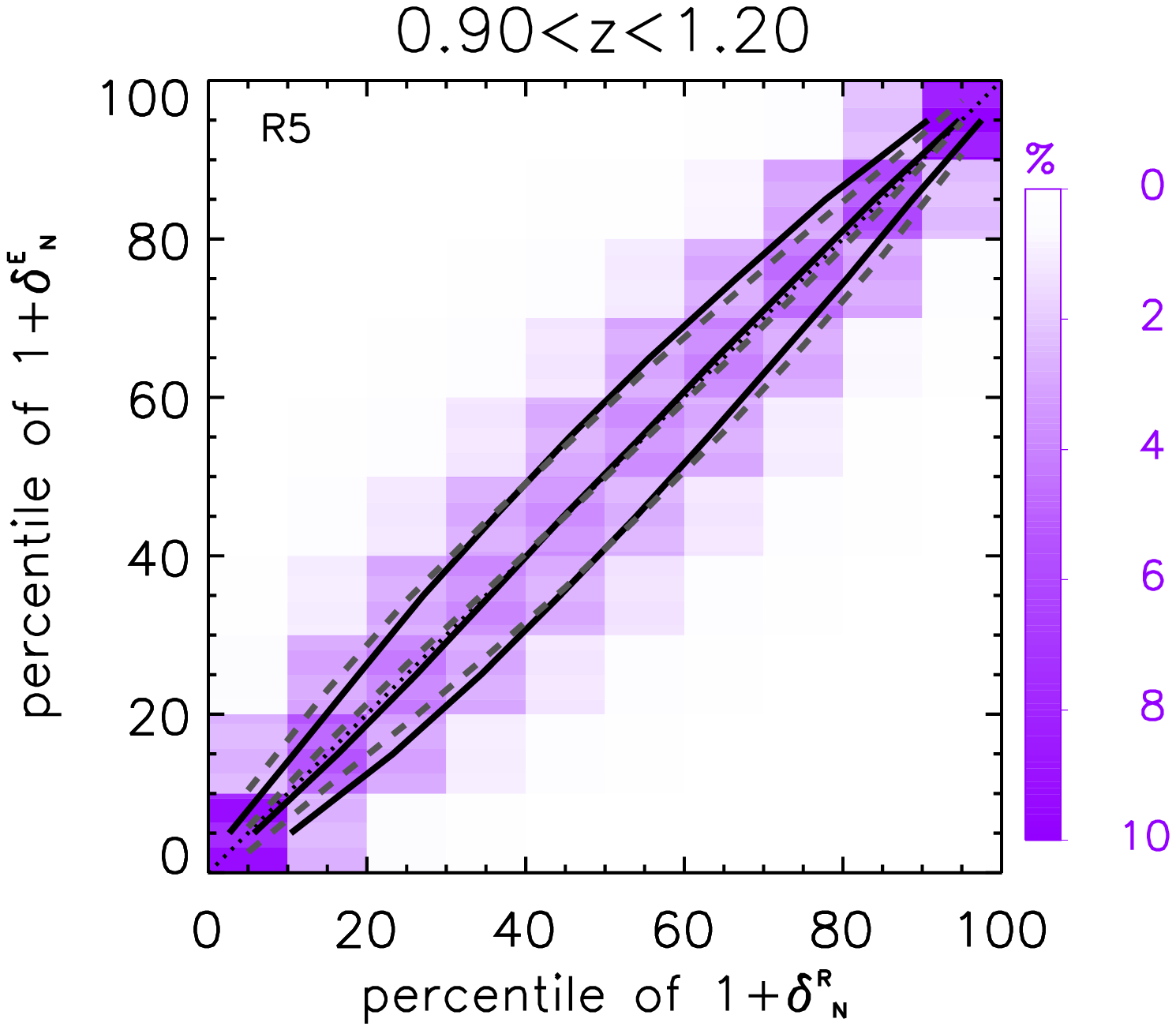}
\includegraphics[width=4.1cm]{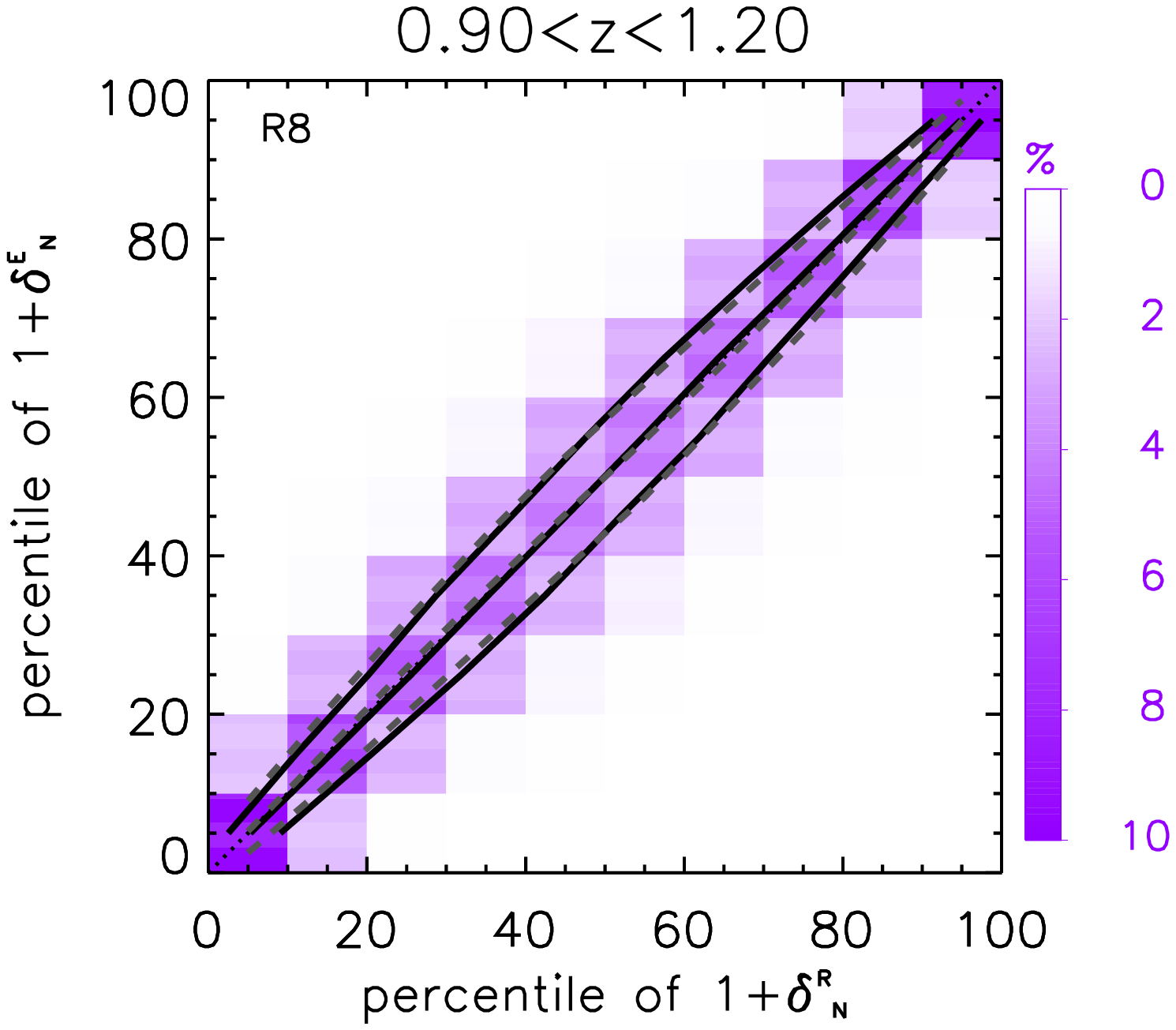}
\includegraphics[width=4.1cm]{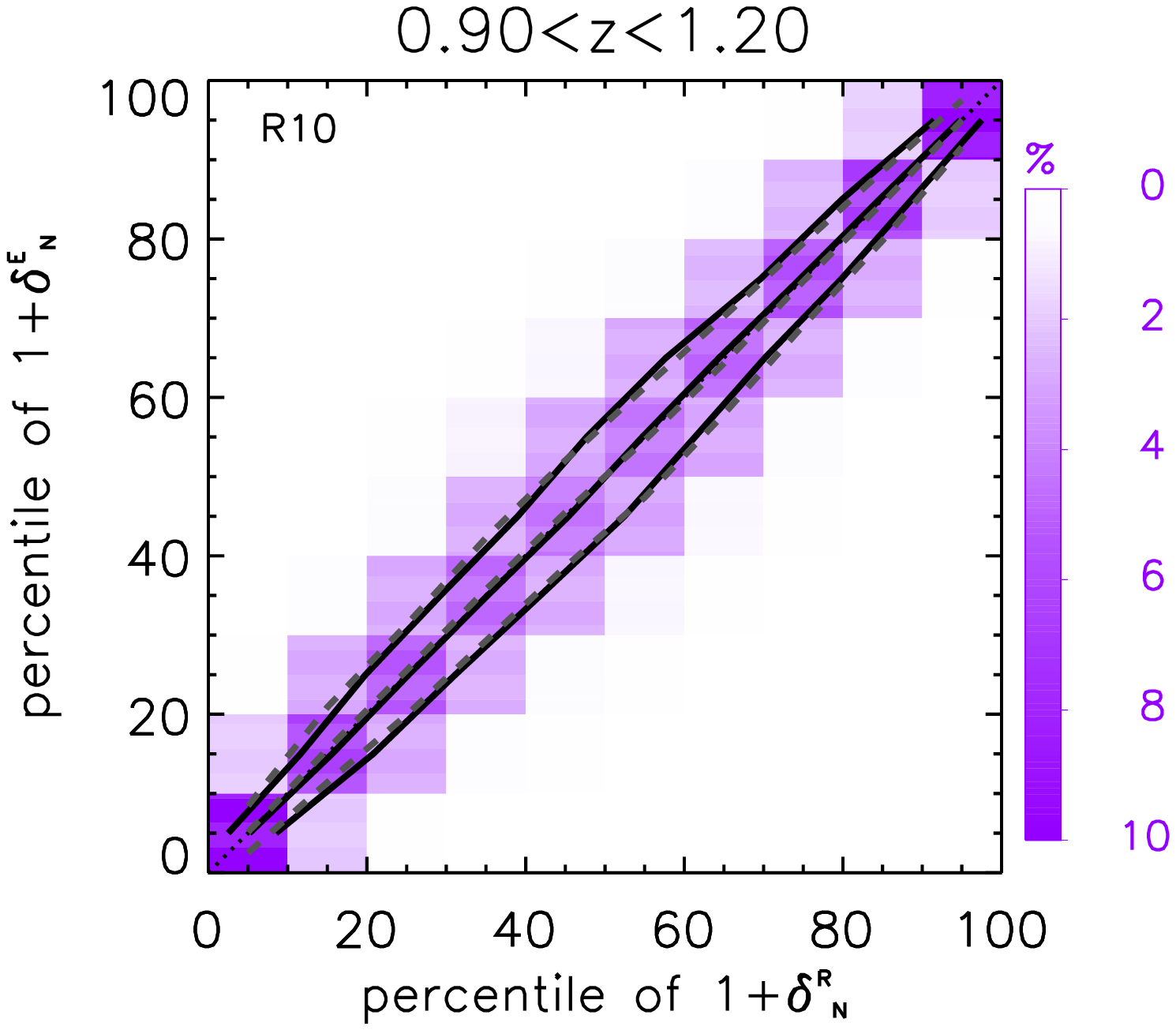}
\caption{As in Fig.~\ref{tails_deep1}, but for the redshift range $0.9<z<1.2$.}    
\label{tails_deep1_otherz} 
\end{figure}

\begin{figure} \centering

\includegraphics[width=4.1cm]{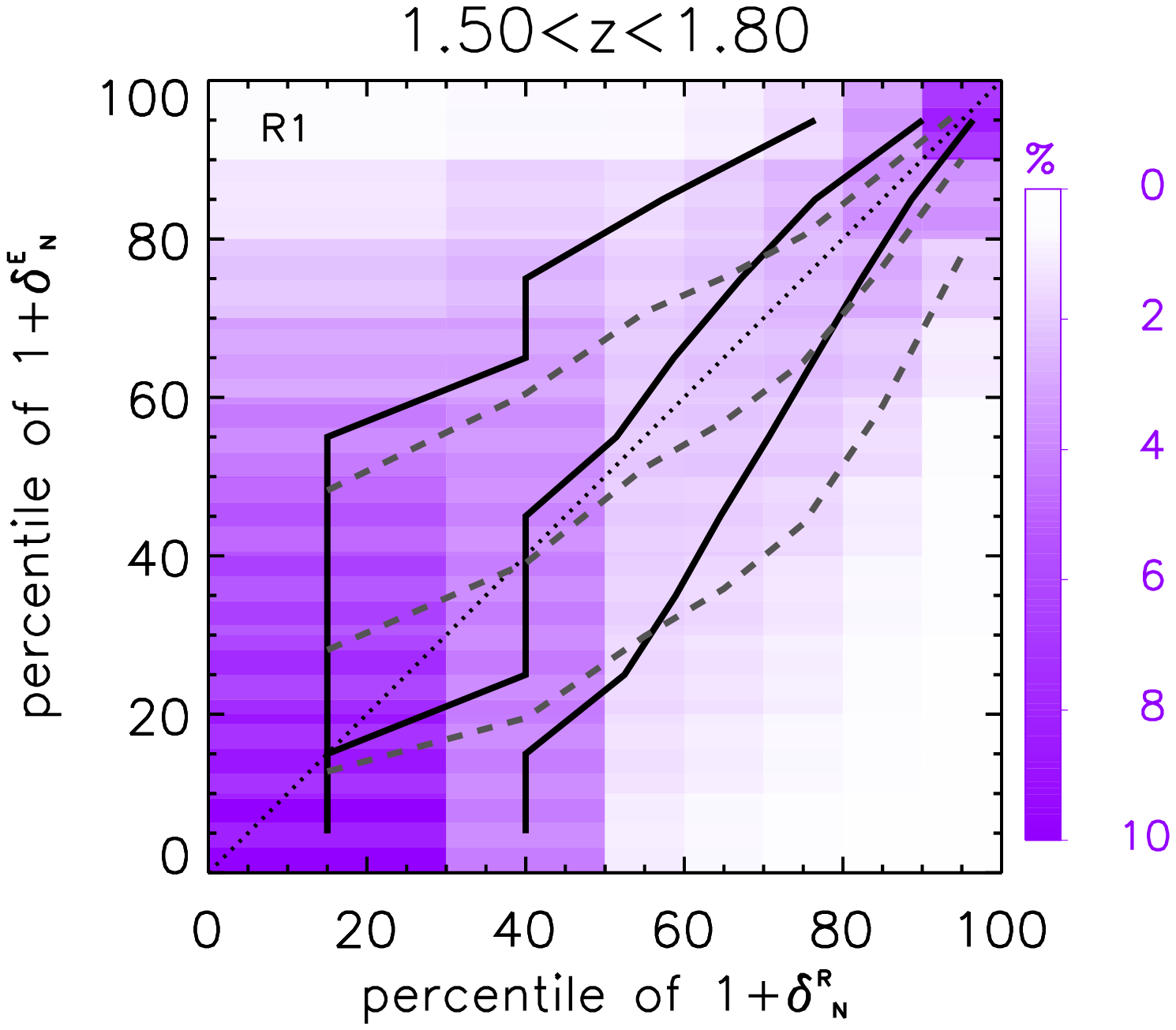}
\includegraphics[width=4.1cm]{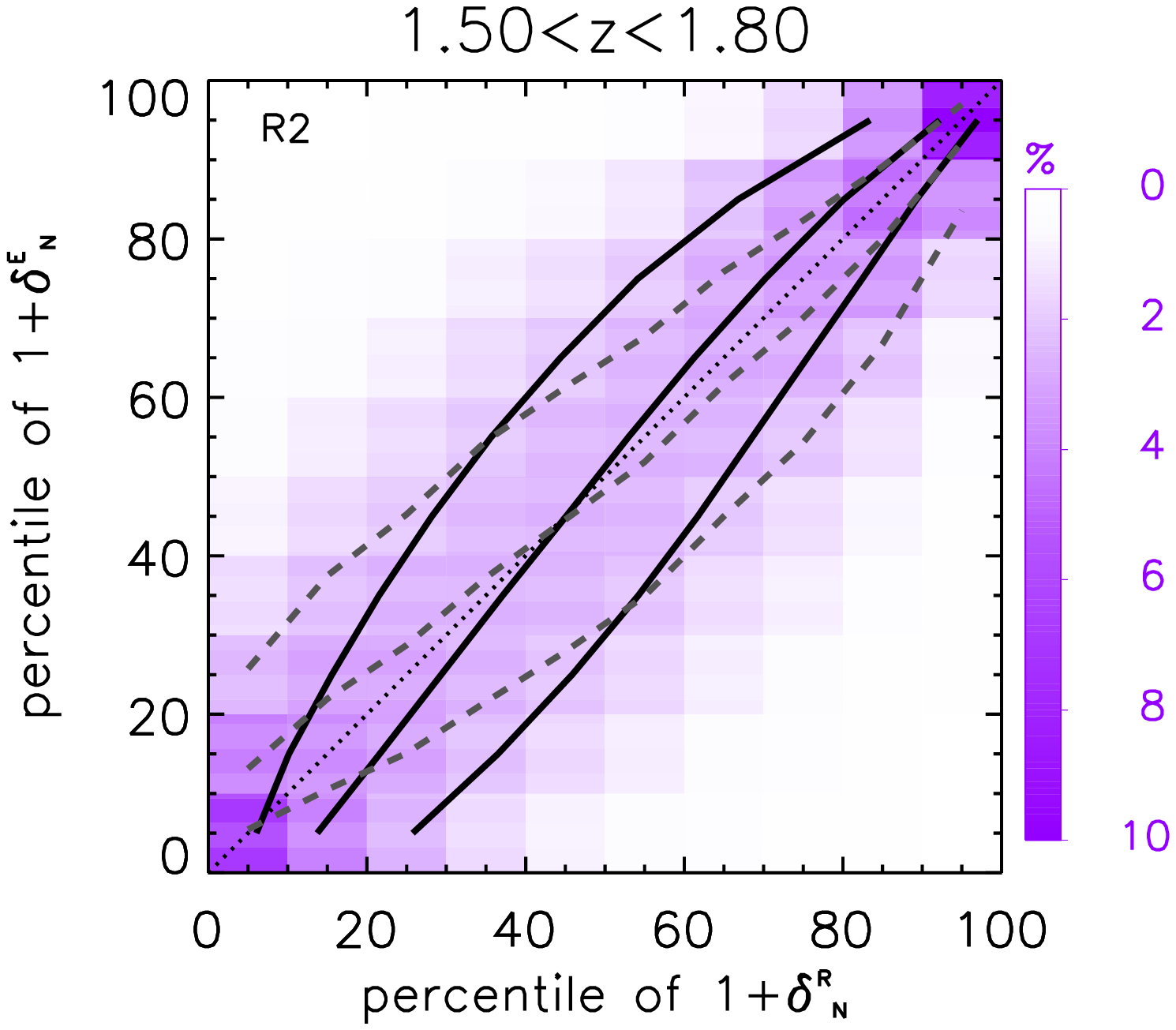}
\includegraphics[width=4.1cm]{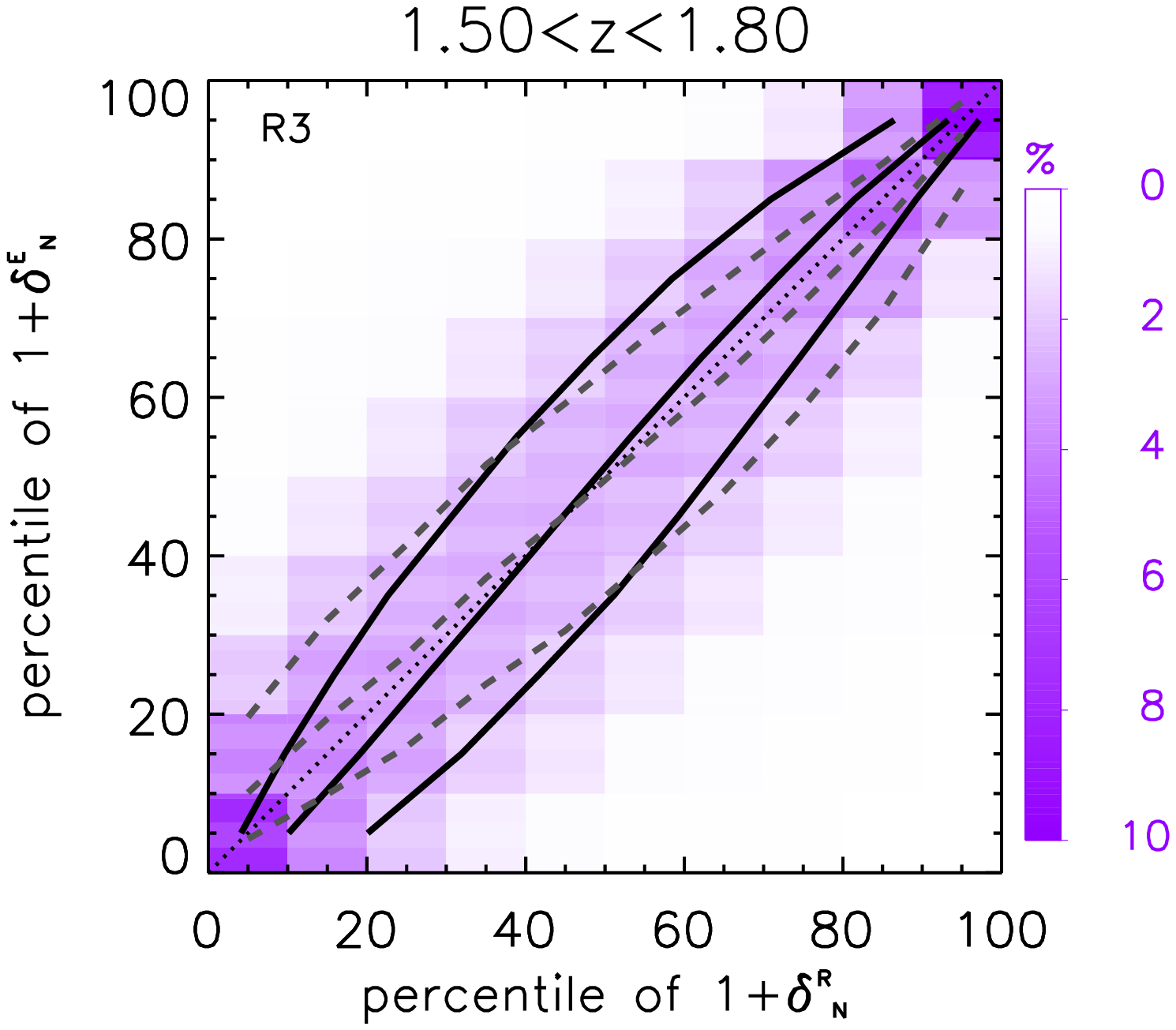}
\includegraphics[width=4.1cm]{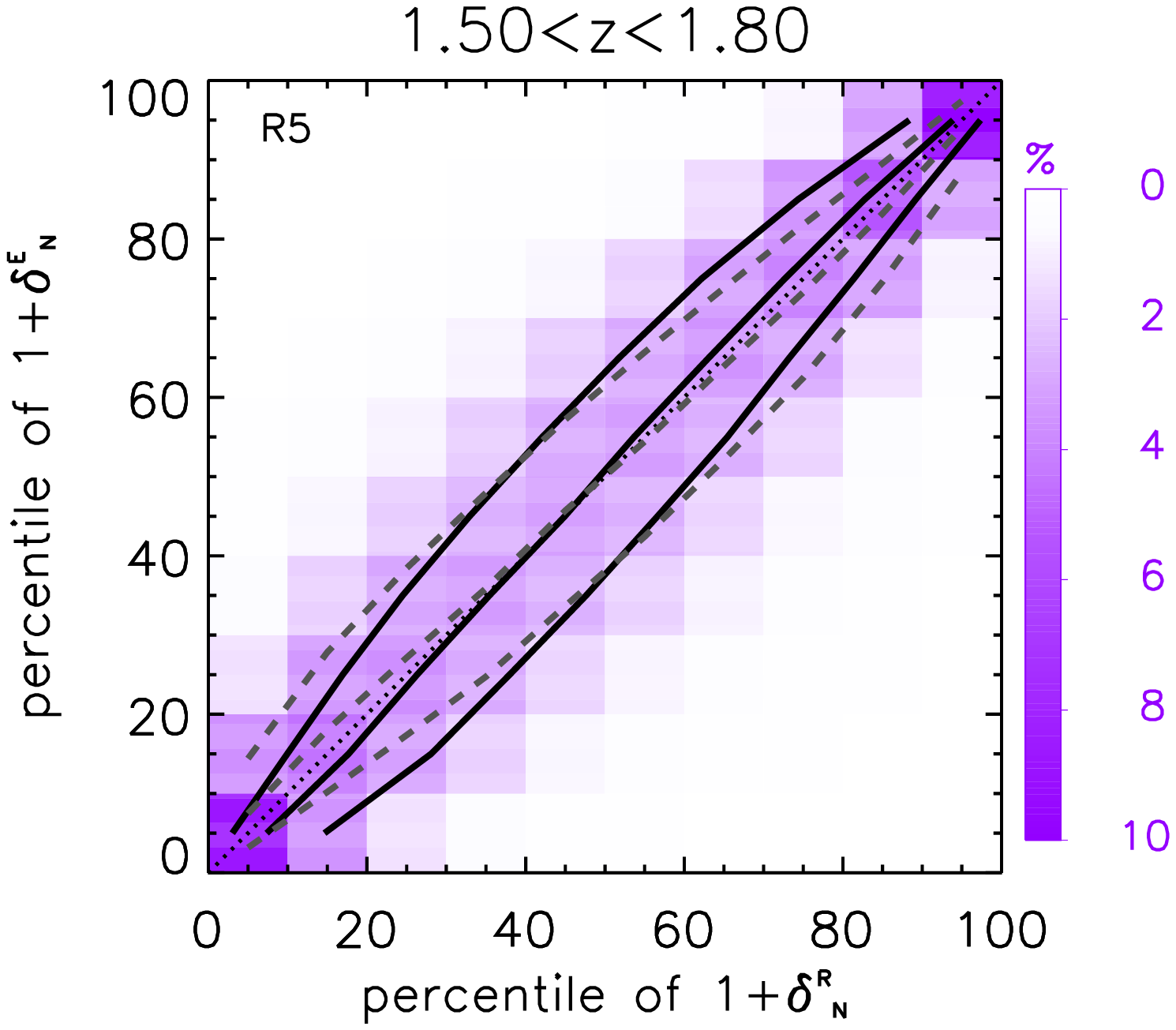}
\includegraphics[width=4.1cm]{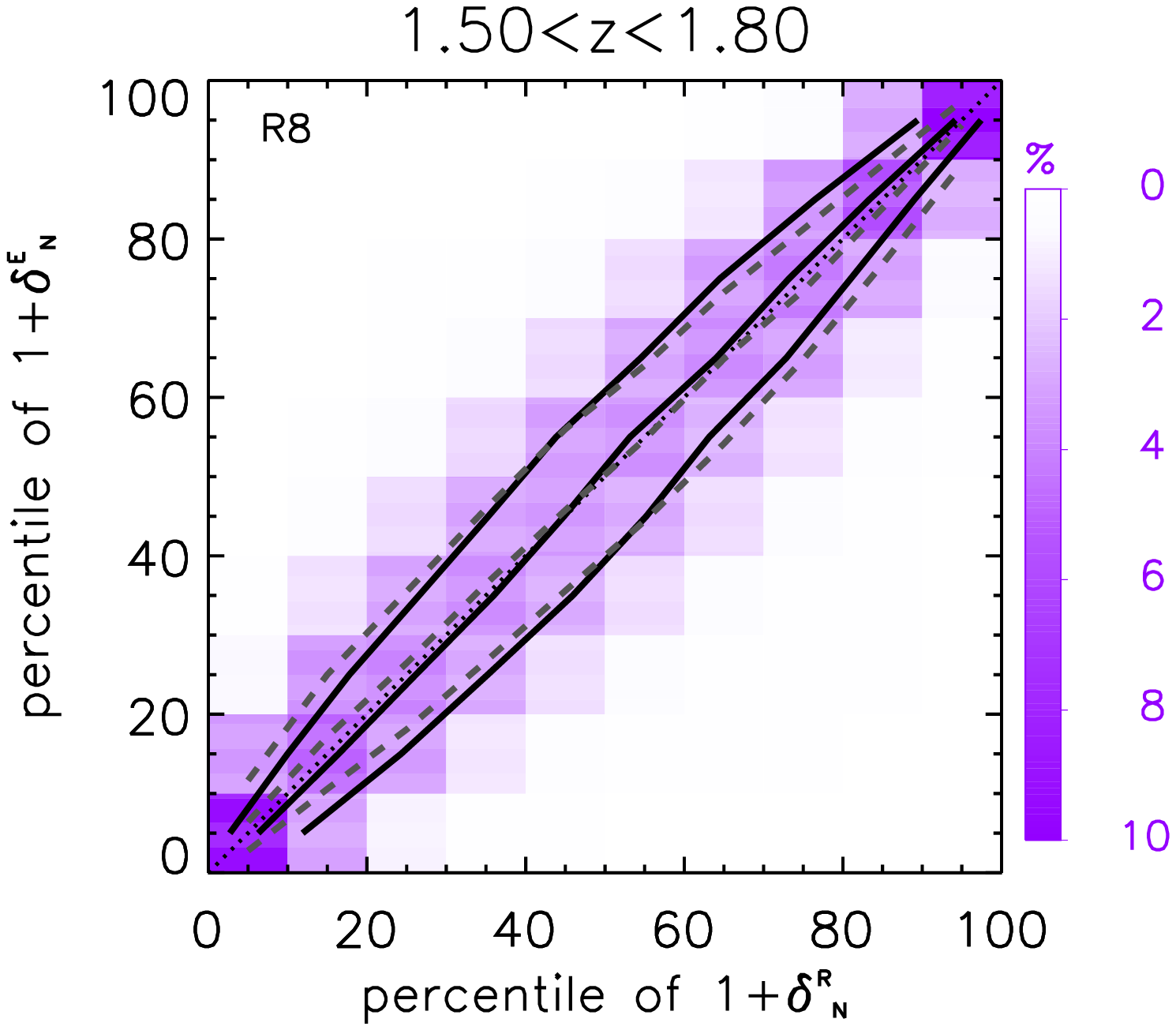}
\includegraphics[width=4.1cm]{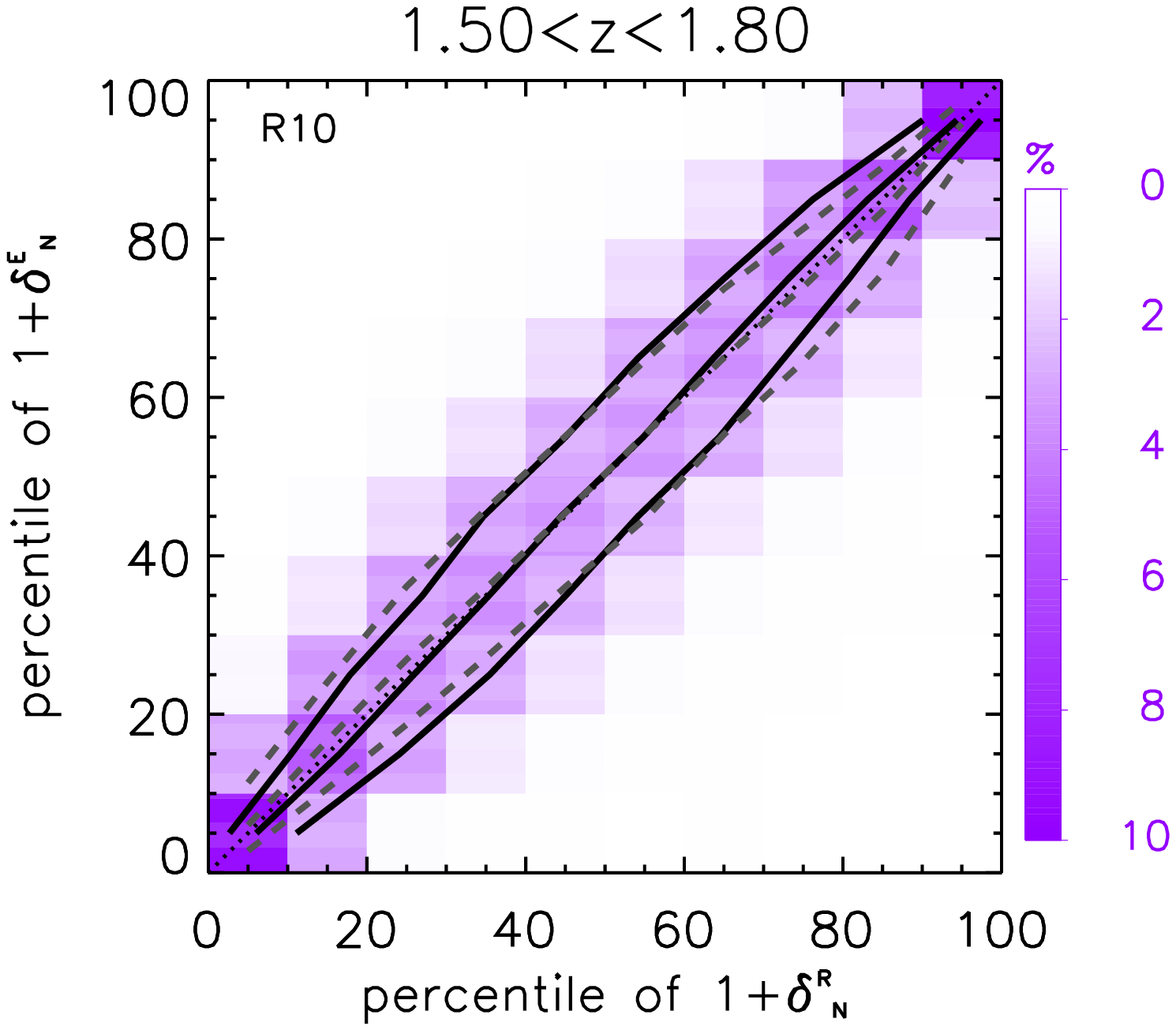}
\caption{As in Fig.~\ref{tails_deep1}, but for the redshift range  $1.5<z<1.8$.}  
\label{tails_deep2_otherz} 
\end{figure}

\begin{figure} \centering
\includegraphics[width=4.1cm]{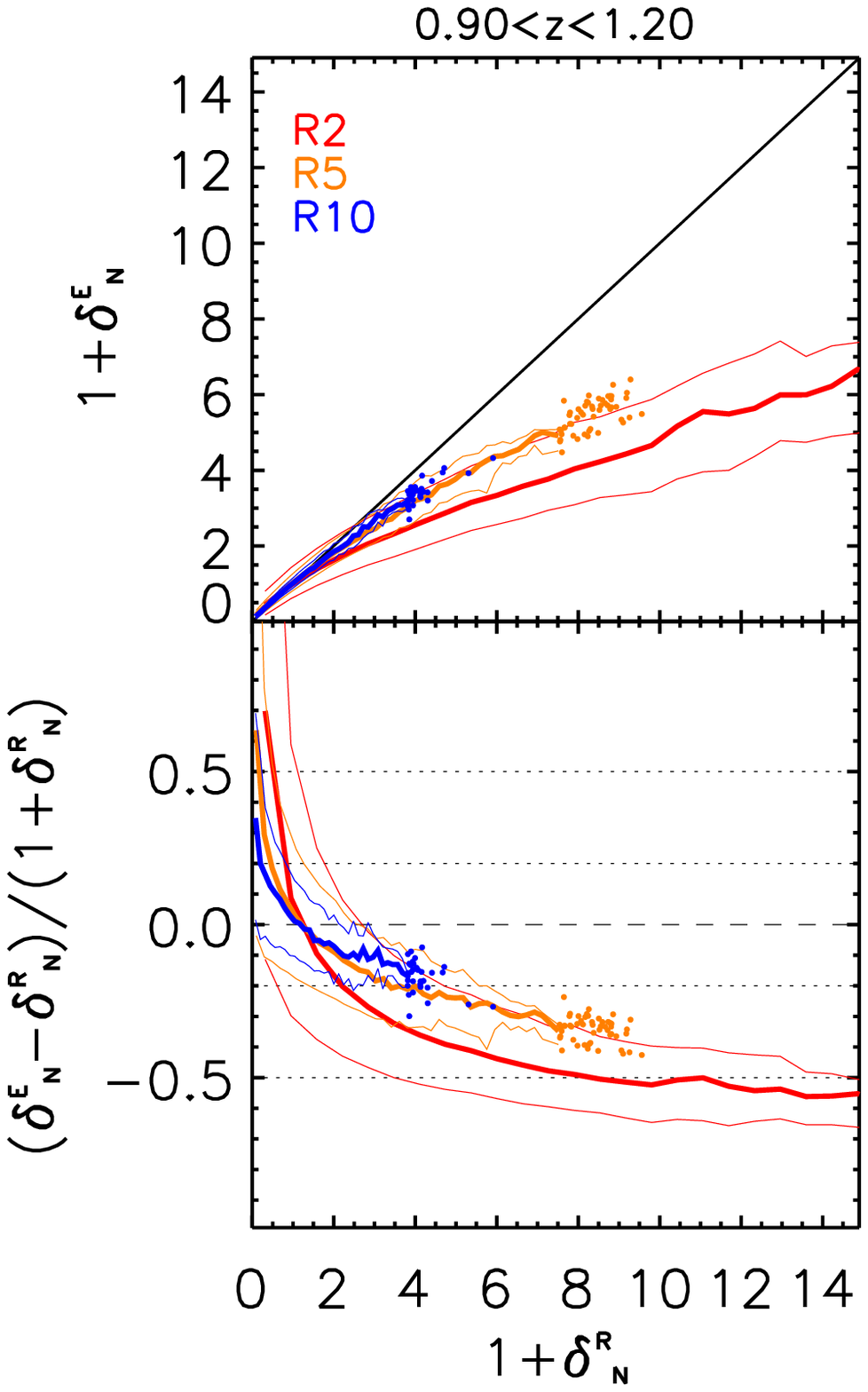}
\includegraphics[width=4.1cm]{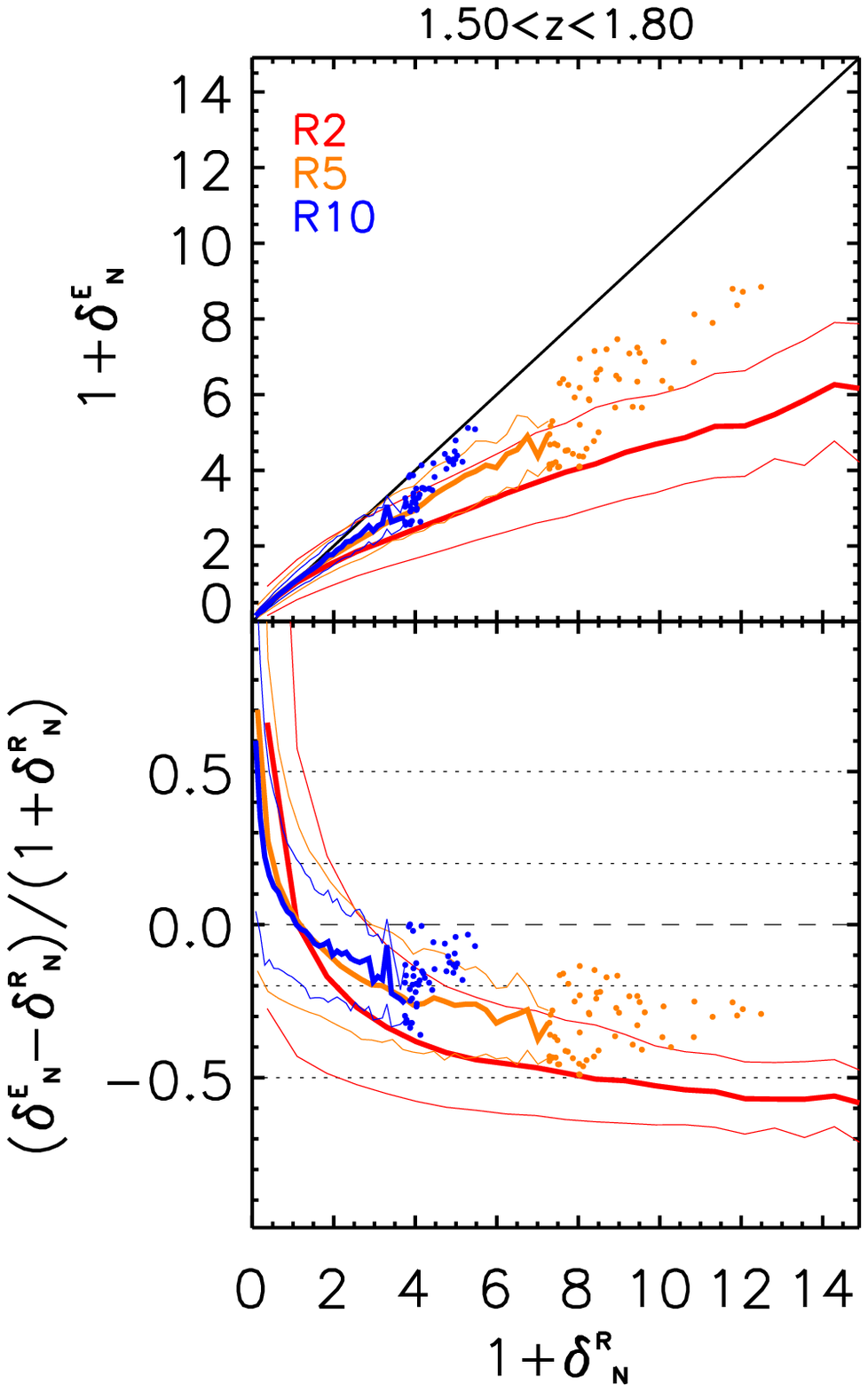}
\caption{As in Fig.~\ref{delta_deep}, but for the redshift 
bins $0.9<z<1.2$ (left) and $1.5<z<1.8$ (right).}
\label{delta_deep_otherz} 
\end{figure}

We show here our results on the density field reconstruction in the
redshift ranges $0.9<z<1.2$ and $1.5<z<1.8$ when using our original
\dsmop.  Figs.~\ref{tails_deep1_otherz} and
\ref{tails_deep2_otherz} are the same as Fig.~\ref{tails_deep1}, and 
Fig.~\ref{delta_deep_otherz} is the same as
Fig.~\ref{delta_deep}, but for these different redshift ranges. Results
are discussed in Sect.~\ref{low_high} and \ref{zade_results},
respectively.

\bsp	
\label{lastpage}
\end{document}